\documentclass[%
 reprint,
 superscriptaddress,
 groupedaddress,
 nofootinbib,
 amsmath,amssymb,
 aps,
 prx,
floatfix,
]{revtex4-2}
\usepackage{amsfonts}
\usepackage{amsthm}
\usepackage{mathtools}
\usepackage{physics}
\usepackage{graphicx}
\usepackage{bm} 
\usepackage{bbold} 
\usepackage{microtype}
\usepackage{enumerate}
\usepackage[hyperfootnotes=true,linkcolor=blue,menucolor=black,colorlinks=true]{hyperref}
\usepackage[capitalise]{cleveref}
\usepackage[acronym,index]{glossaries}
\usepackage{dsfont}
\usepackage{type1cm}
\glsdisablehyper

\newacronym{nisq}{NISQ}{noisy intermediate-scale quantum}
\newacronym{ann}{ANN}{artificial neural network}
\newacronym[longplural={associative memories}]{am}{AM}{associative memory}
\newacronym[longplural={quantum associative memories}]{qam}{QAM}{quantum associative memory}
\newacronym{qrc}{QRC}{quantum reservoir computing}
\newacronym{ml}{ML}{machine learning}
\newacronym{dv}{DV}{discrete variable}
\newacronym{cv}{CV}{continuous variable}
\newacronym{nmse}{NMSE}{normalized mean-square error}

\newcommand{\opa}{\hat{a}}
\newcommand{\opad}{\hat{a}^\dagger}
\newcommand{\opn}{\hat{n}}
\newcommand{\disp}[1]{\mathcal{D}[#1]}

\begin{document}

\title{Neural networks with quantum states of light}

\author{Adrià Labay-Mora}
\email{alabay@ifisc.uib-csic.es}
\thanks{These two authors contributed equally to the work.}
\author{Jorge García-Beni}
\thanks{These two authors contributed equally to the work.}
\author{Gian Luca Giorgi}
\author{Miguel C. Soriano}
\author{Roberta Zambrini}
\email{roberta@ifisc.uib-csic.es}

\affiliation{Institute for Cross-Disciplinary Physics and Complex Systems (IFISC) UIB-CSIC, Campus Universitat Illes Balears, 07122 Palma de Mallorca, Spain.}

\begin{abstract}
Quantum optical networks are instrumental to address fundamental questions and enable applications ranging from communication to computation and, more recently, machine learning. In particular, photonic artificial neural networks offer the opportunity to exploit the advantages of both classical and quantum optics. Photonic neuro-inspired computation and machine learning have been successfully demonstrated in classical settings, while quantum optical networks have triggered breakthrough applications such as teleportation, quantum key distribution and quantum computing. We present a perspective on the state of the art in quantum optical machine learning and the potential advantages of artificial neural networks in circuit designs and beyond, in more general analogue settings characterised by recurrent and coherent complex interactions. We consider two analogue neuro-inspired applications, namely quantum reservoir computing and quantum associative memories, and discuss the enhanced capabilities offered by quantum substrates, highlighting the specific role of light squeezing in this context.
\end{abstract}

\maketitle

Our understanding of the physical properties of light has been transformed by the advent of quantum mechanics in the early 20th century and has played a key role in both the first and second quantum revolutions. The theoretical framework of quantum light \cite{loudon1973quantum} has been growing and experimentally tested in light-matter interaction, with states of light with no classical analogue, enabling several quantum technologies \cite{obrien2009photonic,photQtech2023,Pelucchi2022}. Quantum optics has played an important role in testing the foundations of quantum mechanics as the particle-wave duality and non-locality \cite{shadbolt2014testing}. The first observed non-classical state was demonstrated in quantum optics for the predicted phenomenon of squeezed light \cite{loudon1987squeezed}. This was achieved in the 1980s through experiments that showcased the squeezing of quantum fluctuations in one of the quadratures of the electromagnetic field below the vacuum level, leading to reduced noise in that particular quadrature \cite{slusher1985observation}. Squeezed light has since become a foundational element in numerous applications, playing a crucial role in fields such as precision measurements, quantum information processing, and quantum metrology \cite{Vittorio2004}. For instance, squeezed vacuum states were recently employed in the LIGO interferometer to mitigate the effect of noise on the readout photodetectors, which resulted in a broadband detection enhancement from tens of hertz to several kilohertz \cite{PhysRevX.13.041021}. Quantum optics experiments have also proven instrumental in addressing fundamental questions about the nature of reality, locality, and causality. For instance, tests of Bell's inequalities with photons have been instrumental in laying the grounds for the study of the non-local nature of quantum entanglement \cite{Aspect1982}. 

In the last two decades, single-mode and two-component systems have been successfully scaled to multipartite quantum states, where each mode of the electromagnetic field represents a quantum node that can be coupled with others (via light-matter interaction) to form a complex network. This is the field of multimode quantum optics, which is paving the way to optical quantum networks \cite{multimode}. Successful implementations of these more complex multi-mode quantum light states triggered pioneering applications in quantum teleportation, quantum key distribution, quantum computing and quantum sensing \cite{obrien2009photonic,photQtech2023}. Generally, quantum technologies overcome, if not lack, direct classical counterparts, being based on inherently quantum phenomena. Still, technologies based on light quantum features, often build upon classical optical techniques, leveraging the advantages of both classical and quantum optics. For instance, while quantum techniques such as entanglement-enhanced metrology offer novel capabilities, classical metrology techniques using light-based instruments, such as interferometers, have long been employed for precise measurements. Quantum communication protocols rely on quantum states of light, but transmission using classical light signals in fibre optics is an established technology \cite{Agyrischaotic}. More recently, the potential of quantum optics for neuro-inspired \gls{ml} and \glspl{ann} started to be explored, leveraging the advantages of both classical and quantum optics. 

The scope of this work is to provide a brief perspective of the developments towards quantum optical \glspl{ann}, inspired on the one hand by successful implementations of photonic \gls{ml} in (classical) devices and on the other hand by the unique capabilities of quantum computation and simulation and advances in optical quantum networks. These advances in quantum complex networks and photonic \gls{ml}  set the ground for recent proposals of quantum optical \glspl{ann} and are briefly reviewed. The relevance of quantum optical complex networks in fundamental questions and different applications -- in communication and computation-- is presented in \cref{sec:QOnet}. Photonic \gls{ml} \textit{for} (\cref{sec:for}) and \textit{with} (\cref{sec:with}) quantum optics are then introduced, being the latter our focus. Photonic implementations of computation and \gls{ml} approaches have been successfully demonstrated in classical settings (\cref{sec:with_class}) and several advantages have been reported. Quantum implementations can enhance the capabilities of quantum neural networks, both in circuits and in more general (analogue) settings characterised by recurrent and coherent complex interactions  (\cref{sec:with_quantum}). Two examples of application, namely \gls{qrc} and \gls{qam}, are discussed in \cref{sec:squeez}, highlighting the role of squeezing in enhancing their performance.

\section{Quantum optical networks} \label{sec:QOnet}

\glsresetall

Observations of phenomena such as squeezing, antibunching and entanglement have been instrumental in establishing the fundamental differences between classical and quantum systems, and thus the success and importance of quantum optics \cite{loudon1973quantum}.
While originally considered for one or a few sub-systems, such as two entangled polarisation or frequency components in parametric down-conversion, optical implementations are well suited for scaling to larger multi-mode configurations \cite{loudon1973quantum,multimode}, providing an insightful and diverse realisation of quantum networks \cite{nokkala2023complex}.
The transition to large systems with many interacting (optical) components allows not only a broader approach to fundamental questions and quantum information but also breakthrough applications in quantum computing, quantum information processing, secure communications and precision measurement, as shown in the following selected examples. We will highlight different problems and implementations enabled by quantum optical networks of different nature, that can be relevant in the context of quantum \gls{ann}. A broader review of complex quantum networks beyond optical implementations is presented in Ref.~\cite{nokkala2023complex}, describing quantum dynamics in networks, network representations of states, set-ups, or dynamics (also beyond photonics).

\subsection{Fundamental questions}

(Quantum) optical networks offer a clear advantage with respect to other platforms in terms of scalability. Spatially multimode beams \cite{kolobov} and frequency combs  \cite{diddams2020optical} (broad-spectrum light beams composed of equidistant narrow lines) are outstanding examples. These light fields are very suitable to generate multiphoton entangled states that can also be used as the fundamental brick of a complex network \cite{roslund2014wavelength,reimer2016multiphoton,Cai2017}. Quantum states of light of increasing complexity \cite{brod2019photonic} can also be generated linearly by injecting photons in (large) interferometers, successfully implemented in integrated photonics \cite{wang2020integrated}. All these quantum optical networks can be used to address different fundamental questions, and in the following, we give a few examples, namely the quantumness certification of complex network states,  collective phenomena as spontaneous patterns, synchronisation or time crystals, and controllable simulation of open quantum systems.

Quantum optics is at the heart of the developments of quantum information of the last half-century. A significant challenge in the field of quantum information science is the development of effective methods for certifying the correct functioning of complex quantum devices. In essence, the issue can be summarised as the necessity to guarantee that quantum devices that would be classically intractable perform according to the predictions of quantum physics \cite{eisert2020quantum}. In the context of quantum optical networks, an efficient certification method for multimode pure Gaussian states and non-Gaussian states generated by linear-optical circuits was proposed \cite{aolita2015reliable}. Another remarkable example in this direction can be found looking at the fundamental problem of certifying the non-classicality of an entire network, going beyond the violation of Bell inequalities between one pair of parties. A proof-of-principle experimental realisation of such protocol was reported in Ref.~\cite{wang2023certification}, where full network nonlocality was shown in a photonic star-shaped network consisting of three entangled photon pairs. A proposal for self-testing all entangled states in a network was presented in Ref.~\cite{vsupic2023quantum} which is feasible with current technology.  

Optical systems also provide an outstanding platform for exploring the emergence of collective phenomena addressing the quantum aspects \cite{manzano2013synchronization}. Indeed, combining driving (e.g. by a laser field), dissipation (e.g. in cavities), and (light-matter) interactions in an open quantum system (which can be either a many-body system or a nonlinear oscillator) can give rise to a plethora of phenomena, such as quantum symmetry breaking and dissipative phase transitions. Quantum optics is the natural ground to observe spatio-temporal phenomena such as spontaneous patterns in multimode settings \cite{zambrini2000quantum},  quantum synchronization \cite{PhysRevA.85.052101,cabot2018unveiling}, metastability \cite{cabot2021metastable}, and time crystals  \cite{Else2016,Iemini2018,cabot2023nonequilibrium}). 

Complex quantum networks can also be designed to model engineered and tunable environments, enabling the simulation of a variety of open quantum system dynamics, which allows the study of fundamental issues such as dissipation, decoherence, and measurement. It was shown that typical features of open quantum system dynamics such as the spectral density or quantum non-Markovianity can be implemented in a \gls{cv} optical platform using multimode light, with squeezing and entanglement as resources \cite{vasile2014spectral,Nokkala_2018}. A proof-of-concept of this controllable dissipation implementation has been recently reported in Ref.~\cite{PRXQuantum.4.040310} where such complex networks were experimentally implemented using frequency-combs squeezed light.

\subsection{Quantum internet}

The transmission of information via light represents one of the most widely used methods for communication, with optical fibres or satellites to connect remote places. In recent times, the development of quantum computing has raised the possibility of creating a quantum internet -- an extension of the actual internet, a set of interconnected quantum devices capable of sending and storing information by using quantum mechanics \cite{wehner2018quantum,kimble2008quantum}. The change of paradigm from communication between pairs of users to a fully extended network is the necessary ingredient to make such technology appealing for industry and public institutions, beyond the implication for basic science. The quantum internet aims to be more secure than the current Internet with protocols for quantum key distribution that ensure private communication between two parties \cite{shor2000simple}. Furthermore, it enables the distribution of computation across remote devices, which has applications in quantum sensing \cite{zhang2021distributed} and quantum computation \cite{cirac1999distributed}.

This new approach introduces significant technical and intellectual challenges due to the nature of quantum mechanics and the limitations of \gls{nisq} technology. A primary challenge is photon loss in transmission channels like optical fibres, which increases exponentially with distance. Classical systems mitigate photon loss by using coherent sources and repeaters for amplification. In contrast, the quantum internet transmits qubits encoded in the amplitude or polarization of single photons, which suffer from decoherence during transmission. Additionally, qubits cannot be cloned \cite{wootters1982single}, necessitating new methods to overcome photon loss. One potential solution is the development of all-photonic quantum repeaters, which eliminate the need for matter quantum memories and achieve a communication efficiency that scales polynomially with the channel distance \cite{azuma2015all}.

Similarly, other protocols for quantum communication implemented using squeezed quantum states of light have gained interest recently due to easier experimental viability in the \gls{nisq} era \cite{lloyd2012Gaussianqi}. Compared to discrete variables quantum key distribution, \gls{cv} proposals are expected to be more efficient, attain higher rates and improve the detection using homodyne receivers as opposed to single-photon counters \cite{laudenbach2018cvqkd}. The first proposal used squeezed states \cite{cerf2001cvqkd}, which are secure \cite{gottesman2003secure} and have been shown to have improved
robustness versus the noise of the channel \cite{pirandola2009noiseqkd}.

\subsection{Quantum computation and simulation}
 
Complex quantum states emerge naturally in quantum computation and simulations both in circuit and analogue implementations \cite{deutsch1989quantum,buluta2009quantum,kendon2010quantum}. In this context, successful photonic realizations have been reported, as reviewed in Ref.~\cite{Walmsley:23,aspuru2012photonic}. While quantum computing systems such as superconducting circuits, trapped ions and silicon quantum dots are very popular in state-of-the-art quantum processors, they face challenges in achieving scalable fault tolerance due to fragile quantum states requiring cryogenic or vacuum isolation. Photonic systems, on the other hand, are intrinsically more robust and can be manipulated at room temperature. Photons offer fast propagation and large bandwidth. These properties, together with advanced photon manipulation technologies, position photonic systems as a leading approach for building quantum computers \cite{opticalqc_2007}.

One of the ways of implementing universal photonic quantum computing in \glspl{cv} is measurement-based quantum computing, where cluster states are used as a resource to perform \gls{cv} gates by applying local operations thanks to quantum teleportation protocols \cite{Gu_clusters2009}. This framework enables extremely high scalability and reconfigurability, mainly using temporal and frequency mode multiplexing \cite{furusawa_big,60_frequency_cluster,Cai2017,furusawa_2d_cluster}. Deterministic single and multi-mode gates are within current experimental reach\cite{cluster_gates_1,cluster_gates_2}, which is an advantage compared to probabilistic gates in  discrete variables schemes \cite{kozlowski2019towards}. The main limitation of \gls{cv} measurement-based quantum computing is the implementation of single cubic phase gates, which are a requirement for universality \cite{Gu_clusters2009}. Advances in hybrid discrete-continuous variable implementations have been done to address this limitation \cite{Takeda2013}.

In the search for fault-tolerant quantum computation, quantum error correction is necessary to prevent the propagation of logical errors in operations. Qubit-based quantum error correction techniques are challenging to implement due to the difficulties in scaling up the number of qubits \cite{fowler2012surface}. Photonic quantum computers also allow the implementation of efficient error correction codes that exploit the infinite-dimensional Hilbert space of \glspl{cv} to encode a qubit in single optical modes, thus simplifying the scalability of error correction \cite{cai2021reviewQEC}. Hence, by exploiting the redundancy of the Hilbert space, quantum error correction could be constructed in a single bosonic mode. The most notable ones are Gottesman-Kitaev-Preskil (GKP) \cite{campagne2020quantum}, binomial \cite{hu2019quantum} and cat codes \cite{mirrahimi2014dynamically}. In particular, cat codes can exponentially suppress bit-flip errors \cite{lescanne2020exponential} and squeezed-cat codes can enhance the protection against such errors \cite{labay2023squeezing}. 

Photonic architectures have also been successful in implementing non-universal computing tasks. Boson sampling is one of the most prominent applications of quantum photonic hardware in this regard, as photonic networks generated with large interferometers cannot be efficiently simulated by classical computers \cite{Aaronson-Arkhipov-2013}. Gaussian boson sampling represents a specialised model of photonic quantum computation \cite{Gaussian_BS}. It involves the preparation of a multi-mode squeezed state followed by measurements conducted on the Fock basis. The primary distinction from universal photonic circuits lies in the absence of non-Gaussian gates within Gaussian boson sampling and the limitation of measurements to the Fock basis. Quantum advantage has been successfully achieved in Gaussian boson sampling using specifically built photonic hardware \cite{boson-sampling-1,boson-sampling-2}, as well as in a reconfigurable platform built by Xanadu \cite{boson-sampling-3}. Quantum squeezed states have also found applications in the design of high dimensional coherent Ising machines \cite{ising-machines-1,ising-machines-2}, which are suited to solve complex combinatorial optimisation problems. Another technique to solve such problems is quantum annealing which has been seen to be more robust to noise and allows for all-to-all connectivity when qubits are encoded in the ground-states of Kerr parametric oscillators \cite{puri2017annhealing,nigg2017annhealing}. Furthermore, photonic quantum processing units were used to efficiently solve the quantum phase estimation algorithm through a variational eigensolver algorithm \cite{Peruzzo2014}.

Advances in scalable, efficient, and fast, photonic architectures enable the generation of quantum optical networks that open new frontiers also in the rapidly evolving fields of \gls{ml} and neuro-inspired computation, as presented in \cref{sec:with}. Interestingly,  the same developments in these quantum optical complex architectures have also benefited from the use of \gls{ml} methods, and some examples are given in \cref{sec:for}. 

\section{Machine learning for quantum optics} \label{sec:for}

\glsresetall

The use of machine learning techniques to enhance classical optical systems is rapidly advancing, achieving a high degree of sophistication \cite{genty2021machine, wiecha2021deep}. Machine learning is being employed in various applications, such as controlling experimental instabilities, designing novel devices with ad-hoc functionalities, and generating ultrafast optical pulses. For instance, genetic algorithms can predict and mitigate fluctuations in optical experiments, leading to more stable and reliable results \cite{fu2013high}. Additionally, deep learning techniques enable the design of optical components with customised properties, optimising performance for specific tasks \cite{ma2021deep}. Machine learning also plays a crucial role in the generation and shaping of ultrafast optical pulses, which are essential for applications ranging from telecommunications to medical imaging \cite{genty2021machine}. Overall, the integration of machine learning in classical optics enhances existing technologies and also clears the way for innovative solutions and discoveries.

As photonic hardware for quantum optics advances, optimising optical setups for specific applications becomes increasingly challenging, prompting a shift from manual design by scientists to automated methods using machine learning \cite{krenn2016automated}. Machine learning can significantly enhance quantum photonic protocols by optimising parameters, accelerating measurements, and eliminating systematic artifacts, thus enabling configurations previously untestable due to experimental limitations. Furthermore, such automatic exploration methods have already facilitated the creation of novel quantum photonic setups and the discovery of new photonic phenomena, highlighting their transformative potential in this field \cite{krenn2017entanglement}. In these approaches, quantum photonic experiments are recast into graph representations \cite{krenn2016automated}, allowing the task of finding structures in a given graph to be translated into discovering new experimental setups \cite{cervera2022design}. Ideally, these methods for automated design can also offer a better conceptual understanding \cite{krenn2021conceptual}.

Machine learning can also assist in the development of a quantum internet, highly dependent on the interplay between the various building blocks that make up the network, from the hardware used for quantum computers or transmission channels to the protocols used to distribute entanglement among distant nodes or generate secret keys \cite{kozlowski2019towards}. Finding the optimal hardware and software parameters that allow optimal connectivity is a huge computational problem, especially in the \gls{nisq} era. In this context, analytical solutions are difficult to find and one has to resort to optimisation algorithms to find more general solutions \cite{labay2023reducing,avis2023requirements}. \Gls{ml} can be used to find better protocols, optimise hardware parameters or improve the security and transmission rates of quantum key distribution \cite{mafu2024advances}. 

Another challenge in many quantum technologies is the full reconstruction (tomography) of a quantum state, being one of the most resource-consuming tasks. As shown in Ref.~\cite{Torlai2018}, trained artificial neural networks provide a simple and adaptable method for quantum state tomography, effectively utilising a limited amount of experimental data.
An example of the potential for neural networks to assist in the extraction of relevant features from a multipartite quantum state of light was recently reported in Ref.~\cite{koutny2023deep}. The degree of entanglement was quantified without the need to know the full description of the quantum state, achieving an error of up to an order of magnitude lower than the state-of-the-art quantum tomography. Finally, also quantum imaging has greatly benefited of the application of machine learning algorithms. For instance, deep neural networks have been proven effective in a diversity of imaging applications, ghost imaging, or phase retrieval, as experimentally shown in Ref.~\cite{PhysRevLett.121.243902}.

\section{Quantum neural networks} \label{sec:with}

\glsresetall

Photonic quantum technologies represent a promising opportunity to address the demand for fast processing, high performance and energy efficiency in classical and also quantum data processing. As presented in the following (\cref{sec:with_class}), classical optical systems have already enabled successful implementations of landmark results. Proposals exploring designs and applications of \glspl{ann} based on quantum states of light are then introduced, both in feed-forward circuit approaches and beyond, and in particular in \gls{qrc} and \gls{qam} (\cref{sec:with_quantum}).

\subsection{Machine learning with classical light} \label{sec:with_class}

Optical \gls{ml} and the dominant approach of \glspl{ann} harness the unique properties of light to achieve unprecedented speeds and efficiencies in information processing. Classical optical computing, which exploits principles such as light propagation and interference, is experiencing a resurgence as a powerful alternative to traditional electronic computation \cite{mcmahon2023physics}. This renewed interest is driven by significant advancements in photonic technologies and the increasing demand for more efficient and faster computing methods \cite{won2023power}.

Optical neural networks (ONNs) can exploit the inherent properties of light, such as coherence, interference, and diffraction, to implement physical analogues of \gls{ml} algorithms \cite{Shen2017,lin2018all}. Utilising the vast bandwidth of optical frequencies, ONNs enable parallel processing of large amounts of data, making them suitable for complex \gls{ml} tasks. Recent review and perspective articles highlight foundational principles and recent advancements, underscoring the transformative potential of classical optics in neural network architectures \cite{sui2020review,wetzstein2020inference,shastri2021photonics} based on linear and nonlinear elements such as Mach-Zehnder interferometers and saturable absorbers \cite{Shen2017,Harris:18}.

The advantages of optical systems in \gls{ml} are numerous \cite{mcmahon2023physics}. All-optical computing systems can potentially outperform electronic and optoelectronic counterparts in terms of energy consumption and scalability \cite{matuszewski2024role}. This efficiency stems from the minimal energy dissipation in optical fibres and waveguides compared to electronic circuits. Additionally, the integration of optical components on a large scale enhances scalability, making these systems suitable for extensive neural network implementations \cite{hamerly2019large}. The inherent high-speed nature of light facilitates rapid data processing and low latency, which is crucial for real-time \gls{ml} applications~\cite{ashtiani2022chip}.

\begin{figure}[!ht]
    \centering
    \includegraphics[width=\linewidth]{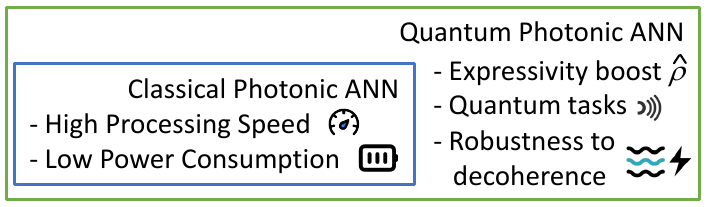}
    \caption{Schematic visualization of the main advantages of classical photonic ANNs, which are inherited by the quantum counterparts. Quantum photonic ANNs have in addition access to a large Hilbert space with the potential of an exponential expressivity boost, exhibit a strong robustness to decoherence, and can efficiently embed quantum states and tasks, as e.g. entanglement detection or quantum sensing.}
    \label{fig:advantages}
\end{figure}

Moving to the quantum regime, the advantages of optics for classical \gls{ml} are inherited as illustrated in \cref{fig:advantages}. Furthermore, as common also in other \gls{ml} settings, the enlarged Hilbert space has the potential to boost the performance (e.g. exponentially increasing the expressivity), and a quantum approach enable the processing of quantum inputs, efficiently executing quantum tasks. Major limitations of quantum technologies caused by decoherence due to the effect of the environment are alleviated for quantum states of light, which exhibit entanglement and noise below the familiar shot noise (squeezing)  even at room temperatures. Among the challenges is the inefficiency of light signal interactions at low intensities, which can be overcome with different strategies paving the way for robust and versatile quantum optical \gls{ml} systems.

\subsection{Quantum neural networks architectures}   \label{sec:with_quantum}

Successful implementations of quantum computation and simulations, landmark experiments in quantum optics (see \cref{sec:QOnet}) and recent advance of \gls{ml}, even with classical light,  hints at the potential high performance of quantum machine learning with quantum optical settings. The first proposed designs of quantum neural networks have been inspired by classical feed-forward neural networks with a sequence of unitary gates acting on different layers \cite{wan2017quantum}.  
These \gls{ann} circuits are based on a sequence of generally local operations.
Alternative design are instead inspired by recurrent neural networks, implemented to the coherent interaction between different components in complex interacting systems, being associated with analogue designs common in quantum simulations. This distinction between circuit and analogue design in quantum neuromorphic computing has been recently reviewed in Ref.~\cite{grollier2020neuromorphic}, comparing approaches based on parameterised quantum circuits,  with others based on quantum oscillator networks to compute, closer to classical neuromorphic computing. 
Focusing on photonic quantum \gls{ann}, there have been recent proposals in both analogue and circuit platforms, encompassing a broad spectrum of designs that also includes both \gls{cv} \cite{lloyd2019cvqnn} and  discrete variables \cite{Steinbrecher2019}. 

Quantum \gls{ann} based on variational quantum circuits have been the primary focus in the quantum \gls{ml} community \cite{cerezo2022challenges,jeswal2019recent} impulsed by the development of state-of-the-art superconducting gate-based circuits \gls{nisq} computers. A variational quantum circuit consists on a sequence of unitary operations with adjustable parameters, typically used within hybrid quantum-classical algorithms, where a classical optimizer iteratively tunes these parameters to minimize a cost function related to a specific problem. In the last few years these circuit-based, feed-forward neural networks have been studied in integrated photonic-based platforms, often structured as regular deep \glspl{ann}. The linear neuron connections are realised with linear optical components while the nonlinear neuron activation often relies on quantum measurement. In Ref.~\cite{Steinbrecher2019}, the concept of a quantum optical neural network was proposed to demonstrate that quantum optical phenomena can be integrated into neural networks. The considered platform consists of inputs in single-photon Fock states, variational linear optics operation circuits, single-site Kerr-type nonlinearities, and photon-number-resolving detectors. This framework has been expanded to quantum convolutional neural networks \cite{qu_CNNs}. Neural networks based on variational quantum circuits within the \gls{cv} architecture have been studied in Ref.~\cite{lloyd2019cvqnn}, where electromagnetic field amplitudes encode quantum information. The layered structure consists of continuously parameterised Gaussian and non-Gaussian gates, universal for \gls{cv} quantum computation. Applications discussed include fraud detection, image generation, and hybrid classical-quantum autoencoders. Circuit-based implementations have also been proposed in the context of \gls{am} as variations of Grover's search algorithm \cite{ventura2000qam}. While these models are useful in certain use cases \cite{quiroz2021hep,shapoval2019hep}, they deviate from the original idea, where a dynamical system is the resource that performs the association process. A more general overview of classical and quantum \gls{am} is presented below.

Beyond sequential circuit architectures, quantum \gls{ann} have been considered in different proposals based on quantum optical networks in recurrent instead of deep feed-forward designs. 
Recurrent neural network  do naturally arise considering the coherent interactions in complex quantum systems and have been recently considered in different proposals \cite{nokkala2023online,deprins2023quantum,pere-opportunities}. In Ref.~\cite{nokkala2023online} optimisation of a network of oscillators in \gls{cv} enables online quantum time series processing. Training all internal interactions it was possible to entangle sequential inputs at different times. Delay lines have been added leading to ‘memory modes’ as a quantum optical recurrent neural network, and a proof-of-concept implementation has been realised \cite{deprins2023quantum} on the photonic processor Borealis \cite{boson-sampling-3}.

A less demanding architecture of quantum optical neural network not restricted to circuit design (but also suitable for quantum computing \cite{PhysRevApplied.14.024065,fry2023optimizing,domingo2022qrcgates}) and requiring reduced optimization resources are extreme learning machines and, more generally, \gls{qrc} \cite{pere-opportunities}.
Pioneered by the work of Fujii and Nakajima in 2017 \cite{fujii-nakajima-2017}, \gls{qrc} has gained traction as a wider range of quantum substrates has been explored \cite{pere-opportunities}. 
Reservoir computing originated as a classical machine learning paradigm that simplifies the training of recurrent neural networks \cite{lukovsevivcius2009reservoir}. 
It has since become an umbrella term for algorithms and physical systems that exploit random (non-optimised) dynamical systems for machine learning tasks \cite{nakajima2020physical}. 
In particular, classical photonic implementations of reservoir computing are becoming increasingly popular for ultra-fast information processing applications being particularly suited for temporal data and memory tasks \cite{brunner2019photonic}.

Quantum optical platforms have also found their applications in this \gls{ml} framework, either in \glspl{dv} \cite{Spagnolo2022,Guillem2023,Dudas2023}, \glspl{cv} \cite{govia2021quantum,Nokkala2021,GBeni2023,GBeni2024} or hybrid qubit-photonic schemes \cite{Angelatos2021,senanian2023microwave}. \gls{qrc} in \glspl{dv} has been proposed in integrated photonic circuits using a novel quantum memristor to add memory and nonlinearities to the quantum dynamics, via the measurement back action \cite{Spagnolo2022}. Considering instead continuously coupled bosonic networks with hopping Hamiltonians, the larger Hilbert space yields performance improvements, exploiting the larger system expressivity \cite{Guillem2023}. Time series processing was also demonstrated in a proposal based on photon detection of a quantum oscillator dimer in Ref.~\cite{Dudas2023}. 

Moving to \gls{cv} platforms, a single quantum oscillator with a Kerr nonlinearity was shown to achieve the estimation of the phase of a classical signal with lower error than a classical setting \cite{govia2021quantum}. The first \gls{cv} proposal based on quantum oscillator networks in Gaussian states and homodyne detection was proposed in Ref.~\cite{Nokkala2021}, in which universality was also demonstrated. This was extended to an experimental design considering multimode pulses recirculating through a closed cavity loop coupled to an external signal \cite{GBeni2023}. Real-time processing through physical ensembles \cite{GBeni2023} and cavity squeezing for noise robustness \cite{GBeni2024} were also studied. Recently, the first experimental implementation of an analogue \gls{qrc} has been reported in a hybrid qubit-\gls{cv} setup \cite{senanian2023microwave}, where microwave signals were used as input to feed a reservoir made of a quantum superconducting circuit comprising a linear oscillator coupled to a single qubit.

Also \gls{am} can be implemented in quantum neural networks that go beyond circuit designs and quantum computation. \gls{am} arise when a system is able to retrieve the correct pre-stored memory or \textit{pattern} once interrogated with corrupted or partial initial information. \glspl{am} are commonly modelled through the (classical) Hopfiled neural network \cite{hopfield1982neural} consisting of an all-to-all network of classical spins, modelling neurons in active (+1) or inactive (-1) states, which evolve to minimise a given energy function through repeated network updates. This drives the system to settle into one of many stable spin configurations, the one associated with a stored memory or pattern. Here, patterns correspond to strings of classical bits which are written, through a proper \textit{learning rule}, in the weights of the neural connections \cite{amit_1989}. In the spirit of Hopfield, analogue implementations of \gls{qam} in open quantum systems allow spanning a manifold of stable states that can be identified as patterns. Here, generalisations of the Hopfield neural network range from binary neurons to qudits, in both closed \cite{inoue2011pattern, glaser2009nuclear} and open quantum systems \cite{rotondo2018open,fiorelli2019accelerated,fiorelli2021potts}. Some analogue approaches deal with the derivation of effective \gls{qam} models that exploit a quantum substrate, e.g., multimode Dicke models  \cite{fiorelli2020signatures}, confocal cavity QED systems \cite{enhancing2021marsh}, to embed patterns via classical learning rules. All previous work relies on the classical Hebbian learning rule which limits the amount of patterns stored in the system \cite{bodeker2022optimal}. However, recent models compatible with generic quantum neural networks seem to identify a potential quantum advantage \cite{sanpera2021capacity,labay2024theoretical}. Retaining Hopfield's original idea, Ref.~\cite{labay2022prl} proposes the use of a single driven-dissipative quantum resonator which increases the storage capacity of classical \gls{am}. Still, these proposals are limited to the storage of classical-like patterns. In \cref{sec:qam}, we will study a system where patterns are encoded in genuine quantum states of light.

\section{Squeezing in quantum photonic artificial neural networks} \label{sec:squeez}

\glsresetall

As discussed in previous sections, squeezing, a quantum phenomenon reducing light field quadrature fluctuations below shot noise levels, has advanced from fundamental tests like EPR paradox experiments to key applications in quantum technologies \cite{loudon1973quantum}. It enhances measurement sensitivity in quantum metrology \cite{Vittorio2004}, clock synchronization \cite{Giovannetti2001}, and gravitational wave detection \cite{PhysRevX.13.041021}, supports quantum cryptography \cite{Madsen2012}, and enables quantum advantage in Gaussian boson sampling \cite{boson-sampling-1,boson-sampling-2,boson-sampling-3}. Moreover, squeezing is vital for universal measurement-based quantum computing in \glspl{cv} \cite{furusawa_big,furusawa_2d_cluster,cluster_gates_1,cluster_gates_2} and enables \gls{cv} quantum \glspl{ann} in quantum \gls{ml} \cite{lloyd2019cvqnn,Nokkala2021}.

We can characterise squeezed states by the quadrature fluctuations $\ev*{(\Delta \hat{X}_\theta)^2} = \ev*{\hat{X}_\theta^2} - \ev*{\hat{X}_\theta}^2$ where $\hat{X}_\theta = [\opa \exp(-i \theta) + \opad \exp(i \theta)]/\sqrt{2}$ \cite{loudon1987squeezed}. Here, the angle $\theta$ is the direction where the quadrature fluctuations are measured. The operators $\opa$ and $\opad$ are the annihilation and creation operators respectively and commonly used quadratures are the position $\hat{x}$ and momentum $\hat{p}$ quadrature -- as $\hat{x} = \hat{X}_{0}$ and $\hat{p} = \hat{X}_{\pi/2}$. If we choose $\theta$ as the angle of minimum fluctuations $\theta^*$, then a state is squeezed if its fluctuation goes below the vacuum variance (shot noise limit) $\ev*{(\Delta \hat{X}_{\theta^*})^2} < s_{\text{vac}}^{2} = 0.5$. Another quantity that can be used to classify squeezed states is Mandel's $Q$ parameter~\cite{loudon1973quantum}
\begin{equation} \label{eq:def_mandel}
    Q = \frac{\ev*{(\Delta\opn)^2} - \ev*{\opn}}{\ev*{\opn}}\ ,
\end{equation}
where $\opn = \opad \opa$ is the photon-number operator. The Mandel parameter classifies quantum states as sub-Poissonian ($-1 \le Q < 0$) and super-Poissonian ($Q > 0$), with coherent states ($Q = 0$) displaying a Poisson distribution with a mean photon number equal to their variance.

In the following, we will explore two use cases of squeezed states for quantum \gls{ml}. In the first, we will present a platform that realises a quantum photonic recurrent \gls{ann}, taking advantage of squeezing to implement nonlinearity through input encoding and improving the performance of a forecasting task in \gls{qrc}. In the second, we analyse how squeezed states perform in \gls{qam} tasks, demonstrating the use of real quantum states to store quantum patterns.

\subsection{Quantum Reservoir Computing} \label{sec:qrc}

Recent advances in analogue \gls{cv} quantum networks applied to \gls{qrc} have demonstrated the potential of using squeezed states of light to improve performance. Input auxiliary squeezed states are a way to introduce nonlinearity in the input encoding of quantum oscillator networks, since the inherent dynamics of Gaussian states are linear in the quadrature operators \cite{Nokkala2021,Mujal_2021}. An advantage of using squeezed state encoding over coherent state encoding is the access to the broader Hilbert space of quantum correlations contained in the covariance matrix of the quantum reservoir \cite{Nokkala2021}. In later proposals consisting of multimode pulses, the encoding of classical signals in the squeezing phase of the input vacuum states was also used to obtain nonlinearities in the output observables \cite{GBeni2023,GBeni2024}. Another advantage of squeezed input states is that, because the information is encoded in the quantum fluctuations, information processing can be performed while the quantum reservoir is in a vacuum state of zero mean field amplitude. Comparing with classical states of zero mean amplitude, such as for instance thermal states with input encoded in the thermal excitations,  only linear memory in the readout observables is displayed while nonlinear memory is achieved in presence of squeezing and either amplitude or phase encoding \cite{Nokkala2021}. Squeezing is not only beneficial for the nonlinearity in the input layer encoding, but it has also been shown to improve the noise robustness of photonic \gls{qrc} platforms when incorporated into the reservoir dynamics \cite{GBeni2024}. In photonic quantum reservoirs with coherent feedback loops in the form of a cavity, the squeezing produced by the cavity has been shown to be a useful resource for improving the noise resilience of the protocol. In realistic scenarios, noise fluctuations are present throughout the protocol, with those affecting readout measurements being the most detrimental to \gls{qrc} performance \cite{Nokkala_noise}.

\begin{figure}[!hb]
    \centering
    \includegraphics[width=.83\linewidth]{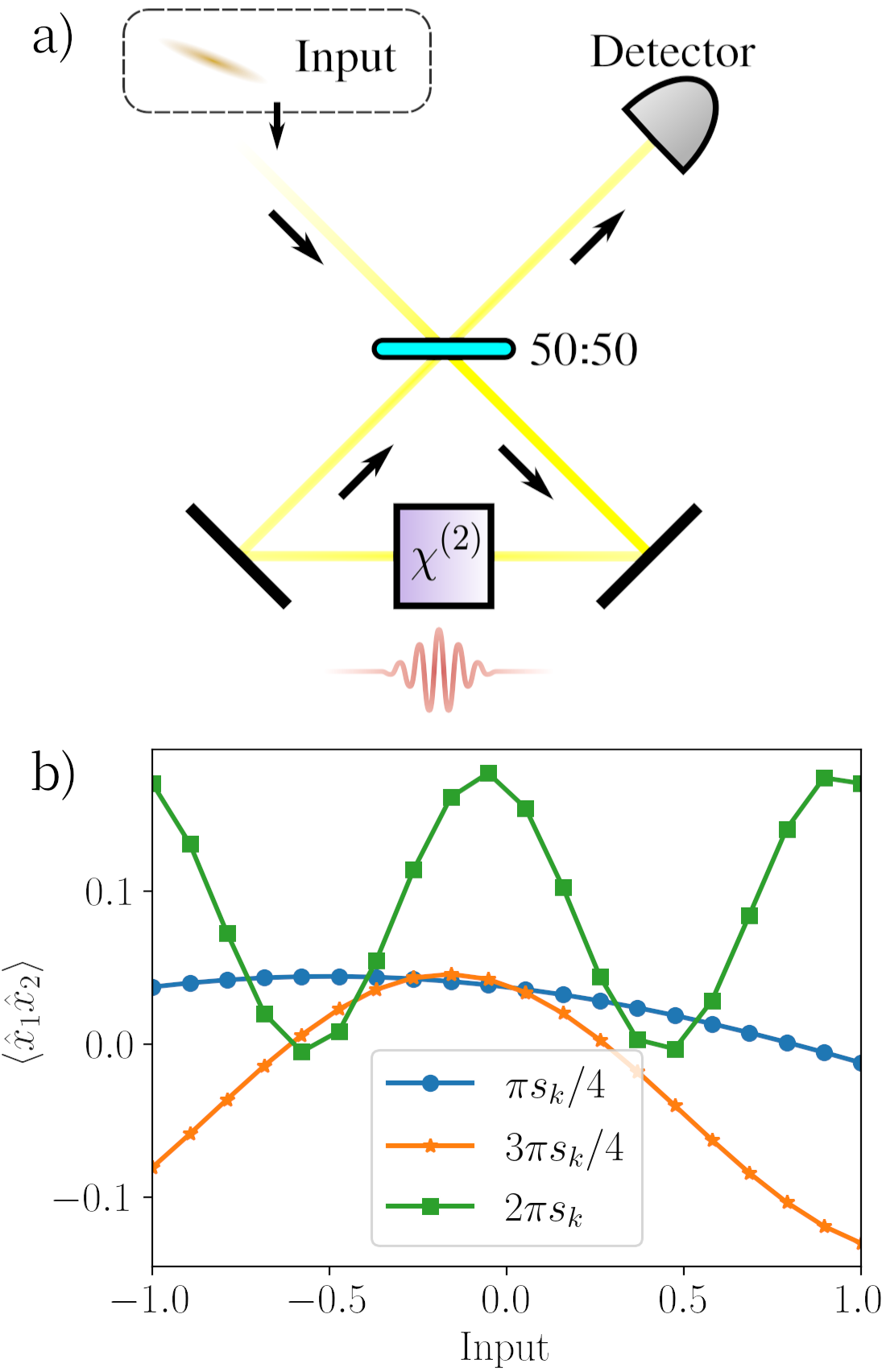}
    \caption{(a) Scheme of a loop-based quantum reservoir. (b) Response of the quantum reservoir to the input at time step $k$. Here the correlations $\langle \hat{x}_{1} \hat{x}_{2} \rangle$ are shown for a random but fixed network and input sequence with an $N = 2$ reservoir. Different lines depict different encodings on the squeezing angle of the input states, $\phi_{k}$. The squeezing present in the input states is set at $\xi_{\text{in}} = 0.75$ in Eq. \eqref{eq_Rin_phase} (around -3.3 dB).}
    \label{qrc_fig}
\end{figure}

The quantum reservoir under consideration is similar to that described in Ref.~\cite{GBeni2024}. The quantum substrate consists of a multimode optical pulse circulating through a closed loop (optical cavity) with a nonlinear crystal that the reservoir pulse passes through on each round trip [as shown in \cref{qrc_fig}(a)]. The cavity is connected to the output detector and to the input information via a 50:50 beam splitter that couples the reservoir to external engineered pulses. The classical input information to be processed is encoded in these external pulses and fed to the reservoir. The remaining optical path of the beam splitter output, with light exiting the cavity, is used to sequentially monitor the state of the quantum reservoir using multimode homodyne detection of the $x$-quadratures \cite{loudon1987homodyne}. The platform has the advantage of having a coherent quantum memory that does not rely on electronic feedback to create a recurrent \gls{ann}.

In \gls{cv} quantum optics, each quantum state is determined by the statistics of its quadrature vector, defined as $\hat{\mathbf{R}} = \left( \hat{x}_{1}, \hat{p}_{1}, \dots, \hat{x}_{N}, \hat{p}_{N} \right)^{\top}$, $\hat{x}_{i}$ and $\hat{p}_{i}$ are the quadrature operators of the $i$-th optical mode contained in the quantum state. Specifically, each of the external engineered pulses that we will couple to the cavity will have a quadrature vector $\hat{\mathbf{R}}_{\text{in}}^{(k)}$, where the label $k$ stands for the order of the input state in the sequence, so it is a temporal label. Each of them will be the product state of single-mode squeezed states with classical information encoded in their phases. In that way, we can write the covariance matrix of every input state as \cite{serafini2017}
\begin{widetext}
\begin{equation} \label{eq_Rin_phase}
    \text{cov} \left[ \hat{\mathbf{R}}_{\text{in}}^{(k)}, \hat{\mathbf{R}}_{\text{in}}^{(k)} \right] = \bigoplus_{i=1}^{N} \left( \begin{array}{cc}
        \cosh(\xi_{\text{in}}) + \cos(\phi_{k}) \sinh(\xi_{\text{in}}) & \sin(\phi_{k}) \sinh(\xi_{\text{in}}) \\
        \sin(\phi_{k}) \sinh(\xi_{\text{in}}) & \cosh(\xi_{\text{in}}) - \cos(\phi_{k}) \sinh(\xi_{\text{in}})
    \end{array} \right) \ ,
\end{equation}
\end{widetext}
where $\xi_{\text{in}}$ stands for the squeezing strength (we consider the same squeezing for every mode), and $\phi_{k}$ is the squeezing phase, which is a function of a classical number $s_{k}$ belonging to the classical input sequence that we want to process. The nonlinear crystal from the cavity applies a quadratic Hamiltonian, that we can write in a generic way as
\begin{equation}
    \hat{H}_{\chi^{(2)}} = \sum_{i,j=1}^{N} \left( \alpha_{ij} \hat{a}_{i}^{\dagger} \hat{a}_{j} + \beta_{ij}\hat{a}_{i}^{\dagger} \hat{a}_{j}^{\dagger} + \text{h.c.} \right), \quad \left(i \geq j\right)
\end{equation}
with coupling parameters $\alpha_{ij}$ and $\beta_{ij}$. If any of the $\beta_{ij}$ terms are different from zero, the crystal generates squeezing and this increases the squeezing present in the circulating reservoir pulse. With this setup, the response of the quantum observables to the classical input fed by the input squeezing phase, $\phi_{k}$ from Eq. \eqref{eq_Rin_phase}, becomes nonlinear, as can be seen in \cref{qrc_fig}(b). As shown in this figure, the non-linearity is not only present but highly tunable by changing the input encoding of the function $\phi_{k}$, which was also illustrated in Ref.~\cite{GBeni2023}. This allows a high reconfigurability of the reservoir for different tasks, and comes at a very low experimental cost, as the network is kept fixed at all times.

Going beyond ideal conditions, noise is to be considered in the measured quantities \cite{mujal2023time,GBeni2023}. Doing so, at every time step, the measured readout values can be written as
\begin{equation}
    \mathbf{O}_{\text{meas}}^{(k)} = \mathbf{O}_{\text{ideal}}^{(k)} + \mathbf{u}\left(0, s_{\text{noise}}^{2}\right) \ ,
\end{equation}
where $\mathbf{O}_{\text{ideal}}^{(k)}$ is a vector containing the expected values of the chosen observables in ideal, noiseless conditions, while $\mathbf{O}_{\text{meas}}^{(k)}$ contains the measured observables. The vector $\mathbf{u}(0,s_{\text{noise}}^{2})$ resembles the readout noise, which we model as additive Gaussian fluctuations with zero mean and variance equal to $s_{\text{noise}}^{2}$ added to the quadrature measurements. In our case, the variance of the vacuum noise is assumed to be $s_{\text{vac}}^{2} = 0.5$, so that we can express the additive noise intensity relative to the vacuum fluctuations. For example, a noise's variance of $s_{\text{noise}}^{2} = 0.1$ would be equivalent to a noise with a relative intensity of 20\% of the vacuum fluctuations.

\begin{figure}
    \centering
    \includegraphics[width=0.85\linewidth]{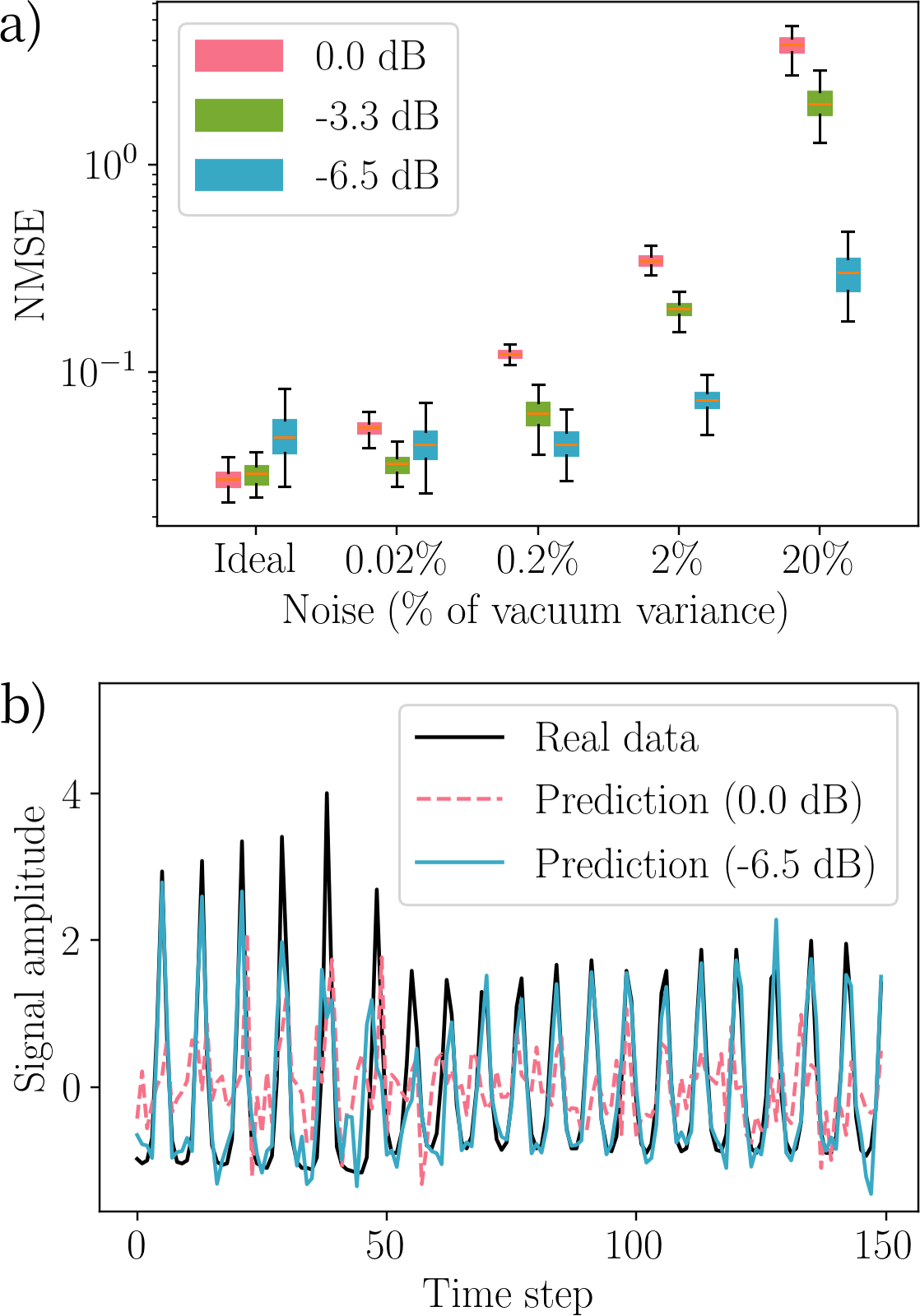}
    \caption{(a) Boxplot of the \gls{nmse} for the chaotic time-series forecasting task as a function of the intensity of the added readout noise relative to the vacuum variance. Different colours show different values of the squeezing generated by the non-linear cavity. Each box contains values from 100 independent realizations. (b) Prediction of the chaotic signal (black) in the presence of high noise (20\% vacuum variance intensity) for a reservoir without cavity squeezing (dashed magenta) and for a reservoir with -6.5 dB of cavity squeezing (blue). In all cases, the reservoir had $N = 12$ modes, the input phase encoding was $\phi_{k} = \pi s_{k}/4$, the input state squeezing is set at around -3.3 dB, and the non-linear cavity squeezing is indicated in the figure legends.}
    \label{fig:santafe}
\end{figure}

For our simulations, the squeezing generated by the Hamiltonian is the same for each mode, which is something we can easily tune using the Bloch-Messiah decomposition \cite{BM_1,BM_2}. We will consider second and fourth-order moments of the measured $x$-quadratures, $\{ 
\ev*{\hat{x}_{i}\hat{x}_{j}}, \ev*{\hat{x}_{i}^{2}\hat{x}_{j}^{2}}, \ev*{\hat{x}_{i}^{3}\hat{x}_{j}} \}_{i,j=1}^{N}$, as outputs for the readout layer ($\langle \cdot \rangle$ stands for the expected values of the observables). The task we consider is the forecasting of the Santa Fe chaotic time series \cite{santafe1,santafe2}. At each time step, we want the reservoir to predict the next step in the series, so the target function is $\bar{y}(s_{k}) = s_{k+1}$.  The data set contains 4000 consecutive values in total, from which we take the training set of 3000 points and the test set of 700 values (the first 300 values are used for the wash-out). We use \gls{nmse} as the performance metric that we want to minimise during training.  It is defined as $\text{NMSE}(\mathbf{y},\bar{\mathbf{y}}) = \langle (\mathbf{y}-\bar{\mathbf{y}})^{2} \rangle/\langle \bar{\mathbf{y}}^{2} \rangle$, where $\bar{\mathbf{y}}$ is the target function vector and $\mathbf{y}$ are the reservoir predictions. Here, the averages $\langle \cdot \rangle$ are taken among the values of either the training or the test set. In \cref{fig:santafe}(a) the \gls{nmse} is shown as a function of the noise intensity (including the ideal noiseless case) for different values of the generated cavity squeezing (different colours). The squeezing seems detrimental to the performance in ideal and very low noise conditions (noise intensities around and below 0.02\% of vacuum noise). However, as the noise increases, the robustness of the reservoir is determined by the amount of squeezing produced by the cavity.  In \cref{fig:santafe}(b) the prediction for a subset of the test values is shown for a single realisation of a reservoir considering noisy measurements either with no cavity squeezing (magenta dashed curve) or with $\sim 6.5$ dB of squeezing (blue curve). The noise conditions are severe (20\% of vacuum fluctuations). The difference in prediction power in both scenarios is clearly seen in the figure. However, with such high noise, it is difficult for the reservoir to predict the abrupt change in oscillation amplitude around time step 50 of the chaotic signal, even with high cavity squeezing. These results are consistent with the ones presented in Ref. \cite{GBeni2024}. Introducing additional squeezing through an active cavity offers enhanced robustness to readout noise. The interplay between linear elements such as the beam splitter setting the number of photons in the loop pulse, and the squeezing effect increasing its energy, determine the final performance and signal to noise ratio in the output layer. Squeezing can counteract the beam-splitter losses and assist in retaining information in the loop pulse for longer, making it more resilient to additive noise.

The proposed platform is not beyond experimental reach with state-of-the-art technologies: necessary tools such as fast squeezing phase setting devices \cite{Tomoda2023,boson-sampling-3}, reconfigurable network structures from nonlinear sources \cite{Cai2017,10.1063/5.0156331} (with some modes reaching squeezing values beyond 6.5 dB \cite{Cai2017}) and multimode homodyne detection of multimode pulses \cite{Cai2021,loudon1987homodyne}, are within current technological capabilities.

\subsection{Quantum Associative Memory} \label{sec:qam}

\Gls{am} models have been explored in the quantum domain with the extension of the classical Hopfield neural network. In these cases, the binary neurons are replaced by two-level quantum systems embedded in a bath, where the classical dynamics is encoded in the jump operators between the system and the bath \cite{rotondo2018open}. Such models give rise to new dynamical phases but cannot improve the memory of classical models \cite{bodeker2022optimal}. Moreover, these systems only allow the retrieval of patterns encoded in classical states. Enabling information to be encoded in quantum states could open up the possibility of improving performance over classical models. In this respect, we have extended the work introduced in Ref.~\cite{labay2022prl}, where patterns are encoded in coherent states, to allow for squeezed states \cite{labay2023squeezing}. We thus demonstrate the use of bona fide quantum states for memory retrieval.

The system under consideration is a single driven-dissipative nonlinear quantum oscillator, where one can exploit its (in principle infinite) number of degrees of freedom, represented by the density matrix of the system, for computational purposes. Following the intuition of Ref.~\cite{PhysRevLett.119.225301}, a density matrix itself can be seen as a complex quantum network, whose links are built through the set of bipartite correlations within degrees of freedom, and the complexity of such network can facilitate the successful achievement of computational tasks \cite{PhysRevApplied.17.064044}.  

The Gorini–Kossakowski–Sudarshan–Lindblad master equation describing the temporal evolution of the oscillator is
\begin{equation} \label{eq:pp_me_nm}
    \pdv{\rho}{t} = -i [\hat{H}_n, \rho] + \gamma_1 \disp{\opa}\rho + \gamma_m \disp{\opa^m} \rho\ ,
\end{equation}
where we have the standard linear (single-photon) damping and an engineered nonlinear (multiphoton) damping \cite{mirrahimi2014dynamically} with rates $\gamma_1$ and $\gamma_m$ respectively. The Hamiltonian, which contains a $n$-order squeezing drive \cite{braunstein1987generalized}, in the rotation frame and after the parametric approximation, is
\begin{equation*} \label{eq:pp_ham_nm}
    \hat{H}_n = \Delta \opad \opa + i \eta \left[\opa^n - (\opad)^n \right]\ .
\end{equation*}
Here, $\Delta = \omega_0 - \omega_s$ is the detuning between the natural oscillator frequency and that of the squeezing force, and $\eta$ is the magnitude of the driving. 

\begin{figure*}[!ht]
\includegraphics[width=0.82\linewidth]{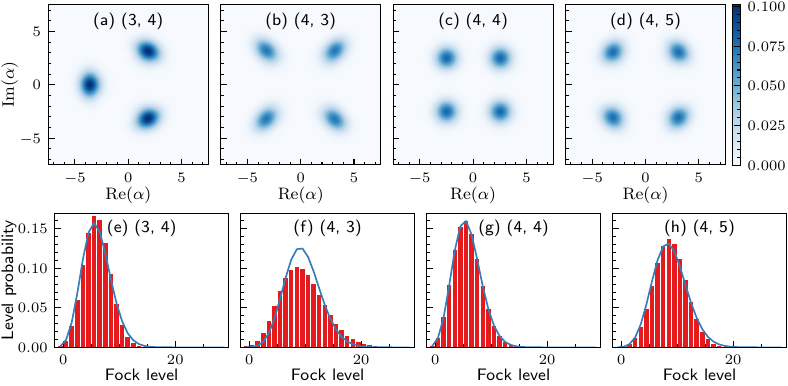}
  \caption{(a-d) Wigner representation of the steady state for different non-linear degrees. (e-h) Probability distribution of the Fock levels (red histogram) corresponding to the steady states in (a-d) respectively. In blue, the Gaussian distribution of a coherent state with the same mean-photon number. All plots use the same value for $\gamma_m / \gamma_1 = 0.2$ and $\Delta/\gamma_1 = 0.4$, the driving strength for each case is: (a,e) $\eta_3 / \gamma_1 = 13.02$, (b,f) $\eta_4/\gamma_1 = 0.8$, (c,g) $\eta_4/\gamma_1 = 3.9$ and (d,h) $\eta_4 / \gamma_1 = 91.13$.}
  \label{fig:lobes}
\end{figure*}

The main ingredient of our approach lies in the nonlinearity which determines the form and phase symmetry of the steady state, changing from coherent states to purely quantum states, depending on the model parameters. In \cref{fig:lobes}(a-d) we can see the Wigner representation of the steady state for different combinations of nonlinear degrees $(n,m)$. Due to the rotational $\mathbb{Z}_n$ symmetry of \cref{eq:pp_me_nm}, the steady state is formed by $n$ symmetrically distributed squeezed-coherent states $\{ \ket{\beta \exp[i(2j + 1) / n],\xi_j} \}_{j=1}^n$ whose shape changes depending on the relation between $n$ and $m$. Only in the case $n=m$, the lobes are well-described by coherent states ($\xi_j = 0$) as we can appreciate in \cref{fig:lobes}(c,g) where the probability distribution of Fock states follows a Gaussian distribution. In all the other cases, we find squeezed states whose distribution is super-Poissonian for $m < n$ [see \cref{fig:lobes}(b)] or sub-Poissonian for $m > n$ [see \cref{fig:lobes}(a,d)], leading to phase-squeezed and amplitude-squeezed states respectively. In our case, the amplitude of the lobes can be determined from the oscillator parameters $\beta \approx (2 n \eta_n / m \gamma_m)^{1/(2m-n)} $ and the strength of the squeezing $\abs{\xi_j}$ has been seen numerically to depend only on the relation of nonlinear degrees $n$ and $m$ \cite{labay2023squeezing}.

In addition, this type of resonator has a metastable dynamical phase where the squeezed lobes $\{ \ket{\beta_j, \xi_j} \}_{j=1}^n$ become metastable states. The metastable phase is a consequence of a large separation in the Liouvillian spectrum, which divides the dynamics into different timescales where the dynamics are confined for a long time, compared to all timescales, in the metastable manifold spanned by the $n$ metastable states \cite{macieszczak2021theory}. Concerning \glspl{am}, metastability allows systems converging to a unique steady state to span a manifold of relevant addressable memories \cite{brinkman2022metastable}. More specifically, within the metastable transient, an initial state will move towards the closest lobe (representing one of the stored memories) and remain there for a long time. Consequently, we can extract information about the corresponding lobe by measuring the state within this regime. 

\begin{figure}[!hb]
    \centering
    \includegraphics[width=\linewidth]{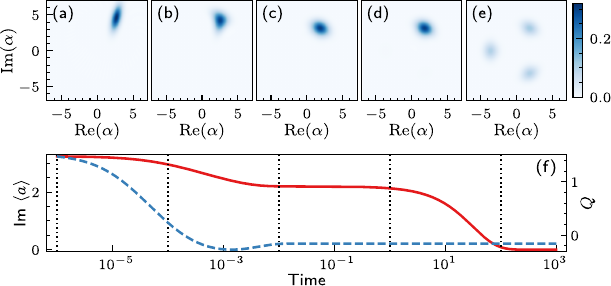}
    \caption{(a-e) Wigner representation of the time evolution of an initial squeezed coherent state with displacement $1.5\abs{\beta_j} \exp(i \pi / 3)$ and squeezing parameter $0.5 \exp(-i 0.15 \pi)$. The times at which the Wigner distribution is calculated correspond to the vertical bars in (f). (f) Evolution of the expectation value of $\opa$ (red, left) and the Mandel parameter $Q$ (blue, right). The resonator parameters are: $n=3$, $m=4$, $\gamma_4/\gamma_1 = 0.2$, $\eta_3 / \gamma_1 = 13.02$ and $\Delta = 0.4$.}
    \label{fig:qam_evolution}
\end{figure}

As an example, in \cref{fig:qam_evolution} we show the time evolution of an initial squeezed-coherent state with a larger amplitude than the lobes. In the panels \cref{fig:qam_evolution}(a-e) we see the Wigner distribution of such state at specific times and in \cref{fig:qam_evolution}(f) the evolution of the imaginary part of the annihilation operator $\opa$ (left axis) characterizing approximately the amplitude of the state and the Mandel's $Q$ parameter (right axis, see \cref{eq:def_mandel}) characterizing the squeezing. 

\begin{figure*}[!ht]
    \centering
    \includegraphics[width=0.8\linewidth]{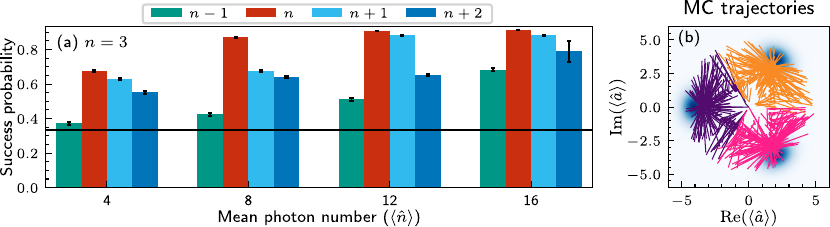}
    \caption{(a) Success probability for measuring the correct lobe starting from an initial squeezed-coherent state with random amplitude and squeezing parameter. The results are shown for different mean photon numbers of the lobes and dissipation powers ($m$) with fixed $n = 3$. Each bar is obtained from an average of over 500 Monte Carlo trajectories with different initial states. We include a horizontal black line that represents the success probability you would obtain by random guessing the lobe, that is, $1 / n$. An example of such trajectories can be seen in (b) for the particular case $(3, 4)$ with $\ev{\hat{n}} =8$. The colors indicate the most similar lobe at $t=0$.}
    \label{fig:success_probability}
\end{figure*}

We can see that the state undergoes three different dynamical regimes. First, an initial decay that kills all Liouvillian modes outside the metastable manifold [see \cref{fig:qam_evolution}(a,b)]. The amplitude of the state decays to the amplitude of the lobes and the Mandel parameter goes from super-Poissonian to sub-Poissonian. The dynamics then freeze and the state remains for a long time in one of the metastable phases [see \cref{fig:qam_evolution}(c,d)]. Finally, the metastable modes decay and the steady state corresponding to \cref{fig:lobes}(a) is reached in \cref{fig:qam_evolution}(e).

We assess the \gls{qam} efficiency by numerically computing the probability that the system is found in the target lobe during the metastable phase. We initialize the system at a random squeezed-coherent state which, by definition, is most similar to one of the $n$ lobes that constitute our system. Hence, the task succeeds if during the metastable phase, the state converges to the \textit{a prior} correct lobe. The results of the success probability for the particular case $n=3$ can be seen in \cref{fig:success_probability} where we compare different levels of squeezing in the memories (different $m$) and increasing mean photon number of the lobes (larger separation).

We appreciate that the largest success probability is achieved in the case $n=m$, i.e. the memories are coherent states. Nevertheless, a high success probability can be achieved by a resonator with $m = n + 1$, which stores the memories in amplitude-squeezed states. This behaviour, however, is not general as for $n = 2$ a higher success probability is achieved with squeezed states instead of coherent states \cite{labay2023squeezing}. When storing more patterns ($n > 2$), one must take into account that the overlap between lobes decreases so a larger mean photon number is needed to correctly discriminate the patterns. Still, as we can see in \cref{fig:success_probability}(b), for a sufficiently large mean photon number the phase space is divided into 3 distinct regions containing all the initial conditions that converge to the lobe inside it. 

The possibility to experimentally realize resonators with high nonlinearities represents an important breakthrough to implement these tasks in actual devices in the near future \cite{svensson2018period,forn2020three}. We believe these types of resonators offer a high versatility to be used for numerous kinds of tasks such as \gls{qam} \cite{labay2022prl,labay2023squeezing}, autonomous quantum error correction \cite{mirrahimi2014dynamically,cai2021reviewQEC} or quantum annealing \cite{puri2017annhealing,nigg2017annhealing}.

\section{Conclusions and outlook}

\glsresetall

Quantum optical networks, which are emerging in various contexts (from integrated photonics to the quantum internet), have the potential to enhance capabilities in communication, computation and \gls{ml}. In the latter, they can exploit the advantages of both classical and quantum optics, namely energy efficiency, high speed, noise immunity, as well as access to a large Hilbert space (with the potential for an exponential increase in expressivity) and to quantum inputs and tasks.
 In particular, quantum photonics offers several advantages, including enhanced computational power through quantum computing, secure communication via quantum cryptography, and highly sensitive detection capabilities for various applications in imaging and sensing. 
 
These neuro-inspired computational methods hold great promise for both feedforward circuits and recurrent neural network designs.  Here, we have focused on two approaches, quantum reservoir computing and associative memories, highlighting the transformative impact of quantum optics in this field and shedding light on the role of the quantum property of squeezing to achieve improved performance. 

In the case of \gls{qrc}, we have illustrated how squeezing increases the range of applicability in realistic (noisy) scenarios. In fact, multimode squeezing enhances the memory available to the system and, as a result, improves performance in several benchmark temporal tasks.
As for \gls{qam}, the introduction of squeezing extends the range of potential solutions for stored patterns to include genuine quantum objects that could not be stored efficiently in classical devices. Moving from a single oscillator to an array will enable storing and retrieving more quantum features, including entangled memories.

These results highlight the potential of using quantum optical techniques to make a paradigm shift in computational methods, paving the way for future innovations in a wide range of fields. By integrating concepts such as squeezing into the design of neural networks, we can achieve significant improvements in efficiency and capability, offering new solutions to complex problems.
 
\begin{acknowledgments}
    This work is dedicated to Rodney Loudon, in memory of his enduring contributions to the field of quantum optics and of his kindness and elegance in research.
We acknowledge the Spanish State Research Agency, through the Mar\'ia de Maeztu project CEX2021-001164-M, through the COQUSY project PID2022-140506NB-C21 and -C22, and through the INFOLANET project PID2022-139409NB-I00, all funded by MCIU/AEI/10.13039/501100011033 and FEDER, UE;
 MINECO through the QUANTUM SPAIN project, and EU through the RTRP - NextGenerationEU within the framework of the Digital Spain 2025 Agenda. ALM is funded by the University of the Balearic Islands through the project BGRH-UIB-2021. JGB is funded by the Conselleria d’Educació, Universitat i Recerca of the Government of the Balearic Islands with grant code FPI/036/2020.
\end{acknowledgments}
\bibliography{references}

\begin{thebibliography}{166}%
\makeatletter
\providecommand \@ifxundefined [1]{%
 \@ifx{#1\undefined}
}%
\providecommand \@ifnum [1]{%
 \ifnum #1\expandafter \@firstoftwo
 \else \expandafter \@secondoftwo
 \fi
}%
\providecommand \@ifx [1]{%
 \ifx #1\expandafter \@firstoftwo
 \else \expandafter \@secondoftwo
 \fi
}%
\providecommand \natexlab [1]{#1}%
\providecommand \enquote  [1]{``#1''}%
\providecommand \bibnamefont  [1]{#1}%
\providecommand \bibfnamefont [1]{#1}%
\providecommand \citenamefont [1]{#1}%
\providecommand \href@noop [0]{\@secondoftwo}%
\providecommand \href [0]{\begingroup \@sanitize@url \@href}%
\providecommand \@href[1]{\@@startlink{#1}\@@href}%
\providecommand \@@href[1]{\endgroup#1\@@endlink}%
\providecommand \@sanitize@url [0]{\catcode `\\12\catcode `\$12\catcode `\&12\catcode `\#12\catcode `\^12\catcode `\_12\catcode `\%12\relax}%
\providecommand \@@startlink[1]{}%
\providecommand \@@endlink[0]{}%
\providecommand \url  [0]{\begingroup\@sanitize@url \@url }%
\providecommand \@url [1]{\endgroup\@href {#1}{\urlprefix }}%
\providecommand \urlprefix  [0]{URL }%
\providecommand \Eprint [0]{\href }%
\providecommand \doibase [0]{https://doi.org/}%
\providecommand \selectlanguage [0]{\@gobble}%
\providecommand \bibinfo  [0]{\@secondoftwo}%
\providecommand \bibfield  [0]{\@secondoftwo}%
\providecommand \translation [1]{[#1]}%
\providecommand \BibitemOpen [0]{}%
\providecommand \bibitemStop [0]{}%
\providecommand \bibitemNoStop [0]{.\EOS\space}%
\providecommand \EOS [0]{\spacefactor3000\relax}%
\providecommand \BibitemShut  [1]{\csname bibitem#1\endcsname}%
\let\auto@bib@innerbib\@empty
\bibitem [{\citenamefont {Loudon}(1973)}]{loudon1973quantum}%
  \BibitemOpen
  \bibfield  {author} {\bibinfo {author} {\bibfnamefont {R.}~\bibnamefont {Loudon}},\ }\href {https://books.google.es/books?id=OHspAQAAMAAJ} {\emph {\bibinfo {title} {The Quantum Theory of Light}}},\ Oxford science publications\ (\bibinfo  {publisher} {Clarendon Press},\ \bibinfo {year} {1973})\ \bibinfo {note} {3rd Edition: Loudon, R. 2000. \textit{The quantum theory of light}. OUP Oxford.}\BibitemShut {Stop}%
\bibitem [{\citenamefont {O'brien}\ \emph {et~al.}(2009)\citenamefont {O'brien}, \citenamefont {Furusawa},\ and\ \citenamefont {Vu{\v{c}}kovi{\'c}}}]{obrien2009photonic}%
  \BibitemOpen
  \bibfield  {author} {\bibinfo {author} {\bibfnamefont {J.~L.}\ \bibnamefont {O'brien}}, \bibinfo {author} {\bibfnamefont {A.}~\bibnamefont {Furusawa}},\ and\ \bibinfo {author} {\bibfnamefont {J.}~\bibnamefont {Vu{\v{c}}kovi{\'c}}},\ }\href@noop {} {\bibfield  {journal} {\bibinfo  {journal} {Nature photonics}\ }\textbf {\bibinfo {volume} {3}},\ \bibinfo {pages} {687} (\bibinfo {year} {2009})}\BibitemShut {NoStop}%
\bibitem [{\citenamefont {Benyoucef}(2023)}]{photQtech2023}%
  \BibitemOpen
  \bibinfo {editor} {\bibfnamefont {M.}~\bibnamefont {Benyoucef}},\ ed.,\ \href@noop {} {\emph {\bibinfo {title} {Photonic Quantum Technologies}}}\ (\bibinfo  {publisher} {John Wiley \& Sons, Ltd},\ \bibinfo {year} {2023})\BibitemShut {NoStop}%
\bibitem [{\citenamefont {Pelucchi}\ \emph {et~al.}(2022)\citenamefont {Pelucchi}, \citenamefont {Fagas}, \citenamefont {Aharonovich}, \citenamefont {Englund}, \citenamefont {Figueroa}, \citenamefont {Gong}, \citenamefont {Hannes}, \citenamefont {Liu}, \citenamefont {Lu}, \citenamefont {Matsuda}, \citenamefont {Pan}, \citenamefont {Schreck}, \citenamefont {Sciarrino}, \citenamefont {Silberhorn}, \citenamefont {Wang},\ and\ \citenamefont {J{\"o}ns}}]{Pelucchi2022}%
  \BibitemOpen
  \bibfield  {author} {\bibinfo {author} {\bibfnamefont {E.}~\bibnamefont {Pelucchi}}, \bibinfo {author} {\bibfnamefont {G.}~\bibnamefont {Fagas}}, \bibinfo {author} {\bibfnamefont {I.}~\bibnamefont {Aharonovich}}, \bibinfo {author} {\bibfnamefont {D.}~\bibnamefont {Englund}}, \bibinfo {author} {\bibfnamefont {E.}~\bibnamefont {Figueroa}}, \bibinfo {author} {\bibfnamefont {Q.}~\bibnamefont {Gong}}, \bibinfo {author} {\bibfnamefont {H.}~\bibnamefont {Hannes}}, \bibinfo {author} {\bibfnamefont {J.}~\bibnamefont {Liu}}, \bibinfo {author} {\bibfnamefont {C.-Y.}\ \bibnamefont {Lu}}, \bibinfo {author} {\bibfnamefont {N.}~\bibnamefont {Matsuda}}, \bibinfo {author} {\bibfnamefont {J.-W.}\ \bibnamefont {Pan}}, \bibinfo {author} {\bibfnamefont {F.}~\bibnamefont {Schreck}}, \bibinfo {author} {\bibfnamefont {F.}~\bibnamefont {Sciarrino}}, \bibinfo {author} {\bibfnamefont {C.}~\bibnamefont {Silberhorn}}, \bibinfo {author} {\bibfnamefont {J.}~\bibnamefont {Wang}},\ and\ \bibinfo {author} {\bibfnamefont {K.~D.}\
  \bibnamefont {J{\"o}ns}},\ }\href {https://doi.org/10.1038/s42254-021-00398-z} {\bibfield  {journal} {\bibinfo  {journal} {Nature Reviews Physics}\ }\textbf {\bibinfo {volume} {4}},\ \bibinfo {pages} {194} (\bibinfo {year} {2022})}\BibitemShut {NoStop}%
\bibitem [{\citenamefont {Shadbolt}\ \emph {et~al.}(2014)\citenamefont {Shadbolt}, \citenamefont {Mathews}, \citenamefont {Laing},\ and\ \citenamefont {O'brien}}]{shadbolt2014testing}%
  \BibitemOpen
  \bibfield  {author} {\bibinfo {author} {\bibfnamefont {P.}~\bibnamefont {Shadbolt}}, \bibinfo {author} {\bibfnamefont {J.~C.}\ \bibnamefont {Mathews}}, \bibinfo {author} {\bibfnamefont {A.}~\bibnamefont {Laing}},\ and\ \bibinfo {author} {\bibfnamefont {J.~L.}\ \bibnamefont {O'brien}},\ }\href@noop {} {\bibfield  {journal} {\bibinfo  {journal} {Nature Physics}\ }\textbf {\bibinfo {volume} {10}},\ \bibinfo {pages} {278} (\bibinfo {year} {2014})}\BibitemShut {NoStop}%
\bibitem [{\citenamefont {Loudon}\ and\ \citenamefont {Knight}(1987)}]{loudon1987squeezed}%
  \BibitemOpen
  \bibfield  {author} {\bibinfo {author} {\bibfnamefont {R.}~\bibnamefont {Loudon}}\ and\ \bibinfo {author} {\bibfnamefont {P.~L.}\ \bibnamefont {Knight}},\ }\href@noop {} {\bibfield  {journal} {\bibinfo  {journal} {Journal of modern optics}\ }\textbf {\bibinfo {volume} {34}},\ \bibinfo {pages} {709} (\bibinfo {year} {1987})}\BibitemShut {NoStop}%
\bibitem [{\citenamefont {Slusher}\ \emph {et~al.}(1985)\citenamefont {Slusher}, \citenamefont {Hollberg}, \citenamefont {Yurke}, \citenamefont {Mertz},\ and\ \citenamefont {Valley}}]{slusher1985observation}%
  \BibitemOpen
  \bibfield  {author} {\bibinfo {author} {\bibfnamefont {R.}~\bibnamefont {Slusher}}, \bibinfo {author} {\bibfnamefont {L.}~\bibnamefont {Hollberg}}, \bibinfo {author} {\bibfnamefont {B.}~\bibnamefont {Yurke}}, \bibinfo {author} {\bibfnamefont {J.}~\bibnamefont {Mertz}},\ and\ \bibinfo {author} {\bibfnamefont {J.}~\bibnamefont {Valley}},\ }\href@noop {} {\bibfield  {journal} {\bibinfo  {journal} {Physical review letters}\ }\textbf {\bibinfo {volume} {55}},\ \bibinfo {pages} {2409} (\bibinfo {year} {1985})}\BibitemShut {NoStop}%
\bibitem [{\citenamefont {Giovannetti}\ \emph {et~al.}(2004)\citenamefont {Giovannetti}, \citenamefont {Lloyd},\ and\ \citenamefont {Maccone}}]{Vittorio2004}%
  \BibitemOpen
  \bibfield  {author} {\bibinfo {author} {\bibfnamefont {V.}~\bibnamefont {Giovannetti}}, \bibinfo {author} {\bibfnamefont {S.}~\bibnamefont {Lloyd}},\ and\ \bibinfo {author} {\bibfnamefont {L.}~\bibnamefont {Maccone}},\ }\href {https://doi.org/10.1126/science.1104149} {\bibfield  {journal} {\bibinfo  {journal} {Science}\ }\textbf {\bibinfo {volume} {306}},\ \bibinfo {pages} {1330} (\bibinfo {year} {2004})},\ \Eprint {https://arxiv.org/abs/https://www.science.org/doi/pdf/10.1126/science.1104149} {https://www.science.org/doi/pdf/10.1126/science.1104149} \BibitemShut {NoStop}%
\bibitem [{\citenamefont {Ganapathy}\ \emph {et~al.}(2023)\citenamefont {Ganapathy} \emph {et~al.}}]{PhysRevX.13.041021}%
  \BibitemOpen
  \bibfield  {author} {\bibinfo {author} {\bibfnamefont {D.}~\bibnamefont {Ganapathy}} \emph {et~al.} (\bibinfo {collaboration} {LIGO O4 Detector Collaboration}),\ }\href {https://doi.org/10.1103/PhysRevX.13.041021} {\bibfield  {journal} {\bibinfo  {journal} {Phys. Rev. X}\ }\textbf {\bibinfo {volume} {13}},\ \bibinfo {pages} {041021} (\bibinfo {year} {2023})}\BibitemShut {NoStop}%
\bibitem [{\citenamefont {Aspect}\ \emph {et~al.}(1982)\citenamefont {Aspect}, \citenamefont {Dalibard},\ and\ \citenamefont {Roger}}]{Aspect1982}%
  \BibitemOpen
  \bibfield  {author} {\bibinfo {author} {\bibfnamefont {A.}~\bibnamefont {Aspect}}, \bibinfo {author} {\bibfnamefont {J.}~\bibnamefont {Dalibard}},\ and\ \bibinfo {author} {\bibfnamefont {G.}~\bibnamefont {Roger}},\ }\href {https://doi.org/10.1103/PhysRevLett.49.1804} {\bibfield  {journal} {\bibinfo  {journal} {Phys. Rev. Lett.}\ }\textbf {\bibinfo {volume} {49}},\ \bibinfo {pages} {1804} (\bibinfo {year} {1982})}\BibitemShut {NoStop}%
\bibitem [{\citenamefont {Fabre}\ and\ \citenamefont {Treps}(2020)}]{multimode}%
  \BibitemOpen
  \bibfield  {author} {\bibinfo {author} {\bibfnamefont {C.}~\bibnamefont {Fabre}}\ and\ \bibinfo {author} {\bibfnamefont {N.}~\bibnamefont {Treps}},\ }\href {https://doi.org/10.1103/RevModPhys.92.035005} {\bibfield  {journal} {\bibinfo  {journal} {Rev. Mod. Phys.}\ }\textbf {\bibinfo {volume} {92}},\ \bibinfo {pages} {035005} (\bibinfo {year} {2020})}\BibitemShut {NoStop}%
\bibitem [{\citenamefont {Argyris}\ \emph {et~al.}(2005)\citenamefont {Argyris}, \citenamefont {Syvridis}, \citenamefont {Larger}, \citenamefont {Annovazzi-Lodi}, \citenamefont {Colet}, \citenamefont {Fischer}, \citenamefont {Garcia-Ojalvo}, \citenamefont {Mirasso}, \citenamefont {Pesquera},\ and\ \citenamefont {Shore}}]{Agyrischaotic}%
  \BibitemOpen
  \bibfield  {author} {\bibinfo {author} {\bibfnamefont {A.}~\bibnamefont {Argyris}}, \bibinfo {author} {\bibfnamefont {D.}~\bibnamefont {Syvridis}}, \bibinfo {author} {\bibfnamefont {L.}~\bibnamefont {Larger}}, \bibinfo {author} {\bibfnamefont {V.}~\bibnamefont {Annovazzi-Lodi}}, \bibinfo {author} {\bibfnamefont {P.}~\bibnamefont {Colet}}, \bibinfo {author} {\bibfnamefont {I.}~\bibnamefont {Fischer}}, \bibinfo {author} {\bibfnamefont {J.}~\bibnamefont {Garcia-Ojalvo}}, \bibinfo {author} {\bibfnamefont {C.~R.}\ \bibnamefont {Mirasso}}, \bibinfo {author} {\bibfnamefont {L.}~\bibnamefont {Pesquera}},\ and\ \bibinfo {author} {\bibfnamefont {K.~A.}\ \bibnamefont {Shore}},\ }\href@noop {} {\bibfield  {journal} {\bibinfo  {journal} {Nature}\ }\textbf {\bibinfo {volume} {438}},\ \bibinfo {pages} {343} (\bibinfo {year} {2005})}\BibitemShut {NoStop}%
\bibitem [{\citenamefont {Nokkala}\ \emph {et~al.}(2023)\citenamefont {Nokkala}, \citenamefont {Piilo},\ and\ \citenamefont {Bianconi}}]{nokkala2023complex}%
  \BibitemOpen
  \bibfield  {author} {\bibinfo {author} {\bibfnamefont {J.}~\bibnamefont {Nokkala}}, \bibinfo {author} {\bibfnamefont {J.}~\bibnamefont {Piilo}},\ and\ \bibinfo {author} {\bibfnamefont {G.}~\bibnamefont {Bianconi}},\ }\href@noop {} {\bibfield  {journal} {\bibinfo  {journal} {Journal of Physics A: Mathematical and Theoretical}\ } (\bibinfo {year} {2023})}\BibitemShut {NoStop}%
\bibitem [{\citenamefont {Kolobov}(2007)}]{kolobov}%
  \BibitemOpen
  \bibfield  {author} {\bibinfo {author} {\bibfnamefont {M.~I.}\ \bibnamefont {Kolobov}},\ }\href {https://books.google.es/books?id=0EUO8872LpUC} {\emph {\bibinfo {title} {Quantum Imaging}}}\ (\bibinfo  {publisher} {Springer New York},\ \bibinfo {year} {2007})\BibitemShut {NoStop}%
\bibitem [{\citenamefont {Diddams}\ \emph {et~al.}(2020)\citenamefont {Diddams}, \citenamefont {Vahala},\ and\ \citenamefont {Udem}}]{diddams2020optical}%
  \BibitemOpen
  \bibfield  {author} {\bibinfo {author} {\bibfnamefont {S.~A.}\ \bibnamefont {Diddams}}, \bibinfo {author} {\bibfnamefont {K.}~\bibnamefont {Vahala}},\ and\ \bibinfo {author} {\bibfnamefont {T.}~\bibnamefont {Udem}},\ }\href@noop {} {\bibfield  {journal} {\bibinfo  {journal} {Science}\ }\textbf {\bibinfo {volume} {369}},\ \bibinfo {pages} {eaay3676} (\bibinfo {year} {2020})}\BibitemShut {NoStop}%
\bibitem [{\citenamefont {Roslund}\ \emph {et~al.}(2014)\citenamefont {Roslund}, \citenamefont {De~Araujo}, \citenamefont {Jiang}, \citenamefont {Fabre},\ and\ \citenamefont {Treps}}]{roslund2014wavelength}%
  \BibitemOpen
  \bibfield  {author} {\bibinfo {author} {\bibfnamefont {J.}~\bibnamefont {Roslund}}, \bibinfo {author} {\bibfnamefont {R.~M.}\ \bibnamefont {De~Araujo}}, \bibinfo {author} {\bibfnamefont {S.}~\bibnamefont {Jiang}}, \bibinfo {author} {\bibfnamefont {C.}~\bibnamefont {Fabre}},\ and\ \bibinfo {author} {\bibfnamefont {N.}~\bibnamefont {Treps}},\ }\href@noop {} {\bibfield  {journal} {\bibinfo  {journal} {Nature Photonics}\ }\textbf {\bibinfo {volume} {8}},\ \bibinfo {pages} {109} (\bibinfo {year} {2014})}\BibitemShut {NoStop}%
\bibitem [{\citenamefont {Reimer}\ \emph {et~al.}(2016)\citenamefont {Reimer}, \citenamefont {Kues}, \citenamefont {Roztocki}, \citenamefont {Wetzel}, \citenamefont {Grazioso}, \citenamefont {Little}, \citenamefont {Chu}, \citenamefont {Johnston}, \citenamefont {Bromberg}, \citenamefont {Caspani}, \citenamefont {Moss},\ and\ \citenamefont {Morandotti}}]{reimer2016multiphoton}%
  \BibitemOpen
  \bibfield  {author} {\bibinfo {author} {\bibfnamefont {C.}~\bibnamefont {Reimer}}, \bibinfo {author} {\bibfnamefont {M.}~\bibnamefont {Kues}}, \bibinfo {author} {\bibfnamefont {P.}~\bibnamefont {Roztocki}}, \bibinfo {author} {\bibfnamefont {B.}~\bibnamefont {Wetzel}}, \bibinfo {author} {\bibfnamefont {F.}~\bibnamefont {Grazioso}}, \bibinfo {author} {\bibfnamefont {B.~E.}\ \bibnamefont {Little}}, \bibinfo {author} {\bibfnamefont {S.~T.}\ \bibnamefont {Chu}}, \bibinfo {author} {\bibfnamefont {T.}~\bibnamefont {Johnston}}, \bibinfo {author} {\bibfnamefont {Y.}~\bibnamefont {Bromberg}}, \bibinfo {author} {\bibfnamefont {L.}~\bibnamefont {Caspani}}, \bibinfo {author} {\bibfnamefont {D.~J.}\ \bibnamefont {Moss}},\ and\ \bibinfo {author} {\bibfnamefont {R.}~\bibnamefont {Morandotti}},\ }\href {https://doi.org/10.1126/science.aad8532} {\bibfield  {journal} {\bibinfo  {journal} {Science}\ }\textbf {\bibinfo {volume} {351}},\ \bibinfo {pages} {1176} (\bibinfo {year} {2016})},\ \Eprint
  {https://arxiv.org/abs/https://www.science.org/doi/pdf/10.1126/science.aad8532} {https://www.science.org/doi/pdf/10.1126/science.aad8532} \BibitemShut {NoStop}%
\bibitem [{\citenamefont {Cai}\ \emph {et~al.}(2017)\citenamefont {Cai}, \citenamefont {Roslund}, \citenamefont {Ferrini}, \citenamefont {Arzani}, \citenamefont {Xu}, \citenamefont {Fabre},\ and\ \citenamefont {Treps}}]{Cai2017}%
  \BibitemOpen
  \bibfield  {author} {\bibinfo {author} {\bibfnamefont {Y.}~\bibnamefont {Cai}}, \bibinfo {author} {\bibfnamefont {J.}~\bibnamefont {Roslund}}, \bibinfo {author} {\bibfnamefont {G.}~\bibnamefont {Ferrini}}, \bibinfo {author} {\bibfnamefont {F.}~\bibnamefont {Arzani}}, \bibinfo {author} {\bibfnamefont {X.}~\bibnamefont {Xu}}, \bibinfo {author} {\bibfnamefont {C.}~\bibnamefont {Fabre}},\ and\ \bibinfo {author} {\bibfnamefont {N.}~\bibnamefont {Treps}},\ }\href {https://doi.org/10.1038/ncomms15645} {\bibfield  {journal} {\bibinfo  {journal} {Nature Communications}\ }\textbf {\bibinfo {volume} {8}},\ \bibinfo {pages} {15645} (\bibinfo {year} {2017})}\BibitemShut {NoStop}%
\bibitem [{\citenamefont {Brod}\ \emph {et~al.}(2019)\citenamefont {Brod}, \citenamefont {Galv{\~a}o}, \citenamefont {Crespi}, \citenamefont {Osellame}, \citenamefont {Spagnolo},\ and\ \citenamefont {Sciarrino}}]{brod2019photonic}%
  \BibitemOpen
  \bibfield  {author} {\bibinfo {author} {\bibfnamefont {D.~J.}\ \bibnamefont {Brod}}, \bibinfo {author} {\bibfnamefont {E.~F.}\ \bibnamefont {Galv{\~a}o}}, \bibinfo {author} {\bibfnamefont {A.}~\bibnamefont {Crespi}}, \bibinfo {author} {\bibfnamefont {R.}~\bibnamefont {Osellame}}, \bibinfo {author} {\bibfnamefont {N.}~\bibnamefont {Spagnolo}},\ and\ \bibinfo {author} {\bibfnamefont {F.}~\bibnamefont {Sciarrino}},\ }\href@noop {} {\bibfield  {journal} {\bibinfo  {journal} {Advanced Photonics}\ }\textbf {\bibinfo {volume} {1}},\ \bibinfo {pages} {034001} (\bibinfo {year} {2019})}\BibitemShut {NoStop}%
\bibitem [{\citenamefont {Wang}\ \emph {et~al.}(2020)\citenamefont {Wang}, \citenamefont {Sciarrino}, \citenamefont {Laing},\ and\ \citenamefont {Thompson}}]{wang2020integrated}%
  \BibitemOpen
  \bibfield  {author} {\bibinfo {author} {\bibfnamefont {J.}~\bibnamefont {Wang}}, \bibinfo {author} {\bibfnamefont {F.}~\bibnamefont {Sciarrino}}, \bibinfo {author} {\bibfnamefont {A.}~\bibnamefont {Laing}},\ and\ \bibinfo {author} {\bibfnamefont {M.~G.}\ \bibnamefont {Thompson}},\ }\href {https://doi.org/10.1038/s41566-019-0532-1} {\bibfield  {journal} {\bibinfo  {journal} {Nature Photonics}\ }\textbf {\bibinfo {volume} {14}},\ \bibinfo {pages} {273} (\bibinfo {year} {2020})}\BibitemShut {NoStop}%
\bibitem [{\citenamefont {Eisert}\ \emph {et~al.}(2020)\citenamefont {Eisert}, \citenamefont {Hangleiter}, \citenamefont {Walk}, \citenamefont {Roth}, \citenamefont {Markham}, \citenamefont {Parekh}, \citenamefont {Chabaud},\ and\ \citenamefont {Kashefi}}]{eisert2020quantum}%
  \BibitemOpen
  \bibfield  {author} {\bibinfo {author} {\bibfnamefont {J.}~\bibnamefont {Eisert}}, \bibinfo {author} {\bibfnamefont {D.}~\bibnamefont {Hangleiter}}, \bibinfo {author} {\bibfnamefont {N.}~\bibnamefont {Walk}}, \bibinfo {author} {\bibfnamefont {I.}~\bibnamefont {Roth}}, \bibinfo {author} {\bibfnamefont {D.}~\bibnamefont {Markham}}, \bibinfo {author} {\bibfnamefont {R.}~\bibnamefont {Parekh}}, \bibinfo {author} {\bibfnamefont {U.}~\bibnamefont {Chabaud}},\ and\ \bibinfo {author} {\bibfnamefont {E.}~\bibnamefont {Kashefi}},\ }\href@noop {} {\bibfield  {journal} {\bibinfo  {journal} {Nature Reviews Physics}\ }\textbf {\bibinfo {volume} {2}},\ \bibinfo {pages} {382} (\bibinfo {year} {2020})}\BibitemShut {NoStop}%
\bibitem [{\citenamefont {Aolita}\ \emph {et~al.}(2015)\citenamefont {Aolita}, \citenamefont {Gogolin}, \citenamefont {Kliesch},\ and\ \citenamefont {Eisert}}]{aolita2015reliable}%
  \BibitemOpen
  \bibfield  {author} {\bibinfo {author} {\bibfnamefont {L.}~\bibnamefont {Aolita}}, \bibinfo {author} {\bibfnamefont {C.}~\bibnamefont {Gogolin}}, \bibinfo {author} {\bibfnamefont {M.}~\bibnamefont {Kliesch}},\ and\ \bibinfo {author} {\bibfnamefont {J.}~\bibnamefont {Eisert}},\ }\href@noop {} {\bibfield  {journal} {\bibinfo  {journal} {Nature communications}\ }\textbf {\bibinfo {volume} {6}},\ \bibinfo {pages} {8498} (\bibinfo {year} {2015})}\BibitemShut {NoStop}%
\bibitem [{\citenamefont {Wang}\ \emph {et~al.}(2023)\citenamefont {Wang}, \citenamefont {Pozas-Kerstjens}, \citenamefont {Zhang}, \citenamefont {Liu}, \citenamefont {Huang}, \citenamefont {Li}, \citenamefont {Guo}, \citenamefont {Gisin},\ and\ \citenamefont {Tavakoli}}]{wang2023certification}%
  \BibitemOpen
  \bibfield  {author} {\bibinfo {author} {\bibfnamefont {N.-N.}\ \bibnamefont {Wang}}, \bibinfo {author} {\bibfnamefont {A.}~\bibnamefont {Pozas-Kerstjens}}, \bibinfo {author} {\bibfnamefont {C.}~\bibnamefont {Zhang}}, \bibinfo {author} {\bibfnamefont {B.-H.}\ \bibnamefont {Liu}}, \bibinfo {author} {\bibfnamefont {Y.-F.}\ \bibnamefont {Huang}}, \bibinfo {author} {\bibfnamefont {C.-F.}\ \bibnamefont {Li}}, \bibinfo {author} {\bibfnamefont {G.-C.}\ \bibnamefont {Guo}}, \bibinfo {author} {\bibfnamefont {N.}~\bibnamefont {Gisin}},\ and\ \bibinfo {author} {\bibfnamefont {A.}~\bibnamefont {Tavakoli}},\ }\href@noop {} {\bibfield  {journal} {\bibinfo  {journal} {Nature Communications}\ }\textbf {\bibinfo {volume} {14}},\ \bibinfo {pages} {2153} (\bibinfo {year} {2023})}\BibitemShut {NoStop}%
\bibitem [{\citenamefont {{\v{S}}upi{\'c}}\ \emph {et~al.}(2023)\citenamefont {{\v{S}}upi{\'c}}, \citenamefont {Bowles}, \citenamefont {Renou}, \citenamefont {Ac{\'\i}n},\ and\ \citenamefont {Hoban}}]{vsupic2023quantum}%
  \BibitemOpen
  \bibfield  {author} {\bibinfo {author} {\bibfnamefont {I.}~\bibnamefont {{\v{S}}upi{\'c}}}, \bibinfo {author} {\bibfnamefont {J.}~\bibnamefont {Bowles}}, \bibinfo {author} {\bibfnamefont {M.-O.}\ \bibnamefont {Renou}}, \bibinfo {author} {\bibfnamefont {A.}~\bibnamefont {Ac{\'\i}n}},\ and\ \bibinfo {author} {\bibfnamefont {M.~J.}\ \bibnamefont {Hoban}},\ }\href@noop {} {\bibfield  {journal} {\bibinfo  {journal} {Nature Physics}\ }\textbf {\bibinfo {volume} {19}},\ \bibinfo {pages} {670} (\bibinfo {year} {2023})}\BibitemShut {NoStop}%
\bibitem [{\citenamefont {Manzano}\ \emph {et~al.}(2013)\citenamefont {Manzano}, \citenamefont {Galve}, \citenamefont {Giorgi}, \citenamefont {Hern{\'a}ndez-Garc{\'\i}a},\ and\ \citenamefont {Zambrini}}]{manzano2013synchronization}%
  \BibitemOpen
  \bibfield  {author} {\bibinfo {author} {\bibfnamefont {G.}~\bibnamefont {Manzano}}, \bibinfo {author} {\bibfnamefont {F.}~\bibnamefont {Galve}}, \bibinfo {author} {\bibfnamefont {G.~L.}\ \bibnamefont {Giorgi}}, \bibinfo {author} {\bibfnamefont {E.}~\bibnamefont {Hern{\'a}ndez-Garc{\'\i}a}},\ and\ \bibinfo {author} {\bibfnamefont {R.}~\bibnamefont {Zambrini}},\ }\href@noop {} {\bibfield  {journal} {\bibinfo  {journal} {Scientific Reports}\ }\textbf {\bibinfo {volume} {3}},\ \bibinfo {pages} {1439} (\bibinfo {year} {2013})}\BibitemShut {NoStop}%
\bibitem [{\citenamefont {Zambrini}\ \emph {et~al.}(2000)\citenamefont {Zambrini}, \citenamefont {Hoyuelos}, \citenamefont {Gatti}, \citenamefont {Colet}, \citenamefont {Lugiato},\ and\ \citenamefont {San~Miguel}}]{zambrini2000quantum}%
  \BibitemOpen
  \bibfield  {author} {\bibinfo {author} {\bibfnamefont {R.}~\bibnamefont {Zambrini}}, \bibinfo {author} {\bibfnamefont {M.}~\bibnamefont {Hoyuelos}}, \bibinfo {author} {\bibfnamefont {A.}~\bibnamefont {Gatti}}, \bibinfo {author} {\bibfnamefont {P.}~\bibnamefont {Colet}}, \bibinfo {author} {\bibfnamefont {L.}~\bibnamefont {Lugiato}},\ and\ \bibinfo {author} {\bibfnamefont {M.}~\bibnamefont {San~Miguel}},\ }\href@noop {} {\bibfield  {journal} {\bibinfo  {journal} {Physical Review A}\ }\textbf {\bibinfo {volume} {62}},\ \bibinfo {pages} {063801} (\bibinfo {year} {2000})}\BibitemShut {NoStop}%
\bibitem [{\citenamefont {Giorgi}\ \emph {et~al.}(2012)\citenamefont {Giorgi}, \citenamefont {Galve}, \citenamefont {Manzano}, \citenamefont {Colet},\ and\ \citenamefont {Zambrini}}]{PhysRevA.85.052101}%
  \BibitemOpen
  \bibfield  {author} {\bibinfo {author} {\bibfnamefont {G.~L.}\ \bibnamefont {Giorgi}}, \bibinfo {author} {\bibfnamefont {F.}~\bibnamefont {Galve}}, \bibinfo {author} {\bibfnamefont {G.}~\bibnamefont {Manzano}}, \bibinfo {author} {\bibfnamefont {P.}~\bibnamefont {Colet}},\ and\ \bibinfo {author} {\bibfnamefont {R.}~\bibnamefont {Zambrini}},\ }\href {https://doi.org/10.1103/PhysRevA.85.052101} {\bibfield  {journal} {\bibinfo  {journal} {Phys. Rev. A}\ }\textbf {\bibinfo {volume} {85}},\ \bibinfo {pages} {052101} (\bibinfo {year} {2012})}\BibitemShut {NoStop}%
\bibitem [{\citenamefont {Cabot}\ \emph {et~al.}(2018)\citenamefont {Cabot}, \citenamefont {Galve}, \citenamefont {Egu{\'\i}luz}, \citenamefont {Klemm}, \citenamefont {Maniscalco},\ and\ \citenamefont {Zambrini}}]{cabot2018unveiling}%
  \BibitemOpen
  \bibfield  {author} {\bibinfo {author} {\bibfnamefont {A.}~\bibnamefont {Cabot}}, \bibinfo {author} {\bibfnamefont {F.}~\bibnamefont {Galve}}, \bibinfo {author} {\bibfnamefont {V.~M.}\ \bibnamefont {Egu{\'\i}luz}}, \bibinfo {author} {\bibfnamefont {K.}~\bibnamefont {Klemm}}, \bibinfo {author} {\bibfnamefont {S.}~\bibnamefont {Maniscalco}},\ and\ \bibinfo {author} {\bibfnamefont {R.}~\bibnamefont {Zambrini}},\ }\href@noop {} {\bibfield  {journal} {\bibinfo  {journal} {npj Quantum Information}\ }\textbf {\bibinfo {volume} {4}},\ \bibinfo {pages} {57} (\bibinfo {year} {2018})}\BibitemShut {NoStop}%
\bibitem [{\citenamefont {Cabot}\ \emph {et~al.}(2021)\citenamefont {Cabot}, \citenamefont {Giorgi},\ and\ \citenamefont {Zambrini}}]{cabot2021metastable}%
  \BibitemOpen
  \bibfield  {author} {\bibinfo {author} {\bibfnamefont {A.}~\bibnamefont {Cabot}}, \bibinfo {author} {\bibfnamefont {G.~L.}\ \bibnamefont {Giorgi}},\ and\ \bibinfo {author} {\bibfnamefont {R.}~\bibnamefont {Zambrini}},\ }\href@noop {} {\bibfield  {journal} {\bibinfo  {journal} {New Journal of Physics}\ } (\bibinfo {year} {2021})}\BibitemShut {NoStop}%
\bibitem [{\citenamefont {Else}\ \emph {et~al.}(2016)\citenamefont {Else}, \citenamefont {Bauer},\ and\ \citenamefont {Nayak}}]{Else2016}%
  \BibitemOpen
  \bibfield  {author} {\bibinfo {author} {\bibfnamefont {D.~V.}\ \bibnamefont {Else}}, \bibinfo {author} {\bibfnamefont {B.}~\bibnamefont {Bauer}},\ and\ \bibinfo {author} {\bibfnamefont {C.}~\bibnamefont {Nayak}},\ }\href {https://doi.org/10.1103/PhysRevLett.117.090402} {\bibfield  {journal} {\bibinfo  {journal} {Phys. Rev. Lett.}\ }\textbf {\bibinfo {volume} {117}},\ \bibinfo {pages} {090402} (\bibinfo {year} {2016})}\BibitemShut {NoStop}%
\bibitem [{\citenamefont {Iemini}\ \emph {et~al.}(2018)\citenamefont {Iemini}, \citenamefont {Russomanno}, \citenamefont {Keeling}, \citenamefont {Schir\`o}, \citenamefont {Dalmonte},\ and\ \citenamefont {Fazio}}]{Iemini2018}%
  \BibitemOpen
  \bibfield  {author} {\bibinfo {author} {\bibfnamefont {F.}~\bibnamefont {Iemini}}, \bibinfo {author} {\bibfnamefont {A.}~\bibnamefont {Russomanno}}, \bibinfo {author} {\bibfnamefont {J.}~\bibnamefont {Keeling}}, \bibinfo {author} {\bibfnamefont {M.}~\bibnamefont {Schir\`o}}, \bibinfo {author} {\bibfnamefont {M.}~\bibnamefont {Dalmonte}},\ and\ \bibinfo {author} {\bibfnamefont {R.}~\bibnamefont {Fazio}},\ }\href {https://doi.org/10.1103/PhysRevLett.121.035301} {\bibfield  {journal} {\bibinfo  {journal} {Phys. Rev. Lett.}\ }\textbf {\bibinfo {volume} {121}},\ \bibinfo {pages} {035301} (\bibinfo {year} {2018})}\BibitemShut {NoStop}%
\bibitem [{\citenamefont {Cabot}\ \emph {et~al.}(2023)\citenamefont {Cabot}, \citenamefont {Giorgi},\ and\ \citenamefont {Zambrini}}]{cabot2023nonequilibrium}%
  \BibitemOpen
  \bibfield  {author} {\bibinfo {author} {\bibfnamefont {A.}~\bibnamefont {Cabot}}, \bibinfo {author} {\bibfnamefont {G.~L.}\ \bibnamefont {Giorgi}},\ and\ \bibinfo {author} {\bibfnamefont {R.}~\bibnamefont {Zambrini}},\ }\href@noop {} {\bibfield  {journal} {\bibinfo  {journal} {PRX Quantum (in press)}\ } (\bibinfo {year} {2023})}\BibitemShut {NoStop}%
\bibitem [{\citenamefont {Vasile}\ \emph {et~al.}(2014)\citenamefont {Vasile}, \citenamefont {Galve},\ and\ \citenamefont {Zambrini}}]{vasile2014spectral}%
  \BibitemOpen
  \bibfield  {author} {\bibinfo {author} {\bibfnamefont {R.}~\bibnamefont {Vasile}}, \bibinfo {author} {\bibfnamefont {F.}~\bibnamefont {Galve}},\ and\ \bibinfo {author} {\bibfnamefont {R.}~\bibnamefont {Zambrini}},\ }\href@noop {} {\bibfield  {journal} {\bibinfo  {journal} {Physical Review A}\ }\textbf {\bibinfo {volume} {89}},\ \bibinfo {pages} {022109} (\bibinfo {year} {2014})}\BibitemShut {NoStop}%
\bibitem [{\citenamefont {Nokkala}\ \emph {et~al.}(2018)\citenamefont {Nokkala}, \citenamefont {Arzani}, \citenamefont {Galve}, \citenamefont {Zambrini}, \citenamefont {Maniscalco}, \citenamefont {Piilo}, \citenamefont {Treps},\ and\ \citenamefont {Parigi}}]{Nokkala_2018}%
  \BibitemOpen
  \bibfield  {author} {\bibinfo {author} {\bibfnamefont {J.}~\bibnamefont {Nokkala}}, \bibinfo {author} {\bibfnamefont {F.}~\bibnamefont {Arzani}}, \bibinfo {author} {\bibfnamefont {F.}~\bibnamefont {Galve}}, \bibinfo {author} {\bibfnamefont {R.}~\bibnamefont {Zambrini}}, \bibinfo {author} {\bibfnamefont {S.}~\bibnamefont {Maniscalco}}, \bibinfo {author} {\bibfnamefont {J.}~\bibnamefont {Piilo}}, \bibinfo {author} {\bibfnamefont {N.}~\bibnamefont {Treps}},\ and\ \bibinfo {author} {\bibfnamefont {V.}~\bibnamefont {Parigi}},\ }\href {https://doi.org/10.1088/1367-2630/aabc77} {\bibfield  {journal} {\bibinfo  {journal} {New Journal of Physics}\ }\textbf {\bibinfo {volume} {20}},\ \bibinfo {pages} {053024} (\bibinfo {year} {2018})}\BibitemShut {NoStop}%
\bibitem [{\citenamefont {Renault}\ \emph {et~al.}(2023)\citenamefont {Renault}, \citenamefont {Nokkala}, \citenamefont {Roeland}, \citenamefont {Joly}, \citenamefont {Zambrini}, \citenamefont {Maniscalco}, \citenamefont {Piilo}, \citenamefont {Treps},\ and\ \citenamefont {Parigi}}]{PRXQuantum.4.040310}%
  \BibitemOpen
  \bibfield  {author} {\bibinfo {author} {\bibfnamefont {P.}~\bibnamefont {Renault}}, \bibinfo {author} {\bibfnamefont {J.}~\bibnamefont {Nokkala}}, \bibinfo {author} {\bibfnamefont {G.}~\bibnamefont {Roeland}}, \bibinfo {author} {\bibfnamefont {N.}~\bibnamefont {Joly}}, \bibinfo {author} {\bibfnamefont {R.}~\bibnamefont {Zambrini}}, \bibinfo {author} {\bibfnamefont {S.}~\bibnamefont {Maniscalco}}, \bibinfo {author} {\bibfnamefont {J.}~\bibnamefont {Piilo}}, \bibinfo {author} {\bibfnamefont {N.}~\bibnamefont {Treps}},\ and\ \bibinfo {author} {\bibfnamefont {V.}~\bibnamefont {Parigi}},\ }\href {https://doi.org/10.1103/PRXQuantum.4.040310} {\bibfield  {journal} {\bibinfo  {journal} {PRX Quantum}\ }\textbf {\bibinfo {volume} {4}},\ \bibinfo {pages} {040310} (\bibinfo {year} {2023})}\BibitemShut {NoStop}%
\bibitem [{\citenamefont {Wehner}\ \emph {et~al.}(2018)\citenamefont {Wehner}, \citenamefont {Elkouss},\ and\ \citenamefont {Hanson}}]{wehner2018quantum}%
  \BibitemOpen
  \bibfield  {author} {\bibinfo {author} {\bibfnamefont {S.}~\bibnamefont {Wehner}}, \bibinfo {author} {\bibfnamefont {D.}~\bibnamefont {Elkouss}},\ and\ \bibinfo {author} {\bibfnamefont {R.}~\bibnamefont {Hanson}},\ }\href@noop {} {\bibfield  {journal} {\bibinfo  {journal} {Science}\ }\textbf {\bibinfo {volume} {362}},\ \bibinfo {pages} {eaam9288} (\bibinfo {year} {2018})}\BibitemShut {NoStop}%
\bibitem [{\citenamefont {Kimble}(2008)}]{kimble2008quantum}%
  \BibitemOpen
  \bibfield  {author} {\bibinfo {author} {\bibfnamefont {H.~J.}\ \bibnamefont {Kimble}},\ }\href@noop {} {\bibfield  {journal} {\bibinfo  {journal} {Nature}\ }\textbf {\bibinfo {volume} {453}},\ \bibinfo {pages} {1023} (\bibinfo {year} {2008})}\BibitemShut {NoStop}%
\bibitem [{\citenamefont {Shor}\ and\ \citenamefont {Preskill}(2000)}]{shor2000simple}%
  \BibitemOpen
  \bibfield  {author} {\bibinfo {author} {\bibfnamefont {P.~W.}\ \bibnamefont {Shor}}\ and\ \bibinfo {author} {\bibfnamefont {J.}~\bibnamefont {Preskill}},\ }\href@noop {} {\bibfield  {journal} {\bibinfo  {journal} {Physical review letters}\ }\textbf {\bibinfo {volume} {85}},\ \bibinfo {pages} {441} (\bibinfo {year} {2000})}\BibitemShut {NoStop}%
\bibitem [{\citenamefont {Zhang}\ and\ \citenamefont {Zhuang}(2021)}]{zhang2021distributed}%
  \BibitemOpen
  \bibfield  {author} {\bibinfo {author} {\bibfnamefont {Z.}~\bibnamefont {Zhang}}\ and\ \bibinfo {author} {\bibfnamefont {Q.}~\bibnamefont {Zhuang}},\ }\href@noop {} {\bibfield  {journal} {\bibinfo  {journal} {Quantum Science and Technology}\ }\textbf {\bibinfo {volume} {6}},\ \bibinfo {pages} {043001} (\bibinfo {year} {2021})}\BibitemShut {NoStop}%
\bibitem [{\citenamefont {Cirac}\ \emph {et~al.}(1999)\citenamefont {Cirac}, \citenamefont {Ekert}, \citenamefont {Huelga},\ and\ \citenamefont {Macchiavello}}]{cirac1999distributed}%
  \BibitemOpen
  \bibfield  {author} {\bibinfo {author} {\bibfnamefont {J.}~\bibnamefont {Cirac}}, \bibinfo {author} {\bibfnamefont {A.}~\bibnamefont {Ekert}}, \bibinfo {author} {\bibfnamefont {S.}~\bibnamefont {Huelga}},\ and\ \bibinfo {author} {\bibfnamefont {C.}~\bibnamefont {Macchiavello}},\ }\href@noop {} {\bibfield  {journal} {\bibinfo  {journal} {Physical Review A}\ }\textbf {\bibinfo {volume} {59}},\ \bibinfo {pages} {4249} (\bibinfo {year} {1999})}\BibitemShut {NoStop}%
\bibitem [{\citenamefont {Wootters}\ and\ \citenamefont {Zurek}(1982)}]{wootters1982single}%
  \BibitemOpen
  \bibfield  {author} {\bibinfo {author} {\bibfnamefont {W.~K.}\ \bibnamefont {Wootters}}\ and\ \bibinfo {author} {\bibfnamefont {W.~H.}\ \bibnamefont {Zurek}},\ }\href@noop {} {\bibfield  {journal} {\bibinfo  {journal} {Nature}\ }\textbf {\bibinfo {volume} {299}},\ \bibinfo {pages} {802} (\bibinfo {year} {1982})}\BibitemShut {NoStop}%
\bibitem [{\citenamefont {Azuma}\ \emph {et~al.}(2015)\citenamefont {Azuma}, \citenamefont {Tamaki},\ and\ \citenamefont {Lo}}]{azuma2015all}%
  \BibitemOpen
  \bibfield  {author} {\bibinfo {author} {\bibfnamefont {K.}~\bibnamefont {Azuma}}, \bibinfo {author} {\bibfnamefont {K.}~\bibnamefont {Tamaki}},\ and\ \bibinfo {author} {\bibfnamefont {H.-K.}\ \bibnamefont {Lo}},\ }\href@noop {} {\bibfield  {journal} {\bibinfo  {journal} {Nature communications}\ }\textbf {\bibinfo {volume} {6}},\ \bibinfo {pages} {1} (\bibinfo {year} {2015})}\BibitemShut {NoStop}%
\bibitem [{\citenamefont {Weedbrook}\ \emph {et~al.}(2012)\citenamefont {Weedbrook}, \citenamefont {Pirandola}, \citenamefont {Garc\'{\i}a-Patr\'on}, \citenamefont {Cerf}, \citenamefont {Ralph}, \citenamefont {Shapiro},\ and\ \citenamefont {Lloyd}}]{lloyd2012Gaussianqi}%
  \BibitemOpen
  \bibfield  {author} {\bibinfo {author} {\bibfnamefont {C.}~\bibnamefont {Weedbrook}}, \bibinfo {author} {\bibfnamefont {S.}~\bibnamefont {Pirandola}}, \bibinfo {author} {\bibfnamefont {R.}~\bibnamefont {Garc\'{\i}a-Patr\'on}}, \bibinfo {author} {\bibfnamefont {N.~J.}\ \bibnamefont {Cerf}}, \bibinfo {author} {\bibfnamefont {T.~C.}\ \bibnamefont {Ralph}}, \bibinfo {author} {\bibfnamefont {J.~H.}\ \bibnamefont {Shapiro}},\ and\ \bibinfo {author} {\bibfnamefont {S.}~\bibnamefont {Lloyd}},\ }\href {https://doi.org/10.1103/RevModPhys.84.621} {\bibfield  {journal} {\bibinfo  {journal} {Rev. Mod. Phys.}\ }\textbf {\bibinfo {volume} {84}},\ \bibinfo {pages} {621} (\bibinfo {year} {2012})}\BibitemShut {NoStop}%
\bibitem [{\citenamefont {Laudenbach}\ \emph {et~al.}(2018)\citenamefont {Laudenbach}, \citenamefont {Pacher}, \citenamefont {Fung}, \citenamefont {Poppe}, \citenamefont {Peev}, \citenamefont {Schrenk}, \citenamefont {Hentschel}, \citenamefont {Walther},\ and\ \citenamefont {H{\"u}bel}}]{laudenbach2018cvqkd}%
  \BibitemOpen
  \bibfield  {author} {\bibinfo {author} {\bibfnamefont {F.}~\bibnamefont {Laudenbach}}, \bibinfo {author} {\bibfnamefont {C.}~\bibnamefont {Pacher}}, \bibinfo {author} {\bibfnamefont {C.-H.~F.}\ \bibnamefont {Fung}}, \bibinfo {author} {\bibfnamefont {A.}~\bibnamefont {Poppe}}, \bibinfo {author} {\bibfnamefont {M.}~\bibnamefont {Peev}}, \bibinfo {author} {\bibfnamefont {B.}~\bibnamefont {Schrenk}}, \bibinfo {author} {\bibfnamefont {M.}~\bibnamefont {Hentschel}}, \bibinfo {author} {\bibfnamefont {P.}~\bibnamefont {Walther}},\ and\ \bibinfo {author} {\bibfnamefont {H.}~\bibnamefont {H{\"u}bel}},\ }\href@noop {} {\bibfield  {journal} {\bibinfo  {journal} {Advanced Quantum Technologies}\ }\textbf {\bibinfo {volume} {1}},\ \bibinfo {pages} {1800011} (\bibinfo {year} {2018})}\BibitemShut {NoStop}%
\bibitem [{\citenamefont {Cerf}\ \emph {et~al.}(2001)\citenamefont {Cerf}, \citenamefont {L\'evy},\ and\ \citenamefont {Assche}}]{cerf2001cvqkd}%
  \BibitemOpen
  \bibfield  {author} {\bibinfo {author} {\bibfnamefont {N.~J.}\ \bibnamefont {Cerf}}, \bibinfo {author} {\bibfnamefont {M.}~\bibnamefont {L\'evy}},\ and\ \bibinfo {author} {\bibfnamefont {G.~V.}\ \bibnamefont {Assche}},\ }\href {https://doi.org/10.1103/PhysRevA.63.052311} {\bibfield  {journal} {\bibinfo  {journal} {Phys. Rev. A}\ }\textbf {\bibinfo {volume} {63}},\ \bibinfo {pages} {052311} (\bibinfo {year} {2001})}\BibitemShut {NoStop}%
\bibitem [{\citenamefont {Gottesman}\ and\ \citenamefont {Preskill}(2003)}]{gottesman2003secure}%
  \BibitemOpen
  \bibfield  {author} {\bibinfo {author} {\bibfnamefont {D.}~\bibnamefont {Gottesman}}\ and\ \bibinfo {author} {\bibfnamefont {J.}~\bibnamefont {Preskill}},\ }\href@noop {} {\bibfield  {journal} {\bibinfo  {journal} {Quantum Information with Continuous Variables}\ ,\ \bibinfo {pages} {317}} (\bibinfo {year} {2003})}\BibitemShut {NoStop}%
\bibitem [{\citenamefont {Pirandola}\ \emph {et~al.}(2009)\citenamefont {Pirandola}, \citenamefont {Garc\'{\i}a-Patr\'on}, \citenamefont {Braunstein},\ and\ \citenamefont {Lloyd}}]{pirandola2009noiseqkd}%
  \BibitemOpen
  \bibfield  {author} {\bibinfo {author} {\bibfnamefont {S.}~\bibnamefont {Pirandola}}, \bibinfo {author} {\bibfnamefont {R.}~\bibnamefont {Garc\'{\i}a-Patr\'on}}, \bibinfo {author} {\bibfnamefont {S.~L.}\ \bibnamefont {Braunstein}},\ and\ \bibinfo {author} {\bibfnamefont {S.}~\bibnamefont {Lloyd}},\ }\href {https://doi.org/10.1103/PhysRevLett.102.050503} {\bibfield  {journal} {\bibinfo  {journal} {Phys. Rev. Lett.}\ }\textbf {\bibinfo {volume} {102}},\ \bibinfo {pages} {050503} (\bibinfo {year} {2009})}\BibitemShut {NoStop}%
\bibitem [{\citenamefont {Deutsch}(1989)}]{deutsch1989quantum}%
  \BibitemOpen
  \bibfield  {author} {\bibinfo {author} {\bibfnamefont {D.~E.}\ \bibnamefont {Deutsch}},\ }\href@noop {} {\bibfield  {journal} {\bibinfo  {journal} {Proceedings of the royal society of London. A. mathematical and physical sciences}\ }\textbf {\bibinfo {volume} {425}},\ \bibinfo {pages} {73} (\bibinfo {year} {1989})}\BibitemShut {NoStop}%
\bibitem [{\citenamefont {Buluta}\ and\ \citenamefont {Nori}(2009)}]{buluta2009quantum}%
  \BibitemOpen
  \bibfield  {author} {\bibinfo {author} {\bibfnamefont {I.}~\bibnamefont {Buluta}}\ and\ \bibinfo {author} {\bibfnamefont {F.}~\bibnamefont {Nori}},\ }\href@noop {} {\bibfield  {journal} {\bibinfo  {journal} {Science}\ }\textbf {\bibinfo {volume} {326}},\ \bibinfo {pages} {108} (\bibinfo {year} {2009})}\BibitemShut {NoStop}%
\bibitem [{\citenamefont {Kendon}\ \emph {et~al.}(2010)\citenamefont {Kendon}, \citenamefont {Nemoto},\ and\ \citenamefont {Munro}}]{kendon2010quantum}%
  \BibitemOpen
  \bibfield  {author} {\bibinfo {author} {\bibfnamefont {V.~M.}\ \bibnamefont {Kendon}}, \bibinfo {author} {\bibfnamefont {K.}~\bibnamefont {Nemoto}},\ and\ \bibinfo {author} {\bibfnamefont {W.~J.}\ \bibnamefont {Munro}},\ }\href@noop {} {\bibfield  {journal} {\bibinfo  {journal} {Philosophical Transactions of the Royal Society A: Mathematical, Physical and Engineering Sciences}\ }\textbf {\bibinfo {volume} {368}},\ \bibinfo {pages} {3609} (\bibinfo {year} {2010})}\BibitemShut {NoStop}%
\bibitem [{\citenamefont {Walmsley}(2023)}]{Walmsley:23}%
  \BibitemOpen
  \bibfield  {author} {\bibinfo {author} {\bibfnamefont {I.}~\bibnamefont {Walmsley}},\ }\href {https://doi.org/10.1364/OPTICAQ.507527} {\bibfield  {journal} {\bibinfo  {journal} {Optica Quantum}\ }\textbf {\bibinfo {volume} {1}},\ \bibinfo {pages} {35} (\bibinfo {year} {2023})}\BibitemShut {NoStop}%
\bibitem [{\citenamefont {Aspuru-Guzik}\ and\ \citenamefont {Walther}(2012)}]{aspuru2012photonic}%
  \BibitemOpen
  \bibfield  {author} {\bibinfo {author} {\bibfnamefont {A.}~\bibnamefont {Aspuru-Guzik}}\ and\ \bibinfo {author} {\bibfnamefont {P.}~\bibnamefont {Walther}},\ }\href@noop {} {\bibfield  {journal} {\bibinfo  {journal} {Nature physics}\ }\textbf {\bibinfo {volume} {8}},\ \bibinfo {pages} {285} (\bibinfo {year} {2012})}\BibitemShut {NoStop}%
\bibitem [{\citenamefont {O'Brien}(2007)}]{opticalqc_2007}%
  \BibitemOpen
  \bibfield  {author} {\bibinfo {author} {\bibfnamefont {J.~L.}\ \bibnamefont {O'Brien}},\ }\href {https://doi.org/10.1126/science.1142892} {\bibfield  {journal} {\bibinfo  {journal} {Science}\ }\textbf {\bibinfo {volume} {318}},\ \bibinfo {pages} {1567} (\bibinfo {year} {2007})},\ \Eprint {https://arxiv.org/abs/https://www.science.org/doi/pdf/10.1126/science.1142892} {https://www.science.org/doi/pdf/10.1126/science.1142892} \BibitemShut {NoStop}%
\bibitem [{\citenamefont {Gu}\ \emph {et~al.}(2009)\citenamefont {Gu}, \citenamefont {Weedbrook}, \citenamefont {Menicucci}, \citenamefont {Ralph},\ and\ \citenamefont {van Loock}}]{Gu_clusters2009}%
  \BibitemOpen
  \bibfield  {author} {\bibinfo {author} {\bibfnamefont {M.}~\bibnamefont {Gu}}, \bibinfo {author} {\bibfnamefont {C.}~\bibnamefont {Weedbrook}}, \bibinfo {author} {\bibfnamefont {N.~C.}\ \bibnamefont {Menicucci}}, \bibinfo {author} {\bibfnamefont {T.~C.}\ \bibnamefont {Ralph}},\ and\ \bibinfo {author} {\bibfnamefont {P.}~\bibnamefont {van Loock}},\ }\href {https://doi.org/10.1103/PhysRevA.79.062318} {\bibfield  {journal} {\bibinfo  {journal} {Phys. Rev. A}\ }\textbf {\bibinfo {volume} {79}},\ \bibinfo {pages} {062318} (\bibinfo {year} {2009})}\BibitemShut {NoStop}%
\bibitem [{\citenamefont {Yoshikawa}\ \emph {et~al.}(2016)\citenamefont {Yoshikawa}, \citenamefont {Yokoyama}, \citenamefont {Kaji}, \citenamefont {Sornphiphatphong}, \citenamefont {Shiozawa}, \citenamefont {Makino},\ and\ \citenamefont {Furusawa}}]{furusawa_big}%
  \BibitemOpen
  \bibfield  {author} {\bibinfo {author} {\bibfnamefont {J.-i.}\ \bibnamefont {Yoshikawa}}, \bibinfo {author} {\bibfnamefont {S.}~\bibnamefont {Yokoyama}}, \bibinfo {author} {\bibfnamefont {T.}~\bibnamefont {Kaji}}, \bibinfo {author} {\bibfnamefont {C.}~\bibnamefont {Sornphiphatphong}}, \bibinfo {author} {\bibfnamefont {Y.}~\bibnamefont {Shiozawa}}, \bibinfo {author} {\bibfnamefont {K.}~\bibnamefont {Makino}},\ and\ \bibinfo {author} {\bibfnamefont {A.}~\bibnamefont {Furusawa}},\ }\href {https://doi.org/10.1063/1.4962732} {\bibfield  {journal} {\bibinfo  {journal} {APL Photonics}\ }\textbf {\bibinfo {volume} {1}},\ \bibinfo {pages} {060801} (\bibinfo {year} {2016})},\ \Eprint {https://arxiv.org/abs/https://pubs.aip.org/aip/app/article-pdf/doi/10.1063/1.4962732/13250765/060801\_1\_online.pdf} {https://pubs.aip.org/aip/app/article-pdf/doi/10.1063/1.4962732/13250765/060801\_1\_online.pdf} \BibitemShut {NoStop}%
\bibitem [{\citenamefont {Chen}\ \emph {et~al.}(2014)\citenamefont {Chen}, \citenamefont {Menicucci},\ and\ \citenamefont {Pfister}}]{60_frequency_cluster}%
  \BibitemOpen
  \bibfield  {author} {\bibinfo {author} {\bibfnamefont {M.}~\bibnamefont {Chen}}, \bibinfo {author} {\bibfnamefont {N.~C.}\ \bibnamefont {Menicucci}},\ and\ \bibinfo {author} {\bibfnamefont {O.}~\bibnamefont {Pfister}},\ }\href {https://doi.org/10.1103/PhysRevLett.112.120505} {\bibfield  {journal} {\bibinfo  {journal} {Phys. Rev. Lett.}\ }\textbf {\bibinfo {volume} {112}},\ \bibinfo {pages} {120505} (\bibinfo {year} {2014})}\BibitemShut {NoStop}%
\bibitem [{\citenamefont {Asavanant}\ \emph {et~al.}(2019)\citenamefont {Asavanant}, \citenamefont {Shiozawa}, \citenamefont {Yokoyama}, \citenamefont {Charoensombutamon}, \citenamefont {Emura}, \citenamefont {Alexander}, \citenamefont {Takeda}, \citenamefont {ichi Yoshikawa}, \citenamefont {Menicucci}, \citenamefont {Yonezawa},\ and\ \citenamefont {Furusawa}}]{furusawa_2d_cluster}%
  \BibitemOpen
  \bibfield  {author} {\bibinfo {author} {\bibfnamefont {W.}~\bibnamefont {Asavanant}}, \bibinfo {author} {\bibfnamefont {Y.}~\bibnamefont {Shiozawa}}, \bibinfo {author} {\bibfnamefont {S.}~\bibnamefont {Yokoyama}}, \bibinfo {author} {\bibfnamefont {B.}~\bibnamefont {Charoensombutamon}}, \bibinfo {author} {\bibfnamefont {H.}~\bibnamefont {Emura}}, \bibinfo {author} {\bibfnamefont {R.~N.}\ \bibnamefont {Alexander}}, \bibinfo {author} {\bibfnamefont {S.}~\bibnamefont {Takeda}}, \bibinfo {author} {\bibfnamefont {J.}~\bibnamefont {ichi Yoshikawa}}, \bibinfo {author} {\bibfnamefont {N.~C.}\ \bibnamefont {Menicucci}}, \bibinfo {author} {\bibfnamefont {H.}~\bibnamefont {Yonezawa}},\ and\ \bibinfo {author} {\bibfnamefont {A.}~\bibnamefont {Furusawa}},\ }\href {https://doi.org/10.1126/science.aay2645} {\bibfield  {journal} {\bibinfo  {journal} {Science}\ }\textbf {\bibinfo {volume} {366}},\ \bibinfo {pages} {373} (\bibinfo {year} {2019})},\ \Eprint
  {https://arxiv.org/abs/https://www.science.org/doi/pdf/10.1126/science.aay2645} {https://www.science.org/doi/pdf/10.1126/science.aay2645} \BibitemShut {NoStop}%
\bibitem [{\citenamefont {Larsen}\ \emph {et~al.}(2021)\citenamefont {Larsen}, \citenamefont {Guo}, \citenamefont {Breum}, \citenamefont {Neergaard-Nielsen},\ and\ \citenamefont {Andersen}}]{cluster_gates_1}%
  \BibitemOpen
  \bibfield  {author} {\bibinfo {author} {\bibfnamefont {M.~V.}\ \bibnamefont {Larsen}}, \bibinfo {author} {\bibfnamefont {X.}~\bibnamefont {Guo}}, \bibinfo {author} {\bibfnamefont {C.~R.}\ \bibnamefont {Breum}}, \bibinfo {author} {\bibfnamefont {J.~S.}\ \bibnamefont {Neergaard-Nielsen}},\ and\ \bibinfo {author} {\bibfnamefont {U.~L.}\ \bibnamefont {Andersen}},\ }\href {https://doi.org/10.1038/s41567-021-01296-y} {\bibfield  {journal} {\bibinfo  {journal} {Nature Physics}\ }\textbf {\bibinfo {volume} {17}},\ \bibinfo {pages} {1018} (\bibinfo {year} {2021})}\BibitemShut {NoStop}%
\bibitem [{\citenamefont {Asavanant}\ \emph {et~al.}(2021)\citenamefont {Asavanant}, \citenamefont {Charoensombutamon}, \citenamefont {Yokoyama}, \citenamefont {Ebihara}, \citenamefont {Nakamura}, \citenamefont {Alexander}, \citenamefont {Endo}, \citenamefont {Yoshikawa}, \citenamefont {Menicucci}, \citenamefont {Yonezawa},\ and\ \citenamefont {Furusawa}}]{cluster_gates_2}%
  \BibitemOpen
  \bibfield  {author} {\bibinfo {author} {\bibfnamefont {W.}~\bibnamefont {Asavanant}}, \bibinfo {author} {\bibfnamefont {B.}~\bibnamefont {Charoensombutamon}}, \bibinfo {author} {\bibfnamefont {S.}~\bibnamefont {Yokoyama}}, \bibinfo {author} {\bibfnamefont {T.}~\bibnamefont {Ebihara}}, \bibinfo {author} {\bibfnamefont {T.}~\bibnamefont {Nakamura}}, \bibinfo {author} {\bibfnamefont {R.~N.}\ \bibnamefont {Alexander}}, \bibinfo {author} {\bibfnamefont {M.}~\bibnamefont {Endo}}, \bibinfo {author} {\bibfnamefont {J.-i.}\ \bibnamefont {Yoshikawa}}, \bibinfo {author} {\bibfnamefont {N.~C.}\ \bibnamefont {Menicucci}}, \bibinfo {author} {\bibfnamefont {H.}~\bibnamefont {Yonezawa}},\ and\ \bibinfo {author} {\bibfnamefont {A.}~\bibnamefont {Furusawa}},\ }\href {https://doi.org/10.1103/PhysRevApplied.16.034005} {\bibfield  {journal} {\bibinfo  {journal} {Phys. Rev. Appl.}\ }\textbf {\bibinfo {volume} {16}},\ \bibinfo {pages} {034005} (\bibinfo {year} {2021})}\BibitemShut {NoStop}%
\bibitem [{\citenamefont {Kozlowski}\ and\ \citenamefont {Wehner}(2019)}]{kozlowski2019towards}%
  \BibitemOpen
  \bibfield  {author} {\bibinfo {author} {\bibfnamefont {W.}~\bibnamefont {Kozlowski}}\ and\ \bibinfo {author} {\bibfnamefont {S.}~\bibnamefont {Wehner}},\ }in\ \href@noop {} {\emph {\bibinfo {booktitle} {Proceedings of the sixth annual ACM international conference on nanoscale computing and communication}}}\ (\bibinfo {year} {2019})\ pp.\ \bibinfo {pages} {1--7}\BibitemShut {NoStop}%
\bibitem [{\citenamefont {Takeda}\ \emph {et~al.}(2013)\citenamefont {Takeda}, \citenamefont {Mizuta}, \citenamefont {Fuwa}, \citenamefont {van Loock},\ and\ \citenamefont {Furusawa}}]{Takeda2013}%
  \BibitemOpen
  \bibfield  {author} {\bibinfo {author} {\bibfnamefont {S.}~\bibnamefont {Takeda}}, \bibinfo {author} {\bibfnamefont {T.}~\bibnamefont {Mizuta}}, \bibinfo {author} {\bibfnamefont {M.}~\bibnamefont {Fuwa}}, \bibinfo {author} {\bibfnamefont {P.}~\bibnamefont {van Loock}},\ and\ \bibinfo {author} {\bibfnamefont {A.}~\bibnamefont {Furusawa}},\ }\href {https://doi.org/10.1038/nature12366} {\bibfield  {journal} {\bibinfo  {journal} {Nature}\ }\textbf {\bibinfo {volume} {500}},\ \bibinfo {pages} {315} (\bibinfo {year} {2013})}\BibitemShut {NoStop}%
\bibitem [{\citenamefont {Fowler}\ \emph {et~al.}(2012)\citenamefont {Fowler}, \citenamefont {Mariantoni}, \citenamefont {Martinis},\ and\ \citenamefont {Cleland}}]{fowler2012surface}%
  \BibitemOpen
  \bibfield  {author} {\bibinfo {author} {\bibfnamefont {A.~G.}\ \bibnamefont {Fowler}}, \bibinfo {author} {\bibfnamefont {M.}~\bibnamefont {Mariantoni}}, \bibinfo {author} {\bibfnamefont {J.~M.}\ \bibnamefont {Martinis}},\ and\ \bibinfo {author} {\bibfnamefont {A.~N.}\ \bibnamefont {Cleland}},\ }\href {https://doi.org/10.1103/PhysRevA.86.032324} {\bibfield  {journal} {\bibinfo  {journal} {Phys. Rev. A}\ }\textbf {\bibinfo {volume} {86}},\ \bibinfo {pages} {032324} (\bibinfo {year} {2012})}\BibitemShut {NoStop}%
\bibitem [{\citenamefont {Cai}\ \emph {et~al.}(2021{\natexlab{a}})\citenamefont {Cai}, \citenamefont {Ma}, \citenamefont {Wang}, \citenamefont {Zou},\ and\ \citenamefont {Sun}}]{cai2021reviewQEC}%
  \BibitemOpen
  \bibfield  {author} {\bibinfo {author} {\bibfnamefont {W.}~\bibnamefont {Cai}}, \bibinfo {author} {\bibfnamefont {Y.}~\bibnamefont {Ma}}, \bibinfo {author} {\bibfnamefont {W.}~\bibnamefont {Wang}}, \bibinfo {author} {\bibfnamefont {C.-L.}\ \bibnamefont {Zou}},\ and\ \bibinfo {author} {\bibfnamefont {L.}~\bibnamefont {Sun}},\ }\href {https://doi.org/10.1016/j.fmre.2020.12.006} {\bibfield  {journal} {\bibinfo  {journal} {Fundamental Research}\ }\textbf {\bibinfo {volume} {1}},\ \bibinfo {pages} {50} (\bibinfo {year} {2021}{\natexlab{a}})}\BibitemShut {NoStop}%
\bibitem [{\citenamefont {Campagne-Ibarcq}\ \emph {et~al.}(2020)\citenamefont {Campagne-Ibarcq}, \citenamefont {Eickbusch}, \citenamefont {Touzard}, \citenamefont {Zalys-Geller}, \citenamefont {Frattini}, \citenamefont {Sivak}, \citenamefont {Reinhold}, \citenamefont {Puri}, \citenamefont {Shankar}, \citenamefont {Schoelkopf} \emph {et~al.}}]{campagne2020quantum}%
  \BibitemOpen
  \bibfield  {author} {\bibinfo {author} {\bibfnamefont {P.}~\bibnamefont {Campagne-Ibarcq}}, \bibinfo {author} {\bibfnamefont {A.}~\bibnamefont {Eickbusch}}, \bibinfo {author} {\bibfnamefont {S.}~\bibnamefont {Touzard}}, \bibinfo {author} {\bibfnamefont {E.}~\bibnamefont {Zalys-Geller}}, \bibinfo {author} {\bibfnamefont {N.~E.}\ \bibnamefont {Frattini}}, \bibinfo {author} {\bibfnamefont {V.~V.}\ \bibnamefont {Sivak}}, \bibinfo {author} {\bibfnamefont {P.}~\bibnamefont {Reinhold}}, \bibinfo {author} {\bibfnamefont {S.}~\bibnamefont {Puri}}, \bibinfo {author} {\bibfnamefont {S.}~\bibnamefont {Shankar}}, \bibinfo {author} {\bibfnamefont {R.~J.}\ \bibnamefont {Schoelkopf}}, \emph {et~al.},\ }\href@noop {} {\bibfield  {journal} {\bibinfo  {journal} {Nature}\ }\textbf {\bibinfo {volume} {584}},\ \bibinfo {pages} {368} (\bibinfo {year} {2020})}\BibitemShut {NoStop}%
\bibitem [{\citenamefont {Hu}\ \emph {et~al.}(2019)\citenamefont {Hu}, \citenamefont {Ma}, \citenamefont {Cai}, \citenamefont {Mu}, \citenamefont {Xu}, \citenamefont {Wang}, \citenamefont {Wu}, \citenamefont {Wang}, \citenamefont {Song}, \citenamefont {Zou} \emph {et~al.}}]{hu2019quantum}%
  \BibitemOpen
  \bibfield  {author} {\bibinfo {author} {\bibfnamefont {L.}~\bibnamefont {Hu}}, \bibinfo {author} {\bibfnamefont {Y.}~\bibnamefont {Ma}}, \bibinfo {author} {\bibfnamefont {W.}~\bibnamefont {Cai}}, \bibinfo {author} {\bibfnamefont {X.}~\bibnamefont {Mu}}, \bibinfo {author} {\bibfnamefont {Y.}~\bibnamefont {Xu}}, \bibinfo {author} {\bibfnamefont {W.}~\bibnamefont {Wang}}, \bibinfo {author} {\bibfnamefont {Y.}~\bibnamefont {Wu}}, \bibinfo {author} {\bibfnamefont {H.}~\bibnamefont {Wang}}, \bibinfo {author} {\bibfnamefont {Y.}~\bibnamefont {Song}}, \bibinfo {author} {\bibfnamefont {C.-L.}\ \bibnamefont {Zou}}, \emph {et~al.},\ }\href@noop {} {\bibfield  {journal} {\bibinfo  {journal} {Nature Physics}\ }\textbf {\bibinfo {volume} {15}},\ \bibinfo {pages} {503} (\bibinfo {year} {2019})}\BibitemShut {NoStop}%
\bibitem [{\citenamefont {Mirrahimi}\ \emph {et~al.}(2014)\citenamefont {Mirrahimi}, \citenamefont {Leghtas}, \citenamefont {Albert}, \citenamefont {Touzard}, \citenamefont {Schoelkopf}, \citenamefont {Jiang},\ and\ \citenamefont {Devoret}}]{mirrahimi2014dynamically}%
  \BibitemOpen
  \bibfield  {author} {\bibinfo {author} {\bibfnamefont {M.}~\bibnamefont {Mirrahimi}}, \bibinfo {author} {\bibfnamefont {Z.}~\bibnamefont {Leghtas}}, \bibinfo {author} {\bibfnamefont {V.~V.}\ \bibnamefont {Albert}}, \bibinfo {author} {\bibfnamefont {S.}~\bibnamefont {Touzard}}, \bibinfo {author} {\bibfnamefont {R.~J.}\ \bibnamefont {Schoelkopf}}, \bibinfo {author} {\bibfnamefont {L.}~\bibnamefont {Jiang}},\ and\ \bibinfo {author} {\bibfnamefont {M.~H.}\ \bibnamefont {Devoret}},\ }\href@noop {} {\bibfield  {journal} {\bibinfo  {journal} {New Journal of Physics}\ }\textbf {\bibinfo {volume} {16}},\ \bibinfo {pages} {045014} (\bibinfo {year} {2014})}\BibitemShut {NoStop}%
\bibitem [{\citenamefont {Lescanne}\ \emph {et~al.}(2020)\citenamefont {Lescanne}, \citenamefont {Villiers}, \citenamefont {Peronnin}, \citenamefont {Sarlette}, \citenamefont {Delbecq}, \citenamefont {Huard}, \citenamefont {Kontos}, \citenamefont {Mirrahimi},\ and\ \citenamefont {Leghtas}}]{lescanne2020exponential}%
  \BibitemOpen
  \bibfield  {author} {\bibinfo {author} {\bibfnamefont {R.}~\bibnamefont {Lescanne}}, \bibinfo {author} {\bibfnamefont {M.}~\bibnamefont {Villiers}}, \bibinfo {author} {\bibfnamefont {T.}~\bibnamefont {Peronnin}}, \bibinfo {author} {\bibfnamefont {A.}~\bibnamefont {Sarlette}}, \bibinfo {author} {\bibfnamefont {M.}~\bibnamefont {Delbecq}}, \bibinfo {author} {\bibfnamefont {B.}~\bibnamefont {Huard}}, \bibinfo {author} {\bibfnamefont {T.}~\bibnamefont {Kontos}}, \bibinfo {author} {\bibfnamefont {M.}~\bibnamefont {Mirrahimi}},\ and\ \bibinfo {author} {\bibfnamefont {Z.}~\bibnamefont {Leghtas}},\ }\href@noop {} {\bibfield  {journal} {\bibinfo  {journal} {Nature Physics}\ }\textbf {\bibinfo {volume} {16}},\ \bibinfo {pages} {509} (\bibinfo {year} {2020})}\BibitemShut {NoStop}%
\bibitem [{\citenamefont {Labay-Mora}\ \emph {et~al.}(2024{\natexlab{a}})\citenamefont {Labay-Mora}, \citenamefont {Zambrini},\ and\ \citenamefont {Giorgi}}]{labay2023squeezing}%
  \BibitemOpen
  \bibfield  {author} {\bibinfo {author} {\bibfnamefont {A.}~\bibnamefont {Labay-Mora}}, \bibinfo {author} {\bibfnamefont {R.}~\bibnamefont {Zambrini}},\ and\ \bibinfo {author} {\bibfnamefont {G.~L.}\ \bibnamefont {Giorgi}},\ }\href {https://doi.org/10.1103/PhysRevA.109.032407} {\bibfield  {journal} {\bibinfo  {journal} {Phys. Rev. A}\ }\textbf {\bibinfo {volume} {109}},\ \bibinfo {pages} {032407} (\bibinfo {year} {2024}{\natexlab{a}})}\BibitemShut {NoStop}%
\bibitem [{\citenamefont {Aaronson}\ and\ \citenamefont {Arkhipov}(2013)}]{Aaronson-Arkhipov-2013}%
  \BibitemOpen
  \bibfield  {author} {\bibinfo {author} {\bibfnamefont {S.}~\bibnamefont {Aaronson}}\ and\ \bibinfo {author} {\bibfnamefont {A.}~\bibnamefont {Arkhipov}},\ }\href {https://doi.org/10.4086/toc.2013.v009a004} {\bibfield  {journal} {\bibinfo  {journal} {Theory of Computing}\ }\textbf {\bibinfo {volume} {9}},\ \bibinfo {pages} {143} (\bibinfo {year} {2013})}\BibitemShut {NoStop}%
\bibitem [{\citenamefont {Hamilton}\ \emph {et~al.}(2017)\citenamefont {Hamilton}, \citenamefont {Kruse}, \citenamefont {Sansoni}, \citenamefont {Barkhofen}, \citenamefont {Silberhorn},\ and\ \citenamefont {Jex}}]{Gaussian_BS}%
  \BibitemOpen
  \bibfield  {author} {\bibinfo {author} {\bibfnamefont {C.~S.}\ \bibnamefont {Hamilton}}, \bibinfo {author} {\bibfnamefont {R.}~\bibnamefont {Kruse}}, \bibinfo {author} {\bibfnamefont {L.}~\bibnamefont {Sansoni}}, \bibinfo {author} {\bibfnamefont {S.}~\bibnamefont {Barkhofen}}, \bibinfo {author} {\bibfnamefont {C.}~\bibnamefont {Silberhorn}},\ and\ \bibinfo {author} {\bibfnamefont {I.}~\bibnamefont {Jex}},\ }\href {https://doi.org/10.1103/PhysRevLett.119.170501} {\bibfield  {journal} {\bibinfo  {journal} {Phys. Rev. Lett.}\ }\textbf {\bibinfo {volume} {119}},\ \bibinfo {pages} {170501} (\bibinfo {year} {2017})}\BibitemShut {NoStop}%
\bibitem [{\citenamefont {Wang}\ \emph {et~al.}(2019)\citenamefont {Wang}, \citenamefont {Qin}, \citenamefont {Ding}, \citenamefont {Chen}, \citenamefont {Chen}, \citenamefont {You}, \citenamefont {He}, \citenamefont {Jiang}, \citenamefont {You}, \citenamefont {Wang}, \citenamefont {Schneider}, \citenamefont {Renema}, \citenamefont {H\"ofling}, \citenamefont {Lu},\ and\ \citenamefont {Pan}}]{boson-sampling-1}%
  \BibitemOpen
  \bibfield  {author} {\bibinfo {author} {\bibfnamefont {H.}~\bibnamefont {Wang}}, \bibinfo {author} {\bibfnamefont {J.}~\bibnamefont {Qin}}, \bibinfo {author} {\bibfnamefont {X.}~\bibnamefont {Ding}}, \bibinfo {author} {\bibfnamefont {M.-C.}\ \bibnamefont {Chen}}, \bibinfo {author} {\bibfnamefont {S.}~\bibnamefont {Chen}}, \bibinfo {author} {\bibfnamefont {X.}~\bibnamefont {You}}, \bibinfo {author} {\bibfnamefont {Y.-M.}\ \bibnamefont {He}}, \bibinfo {author} {\bibfnamefont {X.}~\bibnamefont {Jiang}}, \bibinfo {author} {\bibfnamefont {L.}~\bibnamefont {You}}, \bibinfo {author} {\bibfnamefont {Z.}~\bibnamefont {Wang}}, \bibinfo {author} {\bibfnamefont {C.}~\bibnamefont {Schneider}}, \bibinfo {author} {\bibfnamefont {J.~J.}\ \bibnamefont {Renema}}, \bibinfo {author} {\bibfnamefont {S.}~\bibnamefont {H\"ofling}}, \bibinfo {author} {\bibfnamefont {C.-Y.}\ \bibnamefont {Lu}},\ and\ \bibinfo {author} {\bibfnamefont {J.-W.}\ \bibnamefont {Pan}},\ }\href {https://doi.org/10.1103/PhysRevLett.123.250503} {\bibfield
  {journal} {\bibinfo  {journal} {Phys. Rev. Lett.}\ }\textbf {\bibinfo {volume} {123}},\ \bibinfo {pages} {250503} (\bibinfo {year} {2019})}\BibitemShut {NoStop}%
\bibitem [{\citenamefont {Zhong}\ \emph {et~al.}(2020)\citenamefont {Zhong}, \citenamefont {Wang}, \citenamefont {Deng}, \citenamefont {Chen}, \citenamefont {Peng}, \citenamefont {Luo}, \citenamefont {Qin}, \citenamefont {Wu}, \citenamefont {Ding}, \citenamefont {Hu}, \citenamefont {Hu}, \citenamefont {Yang}, \citenamefont {Zhang}, \citenamefont {Li}, \citenamefont {Li}, \citenamefont {Jiang}, \citenamefont {Gan}, \citenamefont {Yang}, \citenamefont {You}, \citenamefont {Wang}, \citenamefont {Li}, \citenamefont {Liu}, \citenamefont {Lu},\ and\ \citenamefont {Pan}}]{boson-sampling-2}%
  \BibitemOpen
  \bibfield  {author} {\bibinfo {author} {\bibfnamefont {H.-S.}\ \bibnamefont {Zhong}}, \bibinfo {author} {\bibfnamefont {H.}~\bibnamefont {Wang}}, \bibinfo {author} {\bibfnamefont {Y.-H.}\ \bibnamefont {Deng}}, \bibinfo {author} {\bibfnamefont {M.-C.}\ \bibnamefont {Chen}}, \bibinfo {author} {\bibfnamefont {L.-C.}\ \bibnamefont {Peng}}, \bibinfo {author} {\bibfnamefont {Y.-H.}\ \bibnamefont {Luo}}, \bibinfo {author} {\bibfnamefont {J.}~\bibnamefont {Qin}}, \bibinfo {author} {\bibfnamefont {D.}~\bibnamefont {Wu}}, \bibinfo {author} {\bibfnamefont {X.}~\bibnamefont {Ding}}, \bibinfo {author} {\bibfnamefont {Y.}~\bibnamefont {Hu}}, \bibinfo {author} {\bibfnamefont {P.}~\bibnamefont {Hu}}, \bibinfo {author} {\bibfnamefont {X.-Y.}\ \bibnamefont {Yang}}, \bibinfo {author} {\bibfnamefont {W.-J.}\ \bibnamefont {Zhang}}, \bibinfo {author} {\bibfnamefont {H.}~\bibnamefont {Li}}, \bibinfo {author} {\bibfnamefont {Y.}~\bibnamefont {Li}}, \bibinfo {author} {\bibfnamefont {X.}~\bibnamefont {Jiang}}, \bibinfo {author}
  {\bibfnamefont {L.}~\bibnamefont {Gan}}, \bibinfo {author} {\bibfnamefont {G.}~\bibnamefont {Yang}}, \bibinfo {author} {\bibfnamefont {L.}~\bibnamefont {You}}, \bibinfo {author} {\bibfnamefont {Z.}~\bibnamefont {Wang}}, \bibinfo {author} {\bibfnamefont {L.}~\bibnamefont {Li}}, \bibinfo {author} {\bibfnamefont {N.-L.}\ \bibnamefont {Liu}}, \bibinfo {author} {\bibfnamefont {C.-Y.}\ \bibnamefont {Lu}},\ and\ \bibinfo {author} {\bibfnamefont {J.-W.}\ \bibnamefont {Pan}},\ }\href {https://doi.org/10.1126/science.abe8770} {\bibfield  {journal} {\bibinfo  {journal} {Science}\ }\textbf {\bibinfo {volume} {370}},\ \bibinfo {pages} {1460} (\bibinfo {year} {2020})},\ \Eprint {https://arxiv.org/abs/https://www.science.org/doi/pdf/10.1126/science.abe8770} {https://www.science.org/doi/pdf/10.1126/science.abe8770} \BibitemShut {NoStop}%
\bibitem [{\citenamefont {Madsen}\ \emph {et~al.}(2022)\citenamefont {Madsen}, \citenamefont {Laudenbach}, \citenamefont {Askarani}, \citenamefont {Rortais}, \citenamefont {Vincent}, \citenamefont {Bulmer}, \citenamefont {Miatto}, \citenamefont {Neuhaus}, \citenamefont {Helt}, \citenamefont {Collins}, \citenamefont {Lita}, \citenamefont {Gerrits}, \citenamefont {Nam}, \citenamefont {Vaidya}, \citenamefont {Menotti}, \citenamefont {Dhand}, \citenamefont {Vernon}, \citenamefont {Quesada},\ and\ \citenamefont {Lavoie}}]{boson-sampling-3}%
  \BibitemOpen
  \bibfield  {author} {\bibinfo {author} {\bibfnamefont {L.~S.}\ \bibnamefont {Madsen}}, \bibinfo {author} {\bibfnamefont {F.}~\bibnamefont {Laudenbach}}, \bibinfo {author} {\bibfnamefont {M.~F.}\ \bibnamefont {Askarani}}, \bibinfo {author} {\bibfnamefont {F.}~\bibnamefont {Rortais}}, \bibinfo {author} {\bibfnamefont {T.}~\bibnamefont {Vincent}}, \bibinfo {author} {\bibfnamefont {J.~F.~F.}\ \bibnamefont {Bulmer}}, \bibinfo {author} {\bibfnamefont {F.~M.}\ \bibnamefont {Miatto}}, \bibinfo {author} {\bibfnamefont {L.}~\bibnamefont {Neuhaus}}, \bibinfo {author} {\bibfnamefont {L.~G.}\ \bibnamefont {Helt}}, \bibinfo {author} {\bibfnamefont {M.~J.}\ \bibnamefont {Collins}}, \bibinfo {author} {\bibfnamefont {A.~E.}\ \bibnamefont {Lita}}, \bibinfo {author} {\bibfnamefont {T.}~\bibnamefont {Gerrits}}, \bibinfo {author} {\bibfnamefont {S.~W.}\ \bibnamefont {Nam}}, \bibinfo {author} {\bibfnamefont {V.~D.}\ \bibnamefont {Vaidya}}, \bibinfo {author} {\bibfnamefont {M.}~\bibnamefont {Menotti}}, \bibinfo {author}
  {\bibfnamefont {I.}~\bibnamefont {Dhand}}, \bibinfo {author} {\bibfnamefont {Z.}~\bibnamefont {Vernon}}, \bibinfo {author} {\bibfnamefont {N.}~\bibnamefont {Quesada}},\ and\ \bibinfo {author} {\bibfnamefont {J.}~\bibnamefont {Lavoie}},\ }\href {https://doi.org/10.1038/s41586-022-04725-x} {\bibfield  {journal} {\bibinfo  {journal} {Nature}\ }\textbf {\bibinfo {volume} {606}},\ \bibinfo {pages} {75} (\bibinfo {year} {2022})}\BibitemShut {NoStop}%
\bibitem [{\citenamefont {McMahon}\ \emph {et~al.}(2016)\citenamefont {McMahon}, \citenamefont {Marandi}, \citenamefont {Haribara}, \citenamefont {Hamerly}, \citenamefont {Langrock}, \citenamefont {Tamate}, \citenamefont {Inagaki}, \citenamefont {Takesue}, \citenamefont {Utsunomiya}, \citenamefont {Aihara}, \citenamefont {Byer}, \citenamefont {Fejer}, \citenamefont {Mabuchi},\ and\ \citenamefont {Yamamoto}}]{ising-machines-1}%
  \BibitemOpen
  \bibfield  {author} {\bibinfo {author} {\bibfnamefont {P.~L.}\ \bibnamefont {McMahon}}, \bibinfo {author} {\bibfnamefont {A.}~\bibnamefont {Marandi}}, \bibinfo {author} {\bibfnamefont {Y.}~\bibnamefont {Haribara}}, \bibinfo {author} {\bibfnamefont {R.}~\bibnamefont {Hamerly}}, \bibinfo {author} {\bibfnamefont {C.}~\bibnamefont {Langrock}}, \bibinfo {author} {\bibfnamefont {S.}~\bibnamefont {Tamate}}, \bibinfo {author} {\bibfnamefont {T.}~\bibnamefont {Inagaki}}, \bibinfo {author} {\bibfnamefont {H.}~\bibnamefont {Takesue}}, \bibinfo {author} {\bibfnamefont {S.}~\bibnamefont {Utsunomiya}}, \bibinfo {author} {\bibfnamefont {K.}~\bibnamefont {Aihara}}, \bibinfo {author} {\bibfnamefont {R.~L.}\ \bibnamefont {Byer}}, \bibinfo {author} {\bibfnamefont {M.~M.}\ \bibnamefont {Fejer}}, \bibinfo {author} {\bibfnamefont {H.}~\bibnamefont {Mabuchi}},\ and\ \bibinfo {author} {\bibfnamefont {Y.}~\bibnamefont {Yamamoto}},\ }\href {https://doi.org/10.1126/science.aah5178} {\bibfield  {journal} {\bibinfo  {journal}
  {Science}\ }\textbf {\bibinfo {volume} {354}},\ \bibinfo {pages} {614} (\bibinfo {year} {2016})},\ \Eprint {https://arxiv.org/abs/https://www.science.org/doi/pdf/10.1126/science.aah5178} {https://www.science.org/doi/pdf/10.1126/science.aah5178} \BibitemShut {NoStop}%
\bibitem [{\citenamefont {Honjo}\ \emph {et~al.}(2021)\citenamefont {Honjo}, \citenamefont {Sonobe}, \citenamefont {Inaba}, \citenamefont {Inagaki}, \citenamefont {Ikuta}, \citenamefont {Yamada}, \citenamefont {Kazama}, \citenamefont {Enbutsu}, \citenamefont {Umeki}, \citenamefont {Kasahara}, \citenamefont {ichi Kawarabayashi},\ and\ \citenamefont {Takesue}}]{ising-machines-2}%
  \BibitemOpen
  \bibfield  {author} {\bibinfo {author} {\bibfnamefont {T.}~\bibnamefont {Honjo}}, \bibinfo {author} {\bibfnamefont {T.}~\bibnamefont {Sonobe}}, \bibinfo {author} {\bibfnamefont {K.}~\bibnamefont {Inaba}}, \bibinfo {author} {\bibfnamefont {T.}~\bibnamefont {Inagaki}}, \bibinfo {author} {\bibfnamefont {T.}~\bibnamefont {Ikuta}}, \bibinfo {author} {\bibfnamefont {Y.}~\bibnamefont {Yamada}}, \bibinfo {author} {\bibfnamefont {T.}~\bibnamefont {Kazama}}, \bibinfo {author} {\bibfnamefont {K.}~\bibnamefont {Enbutsu}}, \bibinfo {author} {\bibfnamefont {T.}~\bibnamefont {Umeki}}, \bibinfo {author} {\bibfnamefont {R.}~\bibnamefont {Kasahara}}, \bibinfo {author} {\bibfnamefont {K.}~\bibnamefont {ichi Kawarabayashi}},\ and\ \bibinfo {author} {\bibfnamefont {H.}~\bibnamefont {Takesue}},\ }\href {https://doi.org/10.1126/sciadv.abh0952} {\bibfield  {journal} {\bibinfo  {journal} {Science Advances}\ }\textbf {\bibinfo {volume} {7}},\ \bibinfo {pages} {eabh0952} (\bibinfo {year} {2021})},\ \Eprint
  {https://arxiv.org/abs/https://www.science.org/doi/pdf/10.1126/sciadv.abh0952} {https://www.science.org/doi/pdf/10.1126/sciadv.abh0952} \BibitemShut {NoStop}%
\bibitem [{\citenamefont {Puri}\ \emph {et~al.}(2017)\citenamefont {Puri}, \citenamefont {Andersen}, \citenamefont {Grimsmo},\ and\ \citenamefont {Blais}}]{puri2017annhealing}%
  \BibitemOpen
  \bibfield  {author} {\bibinfo {author} {\bibfnamefont {S.}~\bibnamefont {Puri}}, \bibinfo {author} {\bibfnamefont {C.~K.}\ \bibnamefont {Andersen}}, \bibinfo {author} {\bibfnamefont {A.~L.}\ \bibnamefont {Grimsmo}},\ and\ \bibinfo {author} {\bibfnamefont {A.}~\bibnamefont {Blais}},\ }\href@noop {} {\bibfield  {journal} {\bibinfo  {journal} {Nature communications}\ }\textbf {\bibinfo {volume} {8}},\ \bibinfo {pages} {15785} (\bibinfo {year} {2017})}\BibitemShut {NoStop}%
\bibitem [{\citenamefont {Nigg}\ \emph {et~al.}(2017)\citenamefont {Nigg}, \citenamefont {L{\"o}rch},\ and\ \citenamefont {Tiwari}}]{nigg2017annhealing}%
  \BibitemOpen
  \bibfield  {author} {\bibinfo {author} {\bibfnamefont {S.~E.}\ \bibnamefont {Nigg}}, \bibinfo {author} {\bibfnamefont {N.}~\bibnamefont {L{\"o}rch}},\ and\ \bibinfo {author} {\bibfnamefont {R.~P.}\ \bibnamefont {Tiwari}},\ }\href@noop {} {\bibfield  {journal} {\bibinfo  {journal} {Science advances}\ }\textbf {\bibinfo {volume} {3}},\ \bibinfo {pages} {e1602273} (\bibinfo {year} {2017})}\BibitemShut {NoStop}%
\bibitem [{\citenamefont {Peruzzo}\ \emph {et~al.}(2014)\citenamefont {Peruzzo}, \citenamefont {McClean}, \citenamefont {Shadbolt}, \citenamefont {Yung}, \citenamefont {Zhou}, \citenamefont {Love}, \citenamefont {Aspuru-Guzik},\ and\ \citenamefont {O'Brien}}]{Peruzzo2014}%
  \BibitemOpen
  \bibfield  {author} {\bibinfo {author} {\bibfnamefont {A.}~\bibnamefont {Peruzzo}}, \bibinfo {author} {\bibfnamefont {J.}~\bibnamefont {McClean}}, \bibinfo {author} {\bibfnamefont {P.}~\bibnamefont {Shadbolt}}, \bibinfo {author} {\bibfnamefont {M.-H.}\ \bibnamefont {Yung}}, \bibinfo {author} {\bibfnamefont {X.-Q.}\ \bibnamefont {Zhou}}, \bibinfo {author} {\bibfnamefont {P.~J.}\ \bibnamefont {Love}}, \bibinfo {author} {\bibfnamefont {A.}~\bibnamefont {Aspuru-Guzik}},\ and\ \bibinfo {author} {\bibfnamefont {J.~L.}\ \bibnamefont {O'Brien}},\ }\href {https://doi.org/10.1038/ncomms5213} {\bibfield  {journal} {\bibinfo  {journal} {Nature Communications}\ }\textbf {\bibinfo {volume} {5}},\ \bibinfo {pages} {4213} (\bibinfo {year} {2014})}\BibitemShut {NoStop}%
\bibitem [{\citenamefont {Genty}\ \emph {et~al.}(2021)\citenamefont {Genty}, \citenamefont {Salmela}, \citenamefont {Dudley}, \citenamefont {Brunner}, \citenamefont {Kokhanovskiy}, \citenamefont {Kobtsev},\ and\ \citenamefont {Turitsyn}}]{genty2021machine}%
  \BibitemOpen
  \bibfield  {author} {\bibinfo {author} {\bibfnamefont {G.}~\bibnamefont {Genty}}, \bibinfo {author} {\bibfnamefont {L.}~\bibnamefont {Salmela}}, \bibinfo {author} {\bibfnamefont {J.~M.}\ \bibnamefont {Dudley}}, \bibinfo {author} {\bibfnamefont {D.}~\bibnamefont {Brunner}}, \bibinfo {author} {\bibfnamefont {A.}~\bibnamefont {Kokhanovskiy}}, \bibinfo {author} {\bibfnamefont {S.}~\bibnamefont {Kobtsev}},\ and\ \bibinfo {author} {\bibfnamefont {S.~K.}\ \bibnamefont {Turitsyn}},\ }\href@noop {} {\bibfield  {journal} {\bibinfo  {journal} {Nature Photonics}\ }\textbf {\bibinfo {volume} {15}},\ \bibinfo {pages} {91} (\bibinfo {year} {2021})}\BibitemShut {NoStop}%
\bibitem [{\citenamefont {Wiecha}\ \emph {et~al.}(2021)\citenamefont {Wiecha}, \citenamefont {Arbouet}, \citenamefont {Girard},\ and\ \citenamefont {Muskens}}]{wiecha2021deep}%
  \BibitemOpen
  \bibfield  {author} {\bibinfo {author} {\bibfnamefont {P.~R.}\ \bibnamefont {Wiecha}}, \bibinfo {author} {\bibfnamefont {A.}~\bibnamefont {Arbouet}}, \bibinfo {author} {\bibfnamefont {C.}~\bibnamefont {Girard}},\ and\ \bibinfo {author} {\bibfnamefont {O.~L.}\ \bibnamefont {Muskens}},\ }\href@noop {} {\bibfield  {journal} {\bibinfo  {journal} {Photonics Research}\ }\textbf {\bibinfo {volume} {9}},\ \bibinfo {pages} {B182} (\bibinfo {year} {2021})}\BibitemShut {NoStop}%
\bibitem [{\citenamefont {Fu}\ and\ \citenamefont {Kutz}(2013)}]{fu2013high}%
  \BibitemOpen
  \bibfield  {author} {\bibinfo {author} {\bibfnamefont {X.}~\bibnamefont {Fu}}\ and\ \bibinfo {author} {\bibfnamefont {J.~N.}\ \bibnamefont {Kutz}},\ }\href@noop {} {\bibfield  {journal} {\bibinfo  {journal} {Optics express}\ }\textbf {\bibinfo {volume} {21}},\ \bibinfo {pages} {6526} (\bibinfo {year} {2013})}\BibitemShut {NoStop}%
\bibitem [{\citenamefont {Ma}\ \emph {et~al.}(2021)\citenamefont {Ma}, \citenamefont {Liu}, \citenamefont {Kudyshev}, \citenamefont {Boltasseva}, \citenamefont {Cai},\ and\ \citenamefont {Liu}}]{ma2021deep}%
  \BibitemOpen
  \bibfield  {author} {\bibinfo {author} {\bibfnamefont {W.}~\bibnamefont {Ma}}, \bibinfo {author} {\bibfnamefont {Z.}~\bibnamefont {Liu}}, \bibinfo {author} {\bibfnamefont {Z.~A.}\ \bibnamefont {Kudyshev}}, \bibinfo {author} {\bibfnamefont {A.}~\bibnamefont {Boltasseva}}, \bibinfo {author} {\bibfnamefont {W.}~\bibnamefont {Cai}},\ and\ \bibinfo {author} {\bibfnamefont {Y.}~\bibnamefont {Liu}},\ }\href@noop {} {\bibfield  {journal} {\bibinfo  {journal} {Nature Photonics}\ }\textbf {\bibinfo {volume} {15}},\ \bibinfo {pages} {77} (\bibinfo {year} {2021})}\BibitemShut {NoStop}%
\bibitem [{\citenamefont {Krenn}\ \emph {et~al.}(2016)\citenamefont {Krenn}, \citenamefont {Malik}, \citenamefont {Fickler}, \citenamefont {Lapkiewicz},\ and\ \citenamefont {Zeilinger}}]{krenn2016automated}%
  \BibitemOpen
  \bibfield  {author} {\bibinfo {author} {\bibfnamefont {M.}~\bibnamefont {Krenn}}, \bibinfo {author} {\bibfnamefont {M.}~\bibnamefont {Malik}}, \bibinfo {author} {\bibfnamefont {R.}~\bibnamefont {Fickler}}, \bibinfo {author} {\bibfnamefont {R.}~\bibnamefont {Lapkiewicz}},\ and\ \bibinfo {author} {\bibfnamefont {A.}~\bibnamefont {Zeilinger}},\ }\href@noop {} {\bibfield  {journal} {\bibinfo  {journal} {Physical review letters}\ }\textbf {\bibinfo {volume} {116}},\ \bibinfo {pages} {090405} (\bibinfo {year} {2016})}\BibitemShut {NoStop}%
\bibitem [{\citenamefont {Krenn}\ \emph {et~al.}(2017)\citenamefont {Krenn}, \citenamefont {Hochrainer}, \citenamefont {Lahiri},\ and\ \citenamefont {Zeilinger}}]{krenn2017entanglement}%
  \BibitemOpen
  \bibfield  {author} {\bibinfo {author} {\bibfnamefont {M.}~\bibnamefont {Krenn}}, \bibinfo {author} {\bibfnamefont {A.}~\bibnamefont {Hochrainer}}, \bibinfo {author} {\bibfnamefont {M.}~\bibnamefont {Lahiri}},\ and\ \bibinfo {author} {\bibfnamefont {A.}~\bibnamefont {Zeilinger}},\ }\href@noop {} {\bibfield  {journal} {\bibinfo  {journal} {Physical review letters}\ }\textbf {\bibinfo {volume} {118}},\ \bibinfo {pages} {080401} (\bibinfo {year} {2017})}\BibitemShut {NoStop}%
\bibitem [{\citenamefont {Cervera-Lierta}\ \emph {et~al.}(2022)\citenamefont {Cervera-Lierta}, \citenamefont {Krenn},\ and\ \citenamefont {Aspuru-Guzik}}]{cervera2022design}%
  \BibitemOpen
  \bibfield  {author} {\bibinfo {author} {\bibfnamefont {A.}~\bibnamefont {Cervera-Lierta}}, \bibinfo {author} {\bibfnamefont {M.}~\bibnamefont {Krenn}},\ and\ \bibinfo {author} {\bibfnamefont {A.}~\bibnamefont {Aspuru-Guzik}},\ }\href@noop {} {\bibfield  {journal} {\bibinfo  {journal} {Quantum}\ }\textbf {\bibinfo {volume} {6}},\ \bibinfo {pages} {836} (\bibinfo {year} {2022})}\BibitemShut {NoStop}%
\bibitem [{\citenamefont {Krenn}\ \emph {et~al.}(2021)\citenamefont {Krenn}, \citenamefont {Kottmann}, \citenamefont {Tischler},\ and\ \citenamefont {Aspuru-Guzik}}]{krenn2021conceptual}%
  \BibitemOpen
  \bibfield  {author} {\bibinfo {author} {\bibfnamefont {M.}~\bibnamefont {Krenn}}, \bibinfo {author} {\bibfnamefont {J.~S.}\ \bibnamefont {Kottmann}}, \bibinfo {author} {\bibfnamefont {N.}~\bibnamefont {Tischler}},\ and\ \bibinfo {author} {\bibfnamefont {A.}~\bibnamefont {Aspuru-Guzik}},\ }\href@noop {} {\bibfield  {journal} {\bibinfo  {journal} {Physical Review X}\ }\textbf {\bibinfo {volume} {11}},\ \bibinfo {pages} {031044} (\bibinfo {year} {2021})}\BibitemShut {NoStop}%
\bibitem [{\citenamefont {Labay-Mora}\ \emph {et~al.}(2023{\natexlab{a}})\citenamefont {Labay-Mora}, \citenamefont {da~Silva},\ and\ \citenamefont {Wehner}}]{labay2023reducing}%
  \BibitemOpen
  \bibfield  {author} {\bibinfo {author} {\bibfnamefont {A.}~\bibnamefont {Labay-Mora}}, \bibinfo {author} {\bibfnamefont {F.~F.}\ \bibnamefont {da~Silva}},\ and\ \bibinfo {author} {\bibfnamefont {S.}~\bibnamefont {Wehner}},\ }\href@noop {} {\bibfield  {journal} {\bibinfo  {journal} {arXiv preprint arXiv:2309.11448}\ } (\bibinfo {year} {2023}{\natexlab{a}})}\BibitemShut {NoStop}%
\bibitem [{\citenamefont {Avis}\ \emph {et~al.}(2023)\citenamefont {Avis}, \citenamefont {Ferreira~da Silva}, \citenamefont {Coopmans}, \citenamefont {Dahlberg}, \citenamefont {Jirovsk{\'a}}, \citenamefont {Maier}, \citenamefont {Rabbie}, \citenamefont {Torres-Knoop},\ and\ \citenamefont {Wehner}}]{avis2023requirements}%
  \BibitemOpen
  \bibfield  {author} {\bibinfo {author} {\bibfnamefont {G.}~\bibnamefont {Avis}}, \bibinfo {author} {\bibfnamefont {F.}~\bibnamefont {Ferreira~da Silva}}, \bibinfo {author} {\bibfnamefont {T.}~\bibnamefont {Coopmans}}, \bibinfo {author} {\bibfnamefont {A.}~\bibnamefont {Dahlberg}}, \bibinfo {author} {\bibfnamefont {H.}~\bibnamefont {Jirovsk{\'a}}}, \bibinfo {author} {\bibfnamefont {D.}~\bibnamefont {Maier}}, \bibinfo {author} {\bibfnamefont {J.}~\bibnamefont {Rabbie}}, \bibinfo {author} {\bibfnamefont {A.}~\bibnamefont {Torres-Knoop}},\ and\ \bibinfo {author} {\bibfnamefont {S.}~\bibnamefont {Wehner}},\ }\href@noop {} {\bibfield  {journal} {\bibinfo  {journal} {NPJ Quantum Information}\ }\textbf {\bibinfo {volume} {9}},\ \bibinfo {pages} {100} (\bibinfo {year} {2023})}\BibitemShut {NoStop}%
\bibitem [{\citenamefont {Mafu}(2024)}]{mafu2024advances}%
  \BibitemOpen
  \bibfield  {author} {\bibinfo {author} {\bibfnamefont {M.}~\bibnamefont {Mafu}},\ }\href@noop {} {\bibfield  {journal} {\bibinfo  {journal} {IET Quantum Communication}\ } (\bibinfo {year} {2024})}\BibitemShut {NoStop}%
\bibitem [{\citenamefont {Torlai}\ \emph {et~al.}(2018)\citenamefont {Torlai}, \citenamefont {Mazzola}, \citenamefont {Carrasquilla}, \citenamefont {Troyer}, \citenamefont {Melko},\ and\ \citenamefont {Carleo}}]{Torlai2018}%
  \BibitemOpen
  \bibfield  {author} {\bibinfo {author} {\bibfnamefont {G.}~\bibnamefont {Torlai}}, \bibinfo {author} {\bibfnamefont {G.}~\bibnamefont {Mazzola}}, \bibinfo {author} {\bibfnamefont {J.}~\bibnamefont {Carrasquilla}}, \bibinfo {author} {\bibfnamefont {M.}~\bibnamefont {Troyer}}, \bibinfo {author} {\bibfnamefont {R.}~\bibnamefont {Melko}},\ and\ \bibinfo {author} {\bibfnamefont {G.}~\bibnamefont {Carleo}},\ }\href {https://doi.org/10.1038/s41567-018-0048-5} {\bibfield  {journal} {\bibinfo  {journal} {Nature Physics}\ }\textbf {\bibinfo {volume} {14}},\ \bibinfo {pages} {447} (\bibinfo {year} {2018})}\BibitemShut {NoStop}%
\bibitem [{\citenamefont {Koutn{\`y}}\ \emph {et~al.}(2023)\citenamefont {Koutn{\`y}}, \citenamefont {Gin{\'e}s}, \citenamefont {Mocza{\l}a-Dusanowska}, \citenamefont {H{\"o}fling}, \citenamefont {Schneider}, \citenamefont {Predojevi{\'c}},\ and\ \citenamefont {Je{\v{z}}ek}}]{koutny2023deep}%
  \BibitemOpen
  \bibfield  {author} {\bibinfo {author} {\bibfnamefont {D.}~\bibnamefont {Koutn{\`y}}}, \bibinfo {author} {\bibfnamefont {L.}~\bibnamefont {Gin{\'e}s}}, \bibinfo {author} {\bibfnamefont {M.}~\bibnamefont {Mocza{\l}a-Dusanowska}}, \bibinfo {author} {\bibfnamefont {S.}~\bibnamefont {H{\"o}fling}}, \bibinfo {author} {\bibfnamefont {C.}~\bibnamefont {Schneider}}, \bibinfo {author} {\bibfnamefont {A.}~\bibnamefont {Predojevi{\'c}}},\ and\ \bibinfo {author} {\bibfnamefont {M.}~\bibnamefont {Je{\v{z}}ek}},\ }\href@noop {} {\bibfield  {journal} {\bibinfo  {journal} {Science Advances}\ }\textbf {\bibinfo {volume} {9}},\ \bibinfo {pages} {eadd7131} (\bibinfo {year} {2023})}\BibitemShut {NoStop}%
\bibitem [{\citenamefont {Goy}\ \emph {et~al.}(2018)\citenamefont {Goy}, \citenamefont {Arthur}, \citenamefont {Li},\ and\ \citenamefont {Barbastathis}}]{PhysRevLett.121.243902}%
  \BibitemOpen
  \bibfield  {author} {\bibinfo {author} {\bibfnamefont {A.}~\bibnamefont {Goy}}, \bibinfo {author} {\bibfnamefont {K.}~\bibnamefont {Arthur}}, \bibinfo {author} {\bibfnamefont {S.}~\bibnamefont {Li}},\ and\ \bibinfo {author} {\bibfnamefont {G.}~\bibnamefont {Barbastathis}},\ }\href {https://doi.org/10.1103/PhysRevLett.121.243902} {\bibfield  {journal} {\bibinfo  {journal} {Phys. Rev. Lett.}\ }\textbf {\bibinfo {volume} {121}},\ \bibinfo {pages} {243902} (\bibinfo {year} {2018})}\BibitemShut {NoStop}%
\bibitem [{\citenamefont {McMahon}(2023)}]{mcmahon2023physics}%
  \BibitemOpen
  \bibfield  {author} {\bibinfo {author} {\bibfnamefont {P.~L.}\ \bibnamefont {McMahon}},\ }\href@noop {} {\bibfield  {journal} {\bibinfo  {journal} {Nature Reviews Physics}\ }\textbf {\bibinfo {volume} {5}},\ \bibinfo {pages} {717} (\bibinfo {year} {2023})}\BibitemShut {NoStop}%
\bibitem [{\citenamefont {Won}(2023)}]{won2023power}%
  \BibitemOpen
  \bibfield  {author} {\bibinfo {author} {\bibfnamefont {R.}~\bibnamefont {Won}},\ }\href@noop {} {\bibfield  {journal} {\bibinfo  {journal} {Nature Photonics}\ }\textbf {\bibinfo {volume} {17}},\ \bibinfo {pages} {934} (\bibinfo {year} {2023})}\BibitemShut {NoStop}%
\bibitem [{\citenamefont {Shen}\ \emph {et~al.}(2017)\citenamefont {Shen}, \citenamefont {Harris}, \citenamefont {Skirlo}, \citenamefont {Prabhu}, \citenamefont {Baehr-Jones}, \citenamefont {Hochberg}, \citenamefont {Sun}, \citenamefont {Zhao}, \citenamefont {Larochelle}, \citenamefont {Englund},\ and\ \citenamefont {Solja{\v{c}}i{\'{c}}}}]{Shen2017}%
  \BibitemOpen
  \bibfield  {author} {\bibinfo {author} {\bibfnamefont {Y.}~\bibnamefont {Shen}}, \bibinfo {author} {\bibfnamefont {N.~C.}\ \bibnamefont {Harris}}, \bibinfo {author} {\bibfnamefont {S.}~\bibnamefont {Skirlo}}, \bibinfo {author} {\bibfnamefont {M.}~\bibnamefont {Prabhu}}, \bibinfo {author} {\bibfnamefont {T.}~\bibnamefont {Baehr-Jones}}, \bibinfo {author} {\bibfnamefont {M.}~\bibnamefont {Hochberg}}, \bibinfo {author} {\bibfnamefont {X.}~\bibnamefont {Sun}}, \bibinfo {author} {\bibfnamefont {S.}~\bibnamefont {Zhao}}, \bibinfo {author} {\bibfnamefont {H.}~\bibnamefont {Larochelle}}, \bibinfo {author} {\bibfnamefont {D.}~\bibnamefont {Englund}},\ and\ \bibinfo {author} {\bibfnamefont {M.}~\bibnamefont {Solja{\v{c}}i{\'{c}}}},\ }\href {https://doi.org/10.1038/nphoton.2017.93} {\bibfield  {journal} {\bibinfo  {journal} {Nature Photonics}\ }\textbf {\bibinfo {volume} {11}},\ \bibinfo {pages} {441} (\bibinfo {year} {2017})}\BibitemShut {NoStop}%
\bibitem [{\citenamefont {Lin}\ \emph {et~al.}(2018)\citenamefont {Lin}, \citenamefont {Rivenson}, \citenamefont {Yardimci}, \citenamefont {Veli}, \citenamefont {Luo}, \citenamefont {Jarrahi},\ and\ \citenamefont {Ozcan}}]{lin2018all}%
  \BibitemOpen
  \bibfield  {author} {\bibinfo {author} {\bibfnamefont {X.}~\bibnamefont {Lin}}, \bibinfo {author} {\bibfnamefont {Y.}~\bibnamefont {Rivenson}}, \bibinfo {author} {\bibfnamefont {N.~T.}\ \bibnamefont {Yardimci}}, \bibinfo {author} {\bibfnamefont {M.}~\bibnamefont {Veli}}, \bibinfo {author} {\bibfnamefont {Y.}~\bibnamefont {Luo}}, \bibinfo {author} {\bibfnamefont {M.}~\bibnamefont {Jarrahi}},\ and\ \bibinfo {author} {\bibfnamefont {A.}~\bibnamefont {Ozcan}},\ }\href@noop {} {\bibfield  {journal} {\bibinfo  {journal} {Science}\ }\textbf {\bibinfo {volume} {361}},\ \bibinfo {pages} {1004} (\bibinfo {year} {2018})}\BibitemShut {NoStop}%
\bibitem [{\citenamefont {Sui}\ \emph {et~al.}(2020)\citenamefont {Sui}, \citenamefont {Wu}, \citenamefont {Liu}, \citenamefont {Chen},\ and\ \citenamefont {Gu}}]{sui2020review}%
  \BibitemOpen
  \bibfield  {author} {\bibinfo {author} {\bibfnamefont {X.}~\bibnamefont {Sui}}, \bibinfo {author} {\bibfnamefont {Q.}~\bibnamefont {Wu}}, \bibinfo {author} {\bibfnamefont {J.}~\bibnamefont {Liu}}, \bibinfo {author} {\bibfnamefont {Q.}~\bibnamefont {Chen}},\ and\ \bibinfo {author} {\bibfnamefont {G.}~\bibnamefont {Gu}},\ }\href@noop {} {\bibfield  {journal} {\bibinfo  {journal} {IEEE Access}\ }\textbf {\bibinfo {volume} {8}},\ \bibinfo {pages} {70773} (\bibinfo {year} {2020})}\BibitemShut {NoStop}%
\bibitem [{\citenamefont {Wetzstein}\ \emph {et~al.}(2020)\citenamefont {Wetzstein}, \citenamefont {Ozcan}, \citenamefont {Gigan}, \citenamefont {Fan}, \citenamefont {Englund}, \citenamefont {Solja{\v{c}}i{\'c}}, \citenamefont {Denz}, \citenamefont {Miller},\ and\ \citenamefont {Psaltis}}]{wetzstein2020inference}%
  \BibitemOpen
  \bibfield  {author} {\bibinfo {author} {\bibfnamefont {G.}~\bibnamefont {Wetzstein}}, \bibinfo {author} {\bibfnamefont {A.}~\bibnamefont {Ozcan}}, \bibinfo {author} {\bibfnamefont {S.}~\bibnamefont {Gigan}}, \bibinfo {author} {\bibfnamefont {S.}~\bibnamefont {Fan}}, \bibinfo {author} {\bibfnamefont {D.}~\bibnamefont {Englund}}, \bibinfo {author} {\bibfnamefont {M.}~\bibnamefont {Solja{\v{c}}i{\'c}}}, \bibinfo {author} {\bibfnamefont {C.}~\bibnamefont {Denz}}, \bibinfo {author} {\bibfnamefont {D.~A.}\ \bibnamefont {Miller}},\ and\ \bibinfo {author} {\bibfnamefont {D.}~\bibnamefont {Psaltis}},\ }\href@noop {} {\bibfield  {journal} {\bibinfo  {journal} {Nature}\ }\textbf {\bibinfo {volume} {588}},\ \bibinfo {pages} {39} (\bibinfo {year} {2020})}\BibitemShut {NoStop}%
\bibitem [{\citenamefont {Shastri}\ \emph {et~al.}(2021)\citenamefont {Shastri}, \citenamefont {Tait}, \citenamefont {Ferreira~de Lima}, \citenamefont {Pernice}, \citenamefont {Bhaskaran}, \citenamefont {Wright},\ and\ \citenamefont {Prucnal}}]{shastri2021photonics}%
  \BibitemOpen
  \bibfield  {author} {\bibinfo {author} {\bibfnamefont {B.~J.}\ \bibnamefont {Shastri}}, \bibinfo {author} {\bibfnamefont {A.~N.}\ \bibnamefont {Tait}}, \bibinfo {author} {\bibfnamefont {T.}~\bibnamefont {Ferreira~de Lima}}, \bibinfo {author} {\bibfnamefont {W.~H.}\ \bibnamefont {Pernice}}, \bibinfo {author} {\bibfnamefont {H.}~\bibnamefont {Bhaskaran}}, \bibinfo {author} {\bibfnamefont {C.~D.}\ \bibnamefont {Wright}},\ and\ \bibinfo {author} {\bibfnamefont {P.~R.}\ \bibnamefont {Prucnal}},\ }\href@noop {} {\bibfield  {journal} {\bibinfo  {journal} {Nature Photonics}\ }\textbf {\bibinfo {volume} {15}},\ \bibinfo {pages} {102} (\bibinfo {year} {2021})}\BibitemShut {NoStop}%
\bibitem [{\citenamefont {Harris}\ \emph {et~al.}(2018)\citenamefont {Harris}, \citenamefont {Carolan}, \citenamefont {Bunandar}, \citenamefont {Prabhu}, \citenamefont {Hochberg}, \citenamefont {Baehr-Jones}, \citenamefont {Fanto}, \citenamefont {Smith}, \citenamefont {Tison}, \citenamefont {Alsing},\ and\ \citenamefont {Englund}}]{Harris:18}%
  \BibitemOpen
  \bibfield  {author} {\bibinfo {author} {\bibfnamefont {N.~C.}\ \bibnamefont {Harris}}, \bibinfo {author} {\bibfnamefont {J.}~\bibnamefont {Carolan}}, \bibinfo {author} {\bibfnamefont {D.}~\bibnamefont {Bunandar}}, \bibinfo {author} {\bibfnamefont {M.}~\bibnamefont {Prabhu}}, \bibinfo {author} {\bibfnamefont {M.}~\bibnamefont {Hochberg}}, \bibinfo {author} {\bibfnamefont {T.}~\bibnamefont {Baehr-Jones}}, \bibinfo {author} {\bibfnamefont {M.~L.}\ \bibnamefont {Fanto}}, \bibinfo {author} {\bibfnamefont {A.~M.}\ \bibnamefont {Smith}}, \bibinfo {author} {\bibfnamefont {C.~C.}\ \bibnamefont {Tison}}, \bibinfo {author} {\bibfnamefont {P.~M.}\ \bibnamefont {Alsing}},\ and\ \bibinfo {author} {\bibfnamefont {D.}~\bibnamefont {Englund}},\ }\href {https://doi.org/10.1364/OPTICA.5.001623} {\bibfield  {journal} {\bibinfo  {journal} {Optica}\ }\textbf {\bibinfo {volume} {5}},\ \bibinfo {pages} {1623} (\bibinfo {year} {2018})}\BibitemShut {NoStop}%
\bibitem [{\citenamefont {Matuszewski}\ \emph {et~al.}(2024)\citenamefont {Matuszewski}, \citenamefont {Prystupiuk},\ and\ \citenamefont {Opala}}]{matuszewski2024role}%
  \BibitemOpen
  \bibfield  {author} {\bibinfo {author} {\bibfnamefont {M.}~\bibnamefont {Matuszewski}}, \bibinfo {author} {\bibfnamefont {A.}~\bibnamefont {Prystupiuk}},\ and\ \bibinfo {author} {\bibfnamefont {A.}~\bibnamefont {Opala}},\ }\href@noop {} {\bibfield  {journal} {\bibinfo  {journal} {Physical Review Applied}\ }\textbf {\bibinfo {volume} {21}},\ \bibinfo {pages} {014028} (\bibinfo {year} {2024})}\BibitemShut {NoStop}%
\bibitem [{\citenamefont {Hamerly}\ \emph {et~al.}(2019)\citenamefont {Hamerly}, \citenamefont {Bernstein}, \citenamefont {Sludds}, \citenamefont {Solja{\v{c}}i{\'c}},\ and\ \citenamefont {Englund}}]{hamerly2019large}%
  \BibitemOpen
  \bibfield  {author} {\bibinfo {author} {\bibfnamefont {R.}~\bibnamefont {Hamerly}}, \bibinfo {author} {\bibfnamefont {L.}~\bibnamefont {Bernstein}}, \bibinfo {author} {\bibfnamefont {A.}~\bibnamefont {Sludds}}, \bibinfo {author} {\bibfnamefont {M.}~\bibnamefont {Solja{\v{c}}i{\'c}}},\ and\ \bibinfo {author} {\bibfnamefont {D.}~\bibnamefont {Englund}},\ }\href@noop {} {\bibfield  {journal} {\bibinfo  {journal} {Physical Review X}\ }\textbf {\bibinfo {volume} {9}},\ \bibinfo {pages} {021032} (\bibinfo {year} {2019})}\BibitemShut {NoStop}%
\bibitem [{\citenamefont {Ashtiani}\ \emph {et~al.}(2022)\citenamefont {Ashtiani}, \citenamefont {Geers},\ and\ \citenamefont {Aflatouni}}]{ashtiani2022chip}%
  \BibitemOpen
  \bibfield  {author} {\bibinfo {author} {\bibfnamefont {F.}~\bibnamefont {Ashtiani}}, \bibinfo {author} {\bibfnamefont {A.~J.}\ \bibnamefont {Geers}},\ and\ \bibinfo {author} {\bibfnamefont {F.}~\bibnamefont {Aflatouni}},\ }\href@noop {} {\bibfield  {journal} {\bibinfo  {journal} {Nature}\ }\textbf {\bibinfo {volume} {606}},\ \bibinfo {pages} {501} (\bibinfo {year} {2022})}\BibitemShut {NoStop}%
\bibitem [{\citenamefont {Wan}\ \emph {et~al.}(2017)\citenamefont {Wan}, \citenamefont {Dahlsten}, \citenamefont {Kristj{\'a}nsson}, \citenamefont {Gardner},\ and\ \citenamefont {Kim}}]{wan2017quantum}%
  \BibitemOpen
  \bibfield  {author} {\bibinfo {author} {\bibfnamefont {K.~H.}\ \bibnamefont {Wan}}, \bibinfo {author} {\bibfnamefont {O.}~\bibnamefont {Dahlsten}}, \bibinfo {author} {\bibfnamefont {H.}~\bibnamefont {Kristj{\'a}nsson}}, \bibinfo {author} {\bibfnamefont {R.}~\bibnamefont {Gardner}},\ and\ \bibinfo {author} {\bibfnamefont {M.}~\bibnamefont {Kim}},\ }\href@noop {} {\bibfield  {journal} {\bibinfo  {journal} {npj Quantum information}\ }\textbf {\bibinfo {volume} {3}},\ \bibinfo {pages} {36} (\bibinfo {year} {2017})}\BibitemShut {NoStop}%
\bibitem [{\citenamefont {Marković}\ and\ \citenamefont {Grollier}(2020)}]{grollier2020neuromorphic}%
  \BibitemOpen
  \bibfield  {author} {\bibinfo {author} {\bibfnamefont {D.}~\bibnamefont {Marković}}\ and\ \bibinfo {author} {\bibfnamefont {J.}~\bibnamefont {Grollier}},\ }\href {https://doi.org/10.1063/5.0020014} {\bibfield  {journal} {\bibinfo  {journal} {Applied Physics Letters}\ }\textbf {\bibinfo {volume} {117}},\ \bibinfo {pages} {150501} (\bibinfo {year} {2020})},\ \Eprint {https://arxiv.org/abs/https://pubs.aip.org/aip/apl/article-pdf/doi/10.1063/5.0020014/14539374/150501\_1\_online.pdf} {https://pubs.aip.org/aip/apl/article-pdf/doi/10.1063/5.0020014/14539374/150501\_1\_online.pdf} \BibitemShut {NoStop}%
\bibitem [{\citenamefont {Killoran}\ \emph {et~al.}(2019)\citenamefont {Killoran}, \citenamefont {Bromley}, \citenamefont {Arrazola}, \citenamefont {Schuld}, \citenamefont {Quesada},\ and\ \citenamefont {Lloyd}}]{lloyd2019cvqnn}%
  \BibitemOpen
  \bibfield  {author} {\bibinfo {author} {\bibfnamefont {N.}~\bibnamefont {Killoran}}, \bibinfo {author} {\bibfnamefont {T.~R.}\ \bibnamefont {Bromley}}, \bibinfo {author} {\bibfnamefont {J.~M.}\ \bibnamefont {Arrazola}}, \bibinfo {author} {\bibfnamefont {M.}~\bibnamefont {Schuld}}, \bibinfo {author} {\bibfnamefont {N.}~\bibnamefont {Quesada}},\ and\ \bibinfo {author} {\bibfnamefont {S.}~\bibnamefont {Lloyd}},\ }\href {https://doi.org/10.1103/PhysRevResearch.1.033063} {\bibfield  {journal} {\bibinfo  {journal} {Phys. Rev. Res.}\ }\textbf {\bibinfo {volume} {1}},\ \bibinfo {pages} {033063} (\bibinfo {year} {2019})}\BibitemShut {NoStop}%
\bibitem [{\citenamefont {Steinbrecher}\ \emph {et~al.}(2019)\citenamefont {Steinbrecher}, \citenamefont {Olson}, \citenamefont {Englund},\ and\ \citenamefont {Carolan}}]{Steinbrecher2019}%
  \BibitemOpen
  \bibfield  {author} {\bibinfo {author} {\bibfnamefont {G.~R.}\ \bibnamefont {Steinbrecher}}, \bibinfo {author} {\bibfnamefont {J.~P.}\ \bibnamefont {Olson}}, \bibinfo {author} {\bibfnamefont {D.}~\bibnamefont {Englund}},\ and\ \bibinfo {author} {\bibfnamefont {J.}~\bibnamefont {Carolan}},\ }\href {https://doi.org/10.1038/s41534-019-0174-7} {\bibfield  {journal} {\bibinfo  {journal} {npj Quantum Information}\ }\textbf {\bibinfo {volume} {5}},\ \bibinfo {pages} {60} (\bibinfo {year} {2019})}\BibitemShut {NoStop}%
\bibitem [{\citenamefont {Cerezo}\ \emph {et~al.}(2022)\citenamefont {Cerezo}, \citenamefont {Verdon}, \citenamefont {Huang}, \citenamefont {Cincio},\ and\ \citenamefont {Coles}}]{cerezo2022challenges}%
  \BibitemOpen
  \bibfield  {author} {\bibinfo {author} {\bibfnamefont {M.}~\bibnamefont {Cerezo}}, \bibinfo {author} {\bibfnamefont {G.}~\bibnamefont {Verdon}}, \bibinfo {author} {\bibfnamefont {H.-Y.}\ \bibnamefont {Huang}}, \bibinfo {author} {\bibfnamefont {L.}~\bibnamefont {Cincio}},\ and\ \bibinfo {author} {\bibfnamefont {P.~J.}\ \bibnamefont {Coles}},\ }\href@noop {} {\bibfield  {journal} {\bibinfo  {journal} {Nature Computational Science}\ }\textbf {\bibinfo {volume} {2}},\ \bibinfo {pages} {567} (\bibinfo {year} {2022})}\BibitemShut {NoStop}%
\bibitem [{\citenamefont {Jeswal}\ and\ \citenamefont {Chakraverty}(2019)}]{jeswal2019recent}%
  \BibitemOpen
  \bibfield  {author} {\bibinfo {author} {\bibfnamefont {S.}~\bibnamefont {Jeswal}}\ and\ \bibinfo {author} {\bibfnamefont {S.}~\bibnamefont {Chakraverty}},\ }\href@noop {} {\bibfield  {journal} {\bibinfo  {journal} {Archives of Computational Methods in Engineering}\ }\textbf {\bibinfo {volume} {26}},\ \bibinfo {pages} {793} (\bibinfo {year} {2019})}\BibitemShut {NoStop}%
\bibitem [{\citenamefont {Parthasarathy}\ and\ \citenamefont {Bhowmik}(2021)}]{qu_CNNs}%
  \BibitemOpen
  \bibfield  {author} {\bibinfo {author} {\bibfnamefont {R.}~\bibnamefont {Parthasarathy}}\ and\ \bibinfo {author} {\bibfnamefont {R.~T.}\ \bibnamefont {Bhowmik}},\ }\href {https://doi.org/10.1109/ACCESS.2021.3098775} {\bibfield  {journal} {\bibinfo  {journal} {IEEE Access}\ }\textbf {\bibinfo {volume} {9}},\ \bibinfo {pages} {103337} (\bibinfo {year} {2021})}\BibitemShut {NoStop}%
\bibitem [{\citenamefont {Ventura}\ and\ \citenamefont {Martinez}(2000)}]{ventura2000qam}%
  \BibitemOpen
  \bibfield  {author} {\bibinfo {author} {\bibfnamefont {D.}~\bibnamefont {Ventura}}\ and\ \bibinfo {author} {\bibfnamefont {T.}~\bibnamefont {Martinez}},\ }\href@noop {} {\bibfield  {journal} {\bibinfo  {journal} {Information sciences}\ }\textbf {\bibinfo {volume} {124}},\ \bibinfo {pages} {273} (\bibinfo {year} {2000})}\BibitemShut {NoStop}%
\bibitem [{\citenamefont {Quiroz}\ \emph {et~al.}(2021)\citenamefont {Quiroz}, \citenamefont {Ice}, \citenamefont {Delgado},\ and\ \citenamefont {Humble}}]{quiroz2021hep}%
  \BibitemOpen
  \bibfield  {author} {\bibinfo {author} {\bibfnamefont {G.}~\bibnamefont {Quiroz}}, \bibinfo {author} {\bibfnamefont {L.}~\bibnamefont {Ice}}, \bibinfo {author} {\bibfnamefont {A.}~\bibnamefont {Delgado}},\ and\ \bibinfo {author} {\bibfnamefont {T.~S.}\ \bibnamefont {Humble}},\ }\href@noop {} {\bibfield  {journal} {\bibinfo  {journal} {Nuclear Instruments and Methods in Physics Research Section A: Accelerators, Spectrometers, Detectors and Associated Equipment}\ }\textbf {\bibinfo {volume} {1010}},\ \bibinfo {pages} {165557} (\bibinfo {year} {2021})}\BibitemShut {NoStop}%
\bibitem [{\citenamefont {Shapoval}\ and\ \citenamefont {Calafiura}(2019)}]{shapoval2019hep}%
  \BibitemOpen
  \bibfield  {author} {\bibinfo {author} {\bibfnamefont {I.}~\bibnamefont {Shapoval}}\ and\ \bibinfo {author} {\bibfnamefont {P.}~\bibnamefont {Calafiura}},\ }in\ \href@noop {} {\emph {\bibinfo {booktitle} {EPJ Web of Conferences}}},\ Vol.\ \bibinfo {volume} {214}\ (\bibinfo {organization} {EDP Sciences},\ \bibinfo {year} {2019})\ p.\ \bibinfo {pages} {01012}\BibitemShut {NoStop}%
\bibitem [{\citenamefont {Nokkala}(2023)}]{nokkala2023online}%
  \BibitemOpen
  \bibfield  {author} {\bibinfo {author} {\bibfnamefont {J.}~\bibnamefont {Nokkala}},\ }\href@noop {} {\bibfield  {journal} {\bibinfo  {journal} {Scientific Reports}\ }\textbf {\bibinfo {volume} {13}},\ \bibinfo {pages} {7694} (\bibinfo {year} {2023})}\BibitemShut {NoStop}%
\bibitem [{\citenamefont {Prins}\ \emph {et~al.}(2023)\citenamefont {Prins}, \citenamefont {der Sande},\ and\ \citenamefont {Bienstman}}]{deprins2023quantum}%
  \BibitemOpen
  \bibfield  {author} {\bibinfo {author} {\bibfnamefont {R.~D.}\ \bibnamefont {Prins}}, \bibinfo {author} {\bibfnamefont {G.~V.}\ \bibnamefont {der Sande}},\ and\ \bibinfo {author} {\bibfnamefont {P.}~\bibnamefont {Bienstman}},\ }\href@noop {} {\bibinfo {title} {A quantum optical recurrent neural network for online processing of quantum times series}} (\bibinfo {year} {2023}),\ \Eprint {https://arxiv.org/abs/2306.00134} {arXiv:2306.00134 [quant-ph]} \BibitemShut {NoStop}%
\bibitem [{\citenamefont {Mujal}\ \emph {et~al.}(2021{\natexlab{a}})\citenamefont {Mujal}, \citenamefont {Martínez-Peña}, \citenamefont {Nokkala}, \citenamefont {García-Beni}, \citenamefont {Giorgi}, \citenamefont {Soriano},\ and\ \citenamefont {Zambrini}}]{pere-opportunities}%
  \BibitemOpen
  \bibfield  {author} {\bibinfo {author} {\bibfnamefont {P.}~\bibnamefont {Mujal}}, \bibinfo {author} {\bibfnamefont {R.}~\bibnamefont {Martínez-Peña}}, \bibinfo {author} {\bibfnamefont {J.}~\bibnamefont {Nokkala}}, \bibinfo {author} {\bibfnamefont {J.}~\bibnamefont {García-Beni}}, \bibinfo {author} {\bibfnamefont {G.~L.}\ \bibnamefont {Giorgi}}, \bibinfo {author} {\bibfnamefont {M.~C.}\ \bibnamefont {Soriano}},\ and\ \bibinfo {author} {\bibfnamefont {R.}~\bibnamefont {Zambrini}},\ }\href {https://doi.org/https://doi.org/10.1002/qute.202100027} {\bibfield  {journal} {\bibinfo  {journal} {Advanced Quantum Technologies}\ }\textbf {\bibinfo {volume} {4}},\ \bibinfo {pages} {2100027} (\bibinfo {year} {2021}{\natexlab{a}})},\ \Eprint {https://arxiv.org/abs/https://onlinelibrary.wiley.com/doi/pdf/10.1002/qute.202100027} {https://onlinelibrary.wiley.com/doi/pdf/10.1002/qute.202100027} \BibitemShut {NoStop}%
\bibitem [{\citenamefont {Chen}\ \emph {et~al.}(2020)\citenamefont {Chen}, \citenamefont {Nurdin},\ and\ \citenamefont {Yamamoto}}]{PhysRevApplied.14.024065}%
  \BibitemOpen
  \bibfield  {author} {\bibinfo {author} {\bibfnamefont {J.}~\bibnamefont {Chen}}, \bibinfo {author} {\bibfnamefont {H.~I.}\ \bibnamefont {Nurdin}},\ and\ \bibinfo {author} {\bibfnamefont {N.}~\bibnamefont {Yamamoto}},\ }\href {https://doi.org/10.1103/PhysRevApplied.14.024065} {\bibfield  {journal} {\bibinfo  {journal} {Phys. Rev. Appl.}\ }\textbf {\bibinfo {volume} {14}},\ \bibinfo {pages} {024065} (\bibinfo {year} {2020})}\BibitemShut {NoStop}%
\bibitem [{\citenamefont {Fry}\ \emph {et~al.}(2023)\citenamefont {Fry}, \citenamefont {Deshmukh}, \citenamefont {Chen}, \citenamefont {Rastunkov},\ and\ \citenamefont {Markov}}]{fry2023optimizing}%
  \BibitemOpen
  \bibfield  {author} {\bibinfo {author} {\bibfnamefont {D.}~\bibnamefont {Fry}}, \bibinfo {author} {\bibfnamefont {A.}~\bibnamefont {Deshmukh}}, \bibinfo {author} {\bibfnamefont {S.~Y.-C.}\ \bibnamefont {Chen}}, \bibinfo {author} {\bibfnamefont {V.}~\bibnamefont {Rastunkov}},\ and\ \bibinfo {author} {\bibfnamefont {V.}~\bibnamefont {Markov}},\ }\href@noop {} {\bibfield  {journal} {\bibinfo  {journal} {Scientific Reports}\ }\textbf {\bibinfo {volume} {13}},\ \bibinfo {pages} {19326} (\bibinfo {year} {2023})}\BibitemShut {NoStop}%
\bibitem [{\citenamefont {Domingo}\ \emph {et~al.}(2022)\citenamefont {Domingo}, \citenamefont {Carlo},\ and\ \citenamefont {Borondo}}]{domingo2022qrcgates}%
  \BibitemOpen
  \bibfield  {author} {\bibinfo {author} {\bibfnamefont {L.}~\bibnamefont {Domingo}}, \bibinfo {author} {\bibfnamefont {G.}~\bibnamefont {Carlo}},\ and\ \bibinfo {author} {\bibfnamefont {F.}~\bibnamefont {Borondo}},\ }\href {https://doi.org/10.1103/PhysRevE.106.L043301} {\bibfield  {journal} {\bibinfo  {journal} {Phys. Rev. E}\ }\textbf {\bibinfo {volume} {106}},\ \bibinfo {pages} {L043301} (\bibinfo {year} {2022})}\BibitemShut {NoStop}%
\bibitem [{\citenamefont {Fujii}\ and\ \citenamefont {Nakajima}(2017)}]{fujii-nakajima-2017}%
  \BibitemOpen
  \bibfield  {author} {\bibinfo {author} {\bibfnamefont {K.}~\bibnamefont {Fujii}}\ and\ \bibinfo {author} {\bibfnamefont {K.}~\bibnamefont {Nakajima}},\ }\href {https://doi.org/10.1103/PhysRevApplied.8.024030} {\bibfield  {journal} {\bibinfo  {journal} {Phys. Rev. Appl.}\ }\textbf {\bibinfo {volume} {8}},\ \bibinfo {pages} {024030} (\bibinfo {year} {2017})}\BibitemShut {NoStop}%
\bibitem [{\citenamefont {Luko{\v{s}}evi{\v{c}}ius}\ and\ \citenamefont {Jaeger}(2009)}]{lukovsevivcius2009reservoir}%
  \BibitemOpen
  \bibfield  {author} {\bibinfo {author} {\bibfnamefont {M.}~\bibnamefont {Luko{\v{s}}evi{\v{c}}ius}}\ and\ \bibinfo {author} {\bibfnamefont {H.}~\bibnamefont {Jaeger}},\ }\href@noop {} {\bibfield  {journal} {\bibinfo  {journal} {Computer science review}\ }\textbf {\bibinfo {volume} {3}},\ \bibinfo {pages} {127} (\bibinfo {year} {2009})}\BibitemShut {NoStop}%
\bibitem [{\citenamefont {Nakajima}(2020)}]{nakajima2020physical}%
  \BibitemOpen
  \bibfield  {author} {\bibinfo {author} {\bibfnamefont {K.}~\bibnamefont {Nakajima}},\ }\href@noop {} {\bibfield  {journal} {\bibinfo  {journal} {Japanese Journal of Applied Physics}\ }\textbf {\bibinfo {volume} {59}},\ \bibinfo {pages} {060501} (\bibinfo {year} {2020})}\BibitemShut {NoStop}%
\bibitem [{\citenamefont {Brunner}\ \emph {et~al.}(2019)\citenamefont {Brunner}, \citenamefont {Soriano},\ and\ \citenamefont {Van~der Sande}}]{brunner2019photonic}%
  \BibitemOpen
  \bibfield  {author} {\bibinfo {author} {\bibfnamefont {D.}~\bibnamefont {Brunner}}, \bibinfo {author} {\bibfnamefont {M.~C.}\ \bibnamefont {Soriano}},\ and\ \bibinfo {author} {\bibfnamefont {G.}~\bibnamefont {Van~der Sande}},\ }\href@noop {} {\emph {\bibinfo {title} {Photonic Reservoir Computing: Optical Recurrent Neural Networks}}}\ (\bibinfo  {publisher} {Walter de Gruyter GmbH \& Co KG},\ \bibinfo {year} {2019})\BibitemShut {NoStop}%
\bibitem [{\citenamefont {Spagnolo}\ \emph {et~al.}(2022)\citenamefont {Spagnolo}, \citenamefont {Morris}, \citenamefont {Piacentini}, \citenamefont {Antesberger}, \citenamefont {Massa}, \citenamefont {Crespi}, \citenamefont {Ceccarelli}, \citenamefont {Osellame},\ and\ \citenamefont {Walther}}]{Spagnolo2022}%
  \BibitemOpen
  \bibfield  {author} {\bibinfo {author} {\bibfnamefont {M.}~\bibnamefont {Spagnolo}}, \bibinfo {author} {\bibfnamefont {J.}~\bibnamefont {Morris}}, \bibinfo {author} {\bibfnamefont {S.}~\bibnamefont {Piacentini}}, \bibinfo {author} {\bibfnamefont {M.}~\bibnamefont {Antesberger}}, \bibinfo {author} {\bibfnamefont {F.}~\bibnamefont {Massa}}, \bibinfo {author} {\bibfnamefont {A.}~\bibnamefont {Crespi}}, \bibinfo {author} {\bibfnamefont {F.}~\bibnamefont {Ceccarelli}}, \bibinfo {author} {\bibfnamefont {R.}~\bibnamefont {Osellame}},\ and\ \bibinfo {author} {\bibfnamefont {P.}~\bibnamefont {Walther}},\ }\href {https://doi.org/10.1038/s41566-022-00973-5} {\bibfield  {journal} {\bibinfo  {journal} {Nature Photonics}\ }\textbf {\bibinfo {volume} {16}},\ \bibinfo {pages} {318} (\bibinfo {year} {2022})}\BibitemShut {NoStop}%
\bibitem [{\citenamefont {Llodrà}\ \emph {et~al.}(2023)\citenamefont {Llodrà}, \citenamefont {Charalambous}, \citenamefont {Giorgi},\ and\ \citenamefont {Zambrini}}]{Guillem2023}%
  \BibitemOpen
  \bibfield  {author} {\bibinfo {author} {\bibfnamefont {G.}~\bibnamefont {Llodrà}}, \bibinfo {author} {\bibfnamefont {C.}~\bibnamefont {Charalambous}}, \bibinfo {author} {\bibfnamefont {G.~L.}\ \bibnamefont {Giorgi}},\ and\ \bibinfo {author} {\bibfnamefont {R.}~\bibnamefont {Zambrini}},\ }\href {https://doi.org/https://doi.org/10.1002/qute.202200100} {\bibfield  {journal} {\bibinfo  {journal} {Advanced Quantum Technologies}\ }\textbf {\bibinfo {volume} {6}},\ \bibinfo {pages} {2200100} (\bibinfo {year} {2023})},\ \Eprint {https://arxiv.org/abs/https://onlinelibrary.wiley.com/doi/pdf/10.1002/qute.202200100} {https://onlinelibrary.wiley.com/doi/pdf/10.1002/qute.202200100} \BibitemShut {NoStop}%
\bibitem [{\citenamefont {Dudas}\ \emph {et~al.}(2023)\citenamefont {Dudas}, \citenamefont {Carles}, \citenamefont {Plouet}, \citenamefont {Mizrahi}, \citenamefont {Grollier},\ and\ \citenamefont {Markovi{\'{c}}}}]{Dudas2023}%
  \BibitemOpen
  \bibfield  {author} {\bibinfo {author} {\bibfnamefont {J.}~\bibnamefont {Dudas}}, \bibinfo {author} {\bibfnamefont {B.}~\bibnamefont {Carles}}, \bibinfo {author} {\bibfnamefont {E.}~\bibnamefont {Plouet}}, \bibinfo {author} {\bibfnamefont {F.~A.}\ \bibnamefont {Mizrahi}}, \bibinfo {author} {\bibfnamefont {J.}~\bibnamefont {Grollier}},\ and\ \bibinfo {author} {\bibfnamefont {D.}~\bibnamefont {Markovi{\'{c}}}},\ }\href {https://doi.org/10.1038/s41534-023-00734-4} {\bibfield  {journal} {\bibinfo  {journal} {npj Quantum Information}\ }\textbf {\bibinfo {volume} {9}},\ \bibinfo {pages} {64} (\bibinfo {year} {2023})}\BibitemShut {NoStop}%
\bibitem [{\citenamefont {Govia}\ \emph {et~al.}(2021)\citenamefont {Govia}, \citenamefont {Ribeill}, \citenamefont {Rowlands}, \citenamefont {Krovi},\ and\ \citenamefont {Ohki}}]{govia2021quantum}%
  \BibitemOpen
  \bibfield  {author} {\bibinfo {author} {\bibfnamefont {L.~C.~G.}\ \bibnamefont {Govia}}, \bibinfo {author} {\bibfnamefont {G.~J.}\ \bibnamefont {Ribeill}}, \bibinfo {author} {\bibfnamefont {G.~E.}\ \bibnamefont {Rowlands}}, \bibinfo {author} {\bibfnamefont {H.~K.}\ \bibnamefont {Krovi}},\ and\ \bibinfo {author} {\bibfnamefont {T.~A.}\ \bibnamefont {Ohki}},\ }\href {https://doi.org/10.1103/PhysRevResearch.3.013077} {\bibfield  {journal} {\bibinfo  {journal} {Phys. Rev. Research}\ }\textbf {\bibinfo {volume} {3}},\ \bibinfo {pages} {013077} (\bibinfo {year} {2021})}\BibitemShut {NoStop}%
\bibitem [{\citenamefont {Nokkala}\ \emph {et~al.}(2021)\citenamefont {Nokkala}, \citenamefont {Mart{\'i}nez-Pe{\~{n}}a}, \citenamefont {Giorgi}, \citenamefont {Parigi}, \citenamefont {Soriano},\ and\ \citenamefont {Zambrini}}]{Nokkala2021}%
  \BibitemOpen
  \bibfield  {author} {\bibinfo {author} {\bibfnamefont {J.}~\bibnamefont {Nokkala}}, \bibinfo {author} {\bibfnamefont {R.}~\bibnamefont {Mart{\'i}nez-Pe{\~{n}}a}}, \bibinfo {author} {\bibfnamefont {G.~L.}\ \bibnamefont {Giorgi}}, \bibinfo {author} {\bibfnamefont {V.}~\bibnamefont {Parigi}}, \bibinfo {author} {\bibfnamefont {M.~C.}\ \bibnamefont {Soriano}},\ and\ \bibinfo {author} {\bibfnamefont {R.}~\bibnamefont {Zambrini}},\ }\href {https://doi.org/10.1038/s42005-021-00556-w} {\bibfield  {journal} {\bibinfo  {journal} {Communications Physics}\ }\textbf {\bibinfo {volume} {4}},\ \bibinfo {pages} {53} (\bibinfo {year} {2021})}\BibitemShut {NoStop}%
\bibitem [{\citenamefont {Garc\'{\i}a-Beni}\ \emph {et~al.}(2023)\citenamefont {Garc\'{\i}a-Beni}, \citenamefont {Giorgi}, \citenamefont {Soriano},\ and\ \citenamefont {Zambrini}}]{GBeni2023}%
  \BibitemOpen
  \bibfield  {author} {\bibinfo {author} {\bibfnamefont {J.}~\bibnamefont {Garc\'{\i}a-Beni}}, \bibinfo {author} {\bibfnamefont {G.~L.}\ \bibnamefont {Giorgi}}, \bibinfo {author} {\bibfnamefont {M.~C.}\ \bibnamefont {Soriano}},\ and\ \bibinfo {author} {\bibfnamefont {R.}~\bibnamefont {Zambrini}},\ }\href {https://doi.org/10.1103/PhysRevApplied.20.014051} {\bibfield  {journal} {\bibinfo  {journal} {Phys. Rev. Appl.}\ }\textbf {\bibinfo {volume} {20}},\ \bibinfo {pages} {014051} (\bibinfo {year} {2023})}\BibitemShut {NoStop}%
\bibitem [{\citenamefont {Garc\'{i}a-Beni}\ \emph {et~al.}(2024)\citenamefont {Garc\'{i}a-Beni}, \citenamefont {Giorgi}, \citenamefont {Soriano},\ and\ \citenamefont {Zambrini}}]{GBeni2024}%
  \BibitemOpen
  \bibfield  {author} {\bibinfo {author} {\bibfnamefont {J.}~\bibnamefont {Garc\'{i}a-Beni}}, \bibinfo {author} {\bibfnamefont {G.~L.}\ \bibnamefont {Giorgi}}, \bibinfo {author} {\bibfnamefont {M.~C.}\ \bibnamefont {Soriano}},\ and\ \bibinfo {author} {\bibfnamefont {R.}~\bibnamefont {Zambrini}},\ }\href {https://doi.org/10.1364/OE.507684} {\bibfield  {journal} {\bibinfo  {journal} {Opt. Express}\ }\textbf {\bibinfo {volume} {32}},\ \bibinfo {pages} {6733} (\bibinfo {year} {2024})}\BibitemShut {NoStop}%
\bibitem [{\citenamefont {Angelatos}\ \emph {et~al.}(2021)\citenamefont {Angelatos}, \citenamefont {Khan},\ and\ \citenamefont {T\"ureci}}]{Angelatos2021}%
  \BibitemOpen
  \bibfield  {author} {\bibinfo {author} {\bibfnamefont {G.}~\bibnamefont {Angelatos}}, \bibinfo {author} {\bibfnamefont {S.~A.}\ \bibnamefont {Khan}},\ and\ \bibinfo {author} {\bibfnamefont {H.~E.}\ \bibnamefont {T\"ureci}},\ }\href {https://doi.org/10.1103/PhysRevX.11.041062} {\bibfield  {journal} {\bibinfo  {journal} {Phys. Rev. X}\ }\textbf {\bibinfo {volume} {11}},\ \bibinfo {pages} {041062} (\bibinfo {year} {2021})}\BibitemShut {NoStop}%
\bibitem [{\citenamefont {Senanian}\ \emph {et~al.}(2023)\citenamefont {Senanian}, \citenamefont {Prabhu}, \citenamefont {Kremenetski}, \citenamefont {Roy}, \citenamefont {Cao}, \citenamefont {Kline}, \citenamefont {Onodera}, \citenamefont {Wright}, \citenamefont {Wu}, \citenamefont {Fatemi} \emph {et~al.}}]{senanian2023microwave}%
  \BibitemOpen
  \bibfield  {author} {\bibinfo {author} {\bibfnamefont {A.}~\bibnamefont {Senanian}}, \bibinfo {author} {\bibfnamefont {S.}~\bibnamefont {Prabhu}}, \bibinfo {author} {\bibfnamefont {V.}~\bibnamefont {Kremenetski}}, \bibinfo {author} {\bibfnamefont {S.}~\bibnamefont {Roy}}, \bibinfo {author} {\bibfnamefont {Y.}~\bibnamefont {Cao}}, \bibinfo {author} {\bibfnamefont {J.}~\bibnamefont {Kline}}, \bibinfo {author} {\bibfnamefont {T.}~\bibnamefont {Onodera}}, \bibinfo {author} {\bibfnamefont {L.~G.}\ \bibnamefont {Wright}}, \bibinfo {author} {\bibfnamefont {X.}~\bibnamefont {Wu}}, \bibinfo {author} {\bibfnamefont {V.}~\bibnamefont {Fatemi}}, \emph {et~al.},\ }\href@noop {} {\bibfield  {journal} {\bibinfo  {journal} {arXiv preprint arXiv:2312.16166}\ } (\bibinfo {year} {2023})}\BibitemShut {NoStop}%
\bibitem [{\citenamefont {Hopfield}(1982)}]{hopfield1982neural}%
  \BibitemOpen
  \bibfield  {author} {\bibinfo {author} {\bibfnamefont {J.~J.}\ \bibnamefont {Hopfield}},\ }\href@noop {} {\bibfield  {journal} {\bibinfo  {journal} {Proceedings of the national academy of sciences}\ }\textbf {\bibinfo {volume} {79}},\ \bibinfo {pages} {2554} (\bibinfo {year} {1982})}\BibitemShut {NoStop}%
\bibitem [{\citenamefont {Amit}(1989)}]{amit_1989}%
  \BibitemOpen
  \bibfield  {author} {\bibinfo {author} {\bibfnamefont {D.~J.}\ \bibnamefont {Amit}},\ }\href {https://doi.org/10.1017/CBO9780511623257} {\emph {\bibinfo {title} {Modeling Brain Function: The World of Attractor Neural Networks}}}\ (\bibinfo  {publisher} {Cambridge University Press},\ \bibinfo {year} {1989})\BibitemShut {NoStop}%
\bibitem [{\citenamefont {Inoue}(2011)}]{inoue2011pattern}%
  \BibitemOpen
  \bibfield  {author} {\bibinfo {author} {\bibfnamefont {J.-i.}\ \bibnamefont {Inoue}},\ }in\ \href@noop {} {\emph {\bibinfo {booktitle} {Journal of Physics: Conference Series}}},\ Vol.\ \bibinfo {volume} {297}\ (\bibinfo {organization} {IOP Publishing},\ \bibinfo {year} {2011})\ p.\ \bibinfo {pages} {012012}\BibitemShut {NoStop}%
\bibitem [{\citenamefont {Neigovzen}\ \emph {et~al.}(2009)\citenamefont {Neigovzen}, \citenamefont {Neves}, \citenamefont {Sollacher},\ and\ \citenamefont {Glaser}}]{glaser2009nuclear}%
  \BibitemOpen
  \bibfield  {author} {\bibinfo {author} {\bibfnamefont {R.}~\bibnamefont {Neigovzen}}, \bibinfo {author} {\bibfnamefont {J.~L.}\ \bibnamefont {Neves}}, \bibinfo {author} {\bibfnamefont {R.}~\bibnamefont {Sollacher}},\ and\ \bibinfo {author} {\bibfnamefont {S.~J.}\ \bibnamefont {Glaser}},\ }\href {https://doi.org/10.1103/PhysRevA.79.042321} {\bibfield  {journal} {\bibinfo  {journal} {Phys. Rev. A}\ }\textbf {\bibinfo {volume} {79}},\ \bibinfo {pages} {042321} (\bibinfo {year} {2009})}\BibitemShut {NoStop}%
\bibitem [{\citenamefont {Rotondo}\ \emph {et~al.}(2018)\citenamefont {Rotondo}, \citenamefont {Marcuzzi}, \citenamefont {Garrahan}, \citenamefont {Lesanovsky},\ and\ \citenamefont {M{\"u}ller}}]{rotondo2018open}%
  \BibitemOpen
  \bibfield  {author} {\bibinfo {author} {\bibfnamefont {P.}~\bibnamefont {Rotondo}}, \bibinfo {author} {\bibfnamefont {M.}~\bibnamefont {Marcuzzi}}, \bibinfo {author} {\bibfnamefont {J.~P.}\ \bibnamefont {Garrahan}}, \bibinfo {author} {\bibfnamefont {I.}~\bibnamefont {Lesanovsky}},\ and\ \bibinfo {author} {\bibfnamefont {M.}~\bibnamefont {M{\"u}ller}},\ }\href@noop {} {\bibfield  {journal} {\bibinfo  {journal} {Journal of Physics A: Mathematical and Theoretical}\ }\textbf {\bibinfo {volume} {51}},\ \bibinfo {pages} {115301} (\bibinfo {year} {2018})}\BibitemShut {NoStop}%
\bibitem [{\citenamefont {Fiorelli}\ \emph {et~al.}(2019)\citenamefont {Fiorelli}, \citenamefont {Rotondo}, \citenamefont {Marcuzzi}, \citenamefont {Garrahan},\ and\ \citenamefont {Lesanovsky}}]{fiorelli2019accelerated}%
  \BibitemOpen
  \bibfield  {author} {\bibinfo {author} {\bibfnamefont {E.}~\bibnamefont {Fiorelli}}, \bibinfo {author} {\bibfnamefont {P.}~\bibnamefont {Rotondo}}, \bibinfo {author} {\bibfnamefont {M.}~\bibnamefont {Marcuzzi}}, \bibinfo {author} {\bibfnamefont {J.~P.}\ \bibnamefont {Garrahan}},\ and\ \bibinfo {author} {\bibfnamefont {I.}~\bibnamefont {Lesanovsky}},\ }\href {https://doi.org/10.1103/PhysRevA.99.032126} {\bibfield  {journal} {\bibinfo  {journal} {Phys. Rev. A}\ }\textbf {\bibinfo {volume} {99}},\ \bibinfo {pages} {032126} (\bibinfo {year} {2019})}\BibitemShut {NoStop}%
\bibitem [{\citenamefont {Fiorelli}\ \emph {et~al.}(2022)\citenamefont {Fiorelli}, \citenamefont {Lesanovsky},\ and\ \citenamefont {Müller}}]{fiorelli2021potts}%
  \BibitemOpen
  \bibfield  {author} {\bibinfo {author} {\bibfnamefont {E.}~\bibnamefont {Fiorelli}}, \bibinfo {author} {\bibfnamefont {I.}~\bibnamefont {Lesanovsky}},\ and\ \bibinfo {author} {\bibfnamefont {M.}~\bibnamefont {Müller}},\ }\href {https://doi.org/10.1088/1367-2630/ac5490} {\bibfield  {journal} {\bibinfo  {journal} {New Journal of Physics}\ }\textbf {\bibinfo {volume} {24}},\ \bibinfo {pages} {033012} (\bibinfo {year} {2022})}\BibitemShut {NoStop}%
\bibitem [{\citenamefont {Fiorelli}\ \emph {et~al.}(2020)\citenamefont {Fiorelli}, \citenamefont {Marcuzzi}, \citenamefont {Rotondo}, \citenamefont {Carollo},\ and\ \citenamefont {Lesanovsky}}]{fiorelli2020signatures}%
  \BibitemOpen
  \bibfield  {author} {\bibinfo {author} {\bibfnamefont {E.}~\bibnamefont {Fiorelli}}, \bibinfo {author} {\bibfnamefont {M.}~\bibnamefont {Marcuzzi}}, \bibinfo {author} {\bibfnamefont {P.}~\bibnamefont {Rotondo}}, \bibinfo {author} {\bibfnamefont {F.}~\bibnamefont {Carollo}},\ and\ \bibinfo {author} {\bibfnamefont {I.}~\bibnamefont {Lesanovsky}},\ }\href@noop {} {\bibfield  {journal} {\bibinfo  {journal} {Phys. Rev. Lett.}\ }\textbf {\bibinfo {volume} {125}},\ \bibinfo {pages} {070604} (\bibinfo {year} {2020})}\BibitemShut {NoStop}%
\bibitem [{\citenamefont {Marsh}\ \emph {et~al.}(2021)\citenamefont {Marsh}, \citenamefont {Guo}, \citenamefont {Kroeze}, \citenamefont {Gopalakrishnan}, \citenamefont {Ganguli}, \citenamefont {Keeling},\ and\ \citenamefont {Lev}}]{enhancing2021marsh}%
  \BibitemOpen
  \bibfield  {author} {\bibinfo {author} {\bibfnamefont {B.~P.}\ \bibnamefont {Marsh}}, \bibinfo {author} {\bibfnamefont {Y.}~\bibnamefont {Guo}}, \bibinfo {author} {\bibfnamefont {R.~M.}\ \bibnamefont {Kroeze}}, \bibinfo {author} {\bibfnamefont {S.}~\bibnamefont {Gopalakrishnan}}, \bibinfo {author} {\bibfnamefont {S.}~\bibnamefont {Ganguli}}, \bibinfo {author} {\bibfnamefont {J.}~\bibnamefont {Keeling}},\ and\ \bibinfo {author} {\bibfnamefont {B.~L.}\ \bibnamefont {Lev}},\ }\href {https://doi.org/10.1103/PhysRevX.11.021048} {\bibfield  {journal} {\bibinfo  {journal} {Phys. Rev. X}\ }\textbf {\bibinfo {volume} {11}},\ \bibinfo {pages} {021048} (\bibinfo {year} {2021})}\BibitemShut {NoStop}%
\bibitem [{\citenamefont {B{\"o}deker}\ \emph {et~al.}(2022)\citenamefont {B{\"o}deker}, \citenamefont {Fiorelli},\ and\ \citenamefont {M{\"u}ller}}]{bodeker2022optimal}%
  \BibitemOpen
  \bibfield  {author} {\bibinfo {author} {\bibfnamefont {L.}~\bibnamefont {B{\"o}deker}}, \bibinfo {author} {\bibfnamefont {E.}~\bibnamefont {Fiorelli}},\ and\ \bibinfo {author} {\bibfnamefont {M.}~\bibnamefont {M{\"u}ller}},\ }\href {https://doi.org/10.1103/PhysRevResearch.5.023074} {\bibfield  {journal} {\bibinfo  {journal} {arXiv preprint arXiv:2210.07894}\ }\textbf {\bibinfo {volume} {5}},\ \bibinfo {pages} {023074} (\bibinfo {year} {2022})}\BibitemShut {NoStop}%
\bibitem [{\citenamefont {Lewenstein}\ \emph {et~al.}(2021)\citenamefont {Lewenstein}, \citenamefont {Gratsea}, \citenamefont {Riera-Campeny}, \citenamefont {Aloy}, \citenamefont {Kasper},\ and\ \citenamefont {Sanpera}}]{sanpera2021capacity}%
  \BibitemOpen
  \bibfield  {author} {\bibinfo {author} {\bibfnamefont {M.}~\bibnamefont {Lewenstein}}, \bibinfo {author} {\bibfnamefont {A.}~\bibnamefont {Gratsea}}, \bibinfo {author} {\bibfnamefont {A.}~\bibnamefont {Riera-Campeny}}, \bibinfo {author} {\bibfnamefont {A.}~\bibnamefont {Aloy}}, \bibinfo {author} {\bibfnamefont {V.}~\bibnamefont {Kasper}},\ and\ \bibinfo {author} {\bibfnamefont {A.}~\bibnamefont {Sanpera}},\ }\href {https://doi.org/10.1088/2058-9565/ac070f} {\bibfield  {journal} {\bibinfo  {journal} {Quantum Science and Technology}\ }\textbf {\bibinfo {volume} {6}},\ \bibinfo {pages} {045002} (\bibinfo {year} {2021})}\BibitemShut {NoStop}%
\bibitem [{\citenamefont {Labay-Mora}\ \emph {et~al.}(2024{\natexlab{b}})\citenamefont {Labay-Mora}, \citenamefont {Fiorelli}, \citenamefont {Zambrini},\ and\ \citenamefont {Giorgi}}]{labay2024theoretical}%
  \BibitemOpen
  \bibfield  {author} {\bibinfo {author} {\bibfnamefont {A.}~\bibnamefont {Labay-Mora}}, \bibinfo {author} {\bibfnamefont {E.}~\bibnamefont {Fiorelli}}, \bibinfo {author} {\bibfnamefont {R.}~\bibnamefont {Zambrini}},\ and\ \bibinfo {author} {\bibfnamefont {G.~L.}\ \bibnamefont {Giorgi}},\ }\href@noop {} {\bibfield  {journal} {\bibinfo  {journal} {arXiv preprint arXiv:2408.14272}\ } (\bibinfo {year} {2024}{\natexlab{b}})}\BibitemShut {NoStop}%
\bibitem [{\citenamefont {Labay-Mora}\ \emph {et~al.}(2023{\natexlab{b}})\citenamefont {Labay-Mora}, \citenamefont {Zambrini},\ and\ \citenamefont {Giorgi}}]{labay2022prl}%
  \BibitemOpen
  \bibfield  {author} {\bibinfo {author} {\bibfnamefont {A.}~\bibnamefont {Labay-Mora}}, \bibinfo {author} {\bibfnamefont {R.}~\bibnamefont {Zambrini}},\ and\ \bibinfo {author} {\bibfnamefont {G.~L.}\ \bibnamefont {Giorgi}},\ }\href {https://doi.org/10.1103/PhysRevLett.130.190602} {\bibfield  {journal} {\bibinfo  {journal} {Phys. Rev. Lett.}\ }\textbf {\bibinfo {volume} {130}},\ \bibinfo {pages} {190602} (\bibinfo {year} {2023}{\natexlab{b}})}\BibitemShut {NoStop}%
\bibitem [{\citenamefont {Giovannetti}\ \emph {et~al.}(2001)\citenamefont {Giovannetti}, \citenamefont {Lloyd},\ and\ \citenamefont {Maccone}}]{Giovannetti2001}%
  \BibitemOpen
  \bibfield  {author} {\bibinfo {author} {\bibfnamefont {V.}~\bibnamefont {Giovannetti}}, \bibinfo {author} {\bibfnamefont {S.}~\bibnamefont {Lloyd}},\ and\ \bibinfo {author} {\bibfnamefont {L.}~\bibnamefont {Maccone}},\ }\href {https://doi.org/10.1038/35086525} {\bibfield  {journal} {\bibinfo  {journal} {Nature}\ }\textbf {\bibinfo {volume} {412}},\ \bibinfo {pages} {417} (\bibinfo {year} {2001})}\BibitemShut {NoStop}%
\bibitem [{\citenamefont {Madsen}\ \emph {et~al.}(2012)\citenamefont {Madsen}, \citenamefont {Usenko}, \citenamefont {Lassen}, \citenamefont {Filip},\ and\ \citenamefont {Andersen}}]{Madsen2012}%
  \BibitemOpen
  \bibfield  {author} {\bibinfo {author} {\bibfnamefont {L.~S.}\ \bibnamefont {Madsen}}, \bibinfo {author} {\bibfnamefont {V.~C.}\ \bibnamefont {Usenko}}, \bibinfo {author} {\bibfnamefont {M.}~\bibnamefont {Lassen}}, \bibinfo {author} {\bibfnamefont {R.}~\bibnamefont {Filip}},\ and\ \bibinfo {author} {\bibfnamefont {U.~L.}\ \bibnamefont {Andersen}},\ }\href {https://doi.org/10.1038/ncomms2097} {\bibfield  {journal} {\bibinfo  {journal} {Nature Communications}\ }\textbf {\bibinfo {volume} {3}},\ \bibinfo {pages} {1083} (\bibinfo {year} {2012})}\BibitemShut {NoStop}%
\bibitem [{\citenamefont {Mujal}\ \emph {et~al.}(2021{\natexlab{b}})\citenamefont {Mujal}, \citenamefont {Nokkala}, \citenamefont {Martínez-Peña}, \citenamefont {Giorgi}, \citenamefont {Soriano},\ and\ \citenamefont {Zambrini}}]{Mujal_2021}%
  \BibitemOpen
  \bibfield  {author} {\bibinfo {author} {\bibfnamefont {P.}~\bibnamefont {Mujal}}, \bibinfo {author} {\bibfnamefont {J.}~\bibnamefont {Nokkala}}, \bibinfo {author} {\bibfnamefont {R.}~\bibnamefont {Martínez-Peña}}, \bibinfo {author} {\bibfnamefont {G.~L.}\ \bibnamefont {Giorgi}}, \bibinfo {author} {\bibfnamefont {M.~C.}\ \bibnamefont {Soriano}},\ and\ \bibinfo {author} {\bibfnamefont {R.}~\bibnamefont {Zambrini}},\ }\href {https://doi.org/10.1088/2632-072X/ac340e} {\bibfield  {journal} {\bibinfo  {journal} {Journal of Physics: Complexity}\ }\textbf {\bibinfo {volume} {2}},\ \bibinfo {pages} {045008} (\bibinfo {year} {2021}{\natexlab{b}})}\BibitemShut {NoStop}%
\bibitem [{\citenamefont {Nokkala}\ \emph {et~al.}(2022)\citenamefont {Nokkala}, \citenamefont {Martínez-Peña}, \citenamefont {Zambrini},\ and\ \citenamefont {Soriano}}]{Nokkala_noise}%
  \BibitemOpen
  \bibfield  {author} {\bibinfo {author} {\bibfnamefont {J.}~\bibnamefont {Nokkala}}, \bibinfo {author} {\bibfnamefont {R.}~\bibnamefont {Martínez-Peña}}, \bibinfo {author} {\bibfnamefont {R.}~\bibnamefont {Zambrini}},\ and\ \bibinfo {author} {\bibfnamefont {M.~C.}\ \bibnamefont {Soriano}},\ }\href {https://doi.org/10.1109/TNNLS.2021.3105695} {\bibfield  {journal} {\bibinfo  {journal} {IEEE Transactions on Neural Networks and Learning Systems}\ }\textbf {\bibinfo {volume} {33}},\ \bibinfo {pages} {2664} (\bibinfo {year} {2022})}\BibitemShut {NoStop}%
\bibitem [{\citenamefont {Collett}\ \emph {et~al.}(1987)\citenamefont {Collett}, \citenamefont {Loudon},\ and\ \citenamefont {Gardiner}}]{loudon1987homodyne}%
  \BibitemOpen
  \bibfield  {author} {\bibinfo {author} {\bibfnamefont {M.}~\bibnamefont {Collett}}, \bibinfo {author} {\bibfnamefont {R.}~\bibnamefont {Loudon}},\ and\ \bibinfo {author} {\bibfnamefont {C.}~\bibnamefont {Gardiner}},\ }\href@noop {} {\bibfield  {journal} {\bibinfo  {journal} {Journal of Modern Optics}\ }\textbf {\bibinfo {volume} {34}},\ \bibinfo {pages} {881} (\bibinfo {year} {1987})}\BibitemShut {NoStop}%
\bibitem [{\citenamefont {Serafini}(2017)}]{serafini2017}%
  \BibitemOpen
  \bibfield  {author} {\bibinfo {author} {\bibfnamefont {A.}~\bibnamefont {Serafini}},\ }\href@noop {} {\emph {\bibinfo {title} {Quantum continuous variables: a primer of theoretical methods}}}\ (\bibinfo  {publisher} {CRC Press},\ \bibinfo {year} {2017})\BibitemShut {NoStop}%
\bibitem [{\citenamefont {Mujal}\ \emph {et~al.}(2023)\citenamefont {Mujal}, \citenamefont {Mart{\'\i}nez-Pe{\~n}a}, \citenamefont {Giorgi}, \citenamefont {Soriano},\ and\ \citenamefont {Zambrini}}]{mujal2023time}%
  \BibitemOpen
  \bibfield  {author} {\bibinfo {author} {\bibfnamefont {P.}~\bibnamefont {Mujal}}, \bibinfo {author} {\bibfnamefont {R.}~\bibnamefont {Mart{\'\i}nez-Pe{\~n}a}}, \bibinfo {author} {\bibfnamefont {G.~L.}\ \bibnamefont {Giorgi}}, \bibinfo {author} {\bibfnamefont {M.~C.}\ \bibnamefont {Soriano}},\ and\ \bibinfo {author} {\bibfnamefont {R.}~\bibnamefont {Zambrini}},\ }\href@noop {} {\bibfield  {journal} {\bibinfo  {journal} {npj Quantum Information}\ }\textbf {\bibinfo {volume} {9}},\ \bibinfo {pages} {16} (\bibinfo {year} {2023})}\BibitemShut {NoStop}%
\bibitem [{\citenamefont {Braunstein}(2005)}]{BM_1}%
  \BibitemOpen
  \bibfield  {author} {\bibinfo {author} {\bibfnamefont {S.~L.}\ \bibnamefont {Braunstein}},\ }\href {https://doi.org/10.1103/PhysRevA.71.055801} {\bibfield  {journal} {\bibinfo  {journal} {Phys. Rev. A}\ }\textbf {\bibinfo {volume} {71}},\ \bibinfo {pages} {055801} (\bibinfo {year} {2005})}\BibitemShut {NoStop}%
\bibitem [{\citenamefont {Cariolaro}\ and\ \citenamefont {Pierobon}(2016)}]{BM_2}%
  \BibitemOpen
  \bibfield  {author} {\bibinfo {author} {\bibfnamefont {G.}~\bibnamefont {Cariolaro}}\ and\ \bibinfo {author} {\bibfnamefont {G.}~\bibnamefont {Pierobon}},\ }\href {https://doi.org/10.1103/PhysRevA.93.062115} {\bibfield  {journal} {\bibinfo  {journal} {Phys. Rev. A}\ }\textbf {\bibinfo {volume} {93}},\ \bibinfo {pages} {062115} (\bibinfo {year} {2016})}\BibitemShut {NoStop}%
\bibitem [{\citenamefont {H\"ubner}\ \emph {et~al.}(1989)\citenamefont {H\"ubner}, \citenamefont {Abraham},\ and\ \citenamefont {Weiss}}]{santafe1}%
  \BibitemOpen
  \bibfield  {author} {\bibinfo {author} {\bibfnamefont {U.}~\bibnamefont {H\"ubner}}, \bibinfo {author} {\bibfnamefont {N.~B.}\ \bibnamefont {Abraham}},\ and\ \bibinfo {author} {\bibfnamefont {C.~O.}\ \bibnamefont {Weiss}},\ }\href {https://doi.org/10.1103/PhysRevA.40.6354} {\bibfield  {journal} {\bibinfo  {journal} {Phys. Rev. A}\ }\textbf {\bibinfo {volume} {40}},\ \bibinfo {pages} {6354} (\bibinfo {year} {1989})}\BibitemShut {NoStop}%
\bibitem [{\citenamefont {Weigend}\ and\ \citenamefont {Gershenfeld}(1993)}]{santafe2}%
  \BibitemOpen
  \bibfield  {author} {\bibinfo {author} {\bibfnamefont {A.}~\bibnamefont {Weigend}}\ and\ \bibinfo {author} {\bibfnamefont {N.}~\bibnamefont {Gershenfeld}},\ }in\ \href {https://doi.org/10.1109/ICNN.1993.298828} {\emph {\bibinfo {booktitle} {IEEE International Conference on Neural Networks}}}\ (\bibinfo {year} {1993})\ pp.\ \bibinfo {pages} {1786--1793 vol.3}\BibitemShut {NoStop}%
\bibitem [{\citenamefont {Tomoda}\ \emph {et~al.}(2023)\citenamefont {Tomoda}, \citenamefont {Yoshida}, \citenamefont {Kashiwazaki}, \citenamefont {Umeki}, \citenamefont {Enomoto},\ and\ \citenamefont {Takeda}}]{Tomoda2023}%
  \BibitemOpen
  \bibfield  {author} {\bibinfo {author} {\bibfnamefont {H.}~\bibnamefont {Tomoda}}, \bibinfo {author} {\bibfnamefont {T.}~\bibnamefont {Yoshida}}, \bibinfo {author} {\bibfnamefont {T.}~\bibnamefont {Kashiwazaki}}, \bibinfo {author} {\bibfnamefont {T.}~\bibnamefont {Umeki}}, \bibinfo {author} {\bibfnamefont {Y.}~\bibnamefont {Enomoto}},\ and\ \bibinfo {author} {\bibfnamefont {S.}~\bibnamefont {Takeda}},\ }\href {https://doi.org/10.1364/OE.476025} {\bibfield  {journal} {\bibinfo  {journal} {Opt. Express}\ }\textbf {\bibinfo {volume} {31}},\ \bibinfo {pages} {2161} (\bibinfo {year} {2023})}\BibitemShut {NoStop}%
\bibitem [{\citenamefont {Kouadou}\ \emph {et~al.}(2023)\citenamefont {Kouadou}, \citenamefont {Sansavini}, \citenamefont {Ansquer}, \citenamefont {Henaff}, \citenamefont {Treps},\ and\ \citenamefont {Parigi}}]{10.1063/5.0156331}%
  \BibitemOpen
  \bibfield  {author} {\bibinfo {author} {\bibfnamefont {T.}~\bibnamefont {Kouadou}}, \bibinfo {author} {\bibfnamefont {F.}~\bibnamefont {Sansavini}}, \bibinfo {author} {\bibfnamefont {M.}~\bibnamefont {Ansquer}}, \bibinfo {author} {\bibfnamefont {J.}~\bibnamefont {Henaff}}, \bibinfo {author} {\bibfnamefont {N.}~\bibnamefont {Treps}},\ and\ \bibinfo {author} {\bibfnamefont {V.}~\bibnamefont {Parigi}},\ }\href {https://doi.org/10.1063/5.0156331} {\bibfield  {journal} {\bibinfo  {journal} {APL Photonics}\ }\textbf {\bibinfo {volume} {8}},\ \bibinfo {pages} {086113} (\bibinfo {year} {2023})},\ \Eprint {https://arxiv.org/abs/https://pubs.aip.org/aip/app/article-pdf/doi/10.1063/5.0156331/18095208/086113\_1\_5.0156331.pdf} {https://pubs.aip.org/aip/app/article-pdf/doi/10.1063/5.0156331/18095208/086113\_1\_5.0156331.pdf} \BibitemShut {NoStop}%
\bibitem [{\citenamefont {Cai}\ \emph {et~al.}(2021{\natexlab{b}})\citenamefont {Cai}, \citenamefont {Roslund}, \citenamefont {Thiel}, \citenamefont {Fabre},\ and\ \citenamefont {Treps}}]{Cai2021}%
  \BibitemOpen
  \bibfield  {author} {\bibinfo {author} {\bibfnamefont {Y.}~\bibnamefont {Cai}}, \bibinfo {author} {\bibfnamefont {J.}~\bibnamefont {Roslund}}, \bibinfo {author} {\bibfnamefont {V.}~\bibnamefont {Thiel}}, \bibinfo {author} {\bibfnamefont {C.}~\bibnamefont {Fabre}},\ and\ \bibinfo {author} {\bibfnamefont {N.}~\bibnamefont {Treps}},\ }\href {https://doi.org/10.1038/s41534-021-00419-w} {\bibfield  {journal} {\bibinfo  {journal} {npj Quantum Information}\ }\textbf {\bibinfo {volume} {7}},\ \bibinfo {pages} {82} (\bibinfo {year} {2021}{\natexlab{b}})}\BibitemShut {NoStop}%
\bibitem [{\citenamefont {Valdez}\ \emph {et~al.}(2017)\citenamefont {Valdez}, \citenamefont {Jaschke}, \citenamefont {Vargas},\ and\ \citenamefont {Carr}}]{PhysRevLett.119.225301}%
  \BibitemOpen
  \bibfield  {author} {\bibinfo {author} {\bibfnamefont {M.~A.}\ \bibnamefont {Valdez}}, \bibinfo {author} {\bibfnamefont {D.}~\bibnamefont {Jaschke}}, \bibinfo {author} {\bibfnamefont {D.~L.}\ \bibnamefont {Vargas}},\ and\ \bibinfo {author} {\bibfnamefont {L.~D.}\ \bibnamefont {Carr}},\ }\href {https://doi.org/10.1103/PhysRevLett.119.225301} {\bibfield  {journal} {\bibinfo  {journal} {Phys. Rev. Lett.}\ }\textbf {\bibinfo {volume} {119}},\ \bibinfo {pages} {225301} (\bibinfo {year} {2017})}\BibitemShut {NoStop}%
\bibitem [{\citenamefont {Sakurai}\ \emph {et~al.}(2022)\citenamefont {Sakurai}, \citenamefont {Estarellas}, \citenamefont {Munro},\ and\ \citenamefont {Nemoto}}]{PhysRevApplied.17.064044}%
  \BibitemOpen
  \bibfield  {author} {\bibinfo {author} {\bibfnamefont {A.}~\bibnamefont {Sakurai}}, \bibinfo {author} {\bibfnamefont {M.~P.}\ \bibnamefont {Estarellas}}, \bibinfo {author} {\bibfnamefont {W.~J.}\ \bibnamefont {Munro}},\ and\ \bibinfo {author} {\bibfnamefont {K.}~\bibnamefont {Nemoto}},\ }\href {https://doi.org/10.1103/PhysRevApplied.17.064044} {\bibfield  {journal} {\bibinfo  {journal} {Phys. Rev. Appl.}\ }\textbf {\bibinfo {volume} {17}},\ \bibinfo {pages} {064044} (\bibinfo {year} {2022})}\BibitemShut {NoStop}%
\bibitem [{\citenamefont {Braunstein}\ and\ \citenamefont {McLachlan}(1987)}]{braunstein1987generalized}%
  \BibitemOpen
  \bibfield  {author} {\bibinfo {author} {\bibfnamefont {S.~L.}\ \bibnamefont {Braunstein}}\ and\ \bibinfo {author} {\bibfnamefont {R.~I.}\ \bibnamefont {McLachlan}},\ }\href@noop {} {\bibfield  {journal} {\bibinfo  {journal} {Physical Review A}\ }\textbf {\bibinfo {volume} {35}},\ \bibinfo {pages} {1659} (\bibinfo {year} {1987})}\BibitemShut {NoStop}%
\bibitem [{\citenamefont {Macieszczak}\ \emph {et~al.}(2021)\citenamefont {Macieszczak}, \citenamefont {Rose}, \citenamefont {Lesanovsky},\ and\ \citenamefont {Garrahan}}]{macieszczak2021theory}%
  \BibitemOpen
  \bibfield  {author} {\bibinfo {author} {\bibfnamefont {K.}~\bibnamefont {Macieszczak}}, \bibinfo {author} {\bibfnamefont {D.~C.}\ \bibnamefont {Rose}}, \bibinfo {author} {\bibfnamefont {I.}~\bibnamefont {Lesanovsky}},\ and\ \bibinfo {author} {\bibfnamefont {J.~P.}\ \bibnamefont {Garrahan}},\ }\href@noop {} {\bibfield  {journal} {\bibinfo  {journal} {Phys. Rev. Res.}\ }\textbf {\bibinfo {volume} {3}},\ \bibinfo {pages} {033047} (\bibinfo {year} {2021})}\BibitemShut {NoStop}%
\bibitem [{\citenamefont {Brinkman}\ \emph {et~al.}(2022)\citenamefont {Brinkman}, \citenamefont {Yan}, \citenamefont {Maffei}, \citenamefont {Park}, \citenamefont {Fontanini}, \citenamefont {Wang},\ and\ \citenamefont {La~Camera}}]{brinkman2022metastable}%
  \BibitemOpen
  \bibfield  {author} {\bibinfo {author} {\bibfnamefont {B.~A.}\ \bibnamefont {Brinkman}}, \bibinfo {author} {\bibfnamefont {H.}~\bibnamefont {Yan}}, \bibinfo {author} {\bibfnamefont {A.}~\bibnamefont {Maffei}}, \bibinfo {author} {\bibfnamefont {I.~M.}\ \bibnamefont {Park}}, \bibinfo {author} {\bibfnamefont {A.}~\bibnamefont {Fontanini}}, \bibinfo {author} {\bibfnamefont {J.}~\bibnamefont {Wang}},\ and\ \bibinfo {author} {\bibfnamefont {G.}~\bibnamefont {La~Camera}},\ }\href@noop {} {\bibfield  {journal} {\bibinfo  {journal} {Applied Physics Reviews}\ }\textbf {\bibinfo {volume} {9}},\ \bibinfo {pages} {011313} (\bibinfo {year} {2022})}\BibitemShut {NoStop}%
\bibitem [{\citenamefont {Svensson}\ \emph {et~al.}(2018)\citenamefont {Svensson}, \citenamefont {Bengtsson}, \citenamefont {Bylander}, \citenamefont {Shumeiko},\ and\ \citenamefont {Delsing}}]{svensson2018period}%
  \BibitemOpen
  \bibfield  {author} {\bibinfo {author} {\bibfnamefont {I.-M.}\ \bibnamefont {Svensson}}, \bibinfo {author} {\bibfnamefont {A.}~\bibnamefont {Bengtsson}}, \bibinfo {author} {\bibfnamefont {J.}~\bibnamefont {Bylander}}, \bibinfo {author} {\bibfnamefont {V.}~\bibnamefont {Shumeiko}},\ and\ \bibinfo {author} {\bibfnamefont {P.}~\bibnamefont {Delsing}},\ }\href@noop {} {\bibfield  {journal} {\bibinfo  {journal} {App. Phys. Lett.}\ }\textbf {\bibinfo {volume} {113}},\ \bibinfo {pages} {022602} (\bibinfo {year} {2018})}\BibitemShut {NoStop}%
\bibitem [{\citenamefont {Chang}\ \emph {et~al.}(2020)\citenamefont {Chang}, \citenamefont {Sab\'{\i}n}, \citenamefont {Forn-D\'{\i}az}, \citenamefont {Quijandr\'{\i}a}, \citenamefont {Vadiraj}, \citenamefont {Nsanzineza}, \citenamefont {Johansson},\ and\ \citenamefont {Wilson}}]{forn2020three}%
  \BibitemOpen
  \bibfield  {author} {\bibinfo {author} {\bibfnamefont {C.~W.~S.}\ \bibnamefont {Chang}}, \bibinfo {author} {\bibfnamefont {C.}~\bibnamefont {Sab\'{\i}n}}, \bibinfo {author} {\bibfnamefont {P.}~\bibnamefont {Forn-D\'{\i}az}}, \bibinfo {author} {\bibfnamefont {F.}~\bibnamefont {Quijandr\'{\i}a}}, \bibinfo {author} {\bibfnamefont {A.~M.}\ \bibnamefont {Vadiraj}}, \bibinfo {author} {\bibfnamefont {I.}~\bibnamefont {Nsanzineza}}, \bibinfo {author} {\bibfnamefont {G.}~\bibnamefont {Johansson}},\ and\ \bibinfo {author} {\bibfnamefont {C.~M.}\ \bibnamefont {Wilson}},\ }\href {https://doi.org/10.1103/PhysRevX.10.011011} {\bibfield  {journal} {\bibinfo  {journal} {Phys. Rev. X}\ }\textbf {\bibinfo {volume} {10}},\ \bibinfo {pages} {011011} (\bibinfo {year} {2020})}\BibitemShut {NoStop}%
\end{thebibliography}%

\end{document}